\documentclass[12pt]{article}
\usepackage{graphicx}
\usepackage[utf8]{inputenc}
\usepackage{mathtools}
\usepackage{graphicx,psfrag,epsf,color}
\usepackage{amsmath,amssymb,amsfonts}
\usepackage{array}
\usepackage{cite}
\usepackage{float}
\usepackage{enumitem}
\usepackage{appendix}
\usepackage{bm}
\usepackage{ascii}
\usepackage{dsfont}
\usepackage{caption}
\usepackage{subcaption}
\usepackage[colorlinks]{hyperref}
\usepackage[dvipsnames]{xcolor}
\hypersetup{pageanchor=false,linkcolor=NavyBlue,citecolor=NavyBlue,urlcolor=RoyalPurple}

\renewcommand{\Im}{\textrm{Im}}
\renewcommand{\vec}[1]{\bm{#1}}
\newcommand{\cor}[2]{#1\langle #2 #1\rangle}
\newcommand{\p}{\partial}
\newcommand{\vp}{\varphi}
\newcommand{\ve}{\varepsilon}
\newcommand{\der}{\textrm{d}}
\newcommand{\Lo}{\mathcal{O}}
\newcommand{\Ht}{\hat t}
\newcommand{\Ld}{\mathcal{L}}
\newcommand{\Hd}{\mathcal{H}}

\newcommand{\Z}{\mathbb{Z}}

\newcommand{\tmu}{\tilde\mu}

\newcommand{\Li}{\textrm{Li}}

\newcommand{\nn}{\nonumber}
\newcommand{\secref}[1]{Sec.~\ref{#1}}
\newcommand{\figref}[1]{Fig.~\ref{#1}}
\newcommand{\appref}[1]{App.~\ref{#1}}
\renewcommand{\log}{\ln}

\numberwithin{equation}{section}

\begin{document}
\allowdisplaybreaks

\begin{titlepage}

\begin{flushright}
{\small
TUM-HEP-1592/26\\
CERN-TH-2026-020\\
10 March 2026 \\
}
\end{flushright}

\vskip1cm
\begin{center}
{\Large \bf Renormalisation and matching of massless scalar\\[0.2cm] 
correlation functions in Soft de Sitter Effective Theory}
\end{center}
  \vspace{0.5cm}
\begin{center}
{\sc Martin~Beneke,$^{a}$ \sc Patrick~Hager,$^{b}$ and Andrea~F.~Sanfilippo$^{c}$} 
\\[6mm]
{\it ${}^a$Physik Department T31,\\
James-Franck-Stra\ss e~1,
Technische Universit\"at M\"unchen,\\
D--85748 Garching, Germany}\\[0.2cm]
{\it ${}^b$CERN, Theoretical Physics Department, CH-1211 Geneva 23, Switzerland}\\[0.2cm]
{\it ${}^c$Departamento de F\'isica Te\'orica y del Cosmos, Universidad de Granada,}\\
{\it Campus de Fuentenueva, E–18071 Granada, Spain}
\end{center}
\vskip1cm

\begin{abstract}
\noindent 
For light and massless scalar fields, cosmological correlation functions suffer from infrared divergences and secular logarithms. Soft de Sitter Effective Theory (SdSET) has been proposed by Cohen and Green as the effective description of the non-trivial dynamics of long-wavelength modes $k_{\rm phys} < H$ in de Sitter space, which is responsible for the infrared and late-time logarithms, and as a systematic extension of the stochastic approach.  In this article, we construct SdSET  in dimensional regularisation, including an initial-condition functional.  We demonstrate by examples that renormalisation and matching works as for flat-space effective field theories. Adopting massless $\kappa \phi^4$ theory as the UV theory, we match the tree-level trispectrum and six-point function, and the one-loop power spectrum to SdSET, verifying explicitly that SdSET is the appropriate effective field theory for the quantum dynamics of superhorizon modes.
\end{abstract}

\end{titlepage}

\pagenumbering{roman}
{\hypersetup{hidelinks}
\pdfbookmark[1]{Contents}{ToC}
\setcounter{tocdepth}{2}
\small \tableofcontents}
\vspace{6mm}
\newpage
\pagenumbering{arabic}

\section{Introduction}

The perturbative expansions of correlation functions of the massless, minimally cou\-pled scalar field in de Sitter (dS) spacetime are infrared (IR) divergent irrespective of the spacetime dimension.
The divergence is associated with superhorizon modes characterised by physical momenta $k_{\mathrm{phys}} < H$. In comoving coordinates, introducing the conformal time $\eta$, these modes correspond to $-k\eta \to 0$, and the divergence can be understood as a pile-up of red-shifted long-distance modes at late times that can be seen already in the free two-point function. It features the mode integral $\int\der^{d-1}\vec{k}\,\lvert\phi_{k}\rvert^2$, which diverges logarithmically for any $d$ as the comoving momentum $\lvert \vec{k}\rvert \to 0$.
However, the free quantum theory is not interesting in itself, and there is evidence that when turning on self-interactions, the IR divergences cure themselves through ``dynamical mass generation''.
Indeed, turning to Euclidean de Sitter space (EdS), which is simply the sphere, the situation becomes simpler, as the sphere is compact, and consequently there is no late-time pile-up of infrared modes, and in turn no secular divergences.
The remaining IR divergences can be linked to the constant zero-mode which must be treated non-perturbatively in the path-integral~\cite{Rajaraman:2010xd,Beneke:2012kn}.
Once this mode is properly taken into account, one obtains a well-defined weak-coupling expansion. For self-interacting scalar fields, the parameter of this expansion is $\sqrt{\kappa}$ rather than the original scalar-field self-coupling $\kappa$, demonstrating the non-perturbative character of the problem.

The Lorentzian counterpart is more complicated, but its understanding is crucial for enabling precise predictions of late-time cosmological correlators, which correspond to the initial conditions that seed the large-scale structure of the Universe.
The main complication arises from the additional secular logs, i.e.\ divergences that appear at late times $\eta\to 0$, which are absent in EdS.
Unlike the zero-mode in EdS, these secular divergences originate from a continuous set of IR modes characterised by $k\eta \to 0$, and there is currently no way of resumming these using analytic continuation from Euclidean space.
One approach of handling this pile-up of modes has been presented over 40 years ago in the form of stochastic inflation~\cite{Starobinsky:1982ee,Starobinsky:1986fx,Starobinsky:1994bd}.
At the heart of this formalism is the scale separation of super- and subhorizon modes into classical stochastic variables and a Gaussian noise term, where the dynamics are then described using probability densities following a Fokker-Planck equation.
At leading order (LO), this approach successfully sums leading logarithmic 
IR and secular terms of correlation functions to all orders in the coupling, and confirms the Euclidean intuition, as the perturbative expansion is again performed in terms of $\sqrt{\kappa}$.
Following this seminal work, numerous attempts have been made to recover or extend this leading-order result in a quantum field-theory framework in widely different approaches, e.g.\ using explicit diagrams and the Schwinger-Keldysh path integral~\cite{Garbrecht:2014dca,Andersen:2021lii}, the wave-function of the Universe approach~\cite{Collins:2017haz,Cespedes:2023aal}, the density matrix~\cite{Mirbabayi:2019qtx}, the Yang-Feldman equation~\cite{Prokopec:2007ak},  diagrammatic resummation methods~\cite{Baumgart:2019clc}, or otherwise \cite{Moss:2016uix,Tokuda:2017fdh}.
The extension of the stochastic formalism to next-to-leading order (NLO) in logarithms 
and the full set of NLO corrections have been obtained in~\cite{Gorbenko:2019rza,Mirbabayi:2020vyt}. 

The nature of the problem, i.e.\ the need for scale separation and the importance of superhorizon modes suggests an effective field theory (EFT) framework that should provide clear rules on how to compute correlation functions to any accuracy, up to increasing computational but not conceptual difficulty.
The recently developed Soft de Sitter Effective Theory (SdSET)~\cite{Cohen:2020php,Cohen:2021fzf} provides such a physically intuitive framework to address these challenges.
It was used to compute some effects at next-to-next-to leading order (NNLO), showed clearly how structurally new features such as the full Kramers-Moyal equation arise beyond NLO, and identified operator mixing of an infinite number of relevant operators in the massless case as the origin of this equation.
Their treatment so far leaves out NNLO effects that arise from quantum corrections to terms that exist already at LO, such as the correction to the diffusion coefficient $H^3/(8\pi^2)$.
A consistent calculation requires a precise definition of the regularisation and renormalisation scheme in both the full and the effective theory, as well as an understanding of how scheme independence of observable quantities is manifested.

This work is motivated by our attempt to understand SdSET as a bona-fide EFT. We 
provide a derivation of the free theory in arbitrary spacetime dimension using canonical transformations, clarify the power counting, explicitly demonstrate the implementation of initial conditions in the path integral, employ dimensional regularisation, consistently renormalise the EFT, and carry out explicit matching calculations. In this way we obtain full control of scale and scheme dependence.
Our results fully confirm the physical insight and framework developed in the pioneering works~\cite{Cohen:2020php,Cohen:2021fzf}, but we aim to address issues that arise in practical calculations beyond the leading terms, and to 
establish closer connection to the tools used in EFTs in the high-energy particle-physics context, especially dimensional regularisation and the method of regions~\cite{Beneke:1997zp,Beneke:2023wmt}. We hope that this 
makes the approach more accessible to the particle-physics EFT community.

For reasons of length of presentation, this work is split in two parts.
The first part is organised as follows: \secref{sec::fullth} introduces the regularisation and renormalisation scheme in the full theory, while \secref{sec::SdSET} provides a pedagogical introduction to the formalism of SdSET. We present the explicit matching of the tree-level trispectrum in \secref{sec:trispectrum}. In \secref{sec:penta}, we compute and match for the first time the six-point function, demonstrating that both the method of regions and SdSET can indeed be consistently applied in situations featuring multiple vertices.
Finally, we determine the one-loop correction to the power spectrum in \secref{sec:pwr}.
These calculations demonstrate that SdSET possesses all necessary properties required to serve as an effective field theory with predictive power, suitable for precision applications in cosmology. They further determine counterterms and matching 
coefficients of the SdSET action and initial-condition functional, which eventually 
enter the late-time and IR resummation of the correlation functions at NNLO. 

For completeness, the paper includes  appendices~\ref{app::freered}--\ref{app::pwrloops}, which contain the technical details and explicit calculations for the results and derivations presented in the main text, together with the computation of the 
late-time limit of the six-point function in the full theory of the minimally coupled scalar field. 
The second part of this work~\cite{paperII} will be dedicated to the analysis of composite operators and explicit calculations of their matching coefficients. In particular, in that part we will present the quantum correction to the diffusion coefficient in the Fokker-Planck equation of the stochastic approach, which follows from a two-loop composite-operator anomalous dimension in SdSET.


\section{Minimally coupled scalar field}
\label{sec::fullth}

In the following, we consider the minimally coupled, real scalar field on a rigid dS background, which is matched onto Soft de Sitter Effective Theory.
The next sections are dedicated to the careful definition and construction of both the full and effective theory in dimensional regularisation and the respective renormalisation procedures.

In flat space, dimensional regularisation \cite{tHooft:1972tcz,Bollini:1972bi} has established itself as the standard method for handling ultraviolet divergences, as it provides a regulator without breaking any symmetries of the underlying theory or introducing additional scales.
Dimensional regularisation in dS space is more subtle~\cite{Weinberg:2005vy,Senatore:2009cf}, but when implemented in a way which respects the symmetries of dS spacetime, it provides a consistent scheme for the UV regularisation.
The starting point is the $d=4-2\varepsilon$-dimensional line element of dS space
\begin{equation}
    \der s^2 = g_{\mu\nu}(x)\der x^\mu \der x^\nu = \der t^2 - a(t)^2\der \vec{x}^2\,,
\end{equation}
with metric signature $(+,-,\dots,-)$. 
The scale factor $a(t)$ is given by $a(t)=e^{Ht}$, with Hubble parameter $H$. For computations in the full theory, it is often convenient to work in conformal time $\eta$, where $\der t = a(t)\der \eta$ and the scale factor reads $a(\eta) = -\frac{1}{H\eta}$.
As in flat space, the bare, regularised action is then simply the $d$-dimensional action
\begin{align}\label{eq:FullTheoryAction}
    S&=\int\der^dx\;\sqrt{-g}\bigg[\frac{1}{2}g^{\mu\nu}\p_{\mu}\phi_0\p_{\nu}\phi_0-\frac{1}{2}m^2_0\phi^2_0-\frac{\kappa_0}{4!}\phi^4_0\bigg]\,,\nonumber\\
    &=\int \frac{\der^dx}{(-H\eta)^d}\biggl[
    \frac{(-H\eta)^2}{2}\Bigl((\partial_\eta\phi_0)^2-(\partial_i\phi_0)^2\Bigr) - \frac{1}{2}m_0^2\phi_0^2 - \frac{\kappa_0}{4!}\phi_0^4
    \biggr]\,,
\end{align}
where we use the subscript ``0" to denote bare quantities. For the field with mass $m$, the $d$-dimensional mode functions in the Bunch-Davies vacuum read
\begin{equation}
    \phi_{\vec{k}}(\eta) = \sqrt{\frac{\pi}{4H}}\,(-H\eta)^{\frac{d-1}{2}} H_\nu^{(1)}(-k\eta)\,,
\end{equation}
where $H_\nu^{(1)}$ denotes the Hankel function of the first kind with parameter
 \begin{equation}
    \nu\equiv\sqrt{\bigg(\frac{d-1}{2}\bigg)^2-\frac{m^2}{H^2}}\,.
    \label{eq::nudef}
\end{equation}
The momentum-space Wightman function can be determined from the respective mode functions and reads
\begin{equation}
\cor{}{\phi(\eta,\vec k)\phi(\eta',-\vec k)}'=\frac{\pi}{4H}(-H\eta)^{\frac{d-1}{2}}(-H\eta')^{\frac{d-1}{2}}H^{(1)}_{\nu}(-k\eta)H^{(1)*}_{\nu}(-k\eta')\,,
\label{eq::mphiprop}
\end{equation}
where we are using the primed notation
\begin{equation}
\cor{}{\phi(\eta,\vec k_1)...\phi(\eta,\vec k_n)}\equiv(2\pi)^{d-1}\delta^{(d-1)}\bigg(\sum_{i=1}^n\vec k_i\bigg)\cor{}{\phi(\eta,\vec k_1)...\phi(\eta,\vec k_n)}'
\label{eq::corprime}
\end{equation}
that strips off the overall momentum-conserving delta function. We employ this notation when computing SdSET correlation functions as well. In the equal- and late-time limit $-k\eta\rightarrow 0$, \eqref{eq::mphiprop} reads 
\begin{equation}
    \lim\limits_{-k\eta\rightarrow0}\cor{}{\phi(\eta,\vec k)\phi(\eta,-\vec k)}'=H^{d-2}\bigg[(-\eta)^{2\alpha}\frac{C_{\nu}}{k^{2\nu}}-(-\eta)^{d-1}\frac{\cot(\pi\nu)}{2\nu}+(-\eta)^{2\beta}\frac{C_{-\nu}}{k^{-2\nu}}\bigg]\,,
    \label{eq::lateps}
\end{equation}
where we dropped $\mathcal{O}(\eta)$ corrections to each of the three non-analytic dependencies on $(-\eta)$ and defined
\begin{equation}\label{eq::abdef}
    \alpha\equiv\frac{d-1}{2}-\nu\,,\quad\beta\equiv\frac{d-1}{2}+\nu\,,\quad C_{\nu}\equiv\frac{4^{\nu-1}\Gamma(\nu)^2}{\pi}\,.
\end{equation}

The renormalisation procedure is standard, and the renormalised field, mass, and coupling are related to their bare counterparts as
\begin{equation}
    \phi_0(x)=\sqrt{Z_{\phi}}\phi(x)\,,\quad m^2_0=m^2+\delta m^2\,,\quad \kappa_0=\tmu_f^{4-d}Z_{\kappa}\kappa\,,
\end{equation}
respectively. 
In the above equation,
\begin{equation}
\tmu_f=\sqrt{\frac{e^{\gamma_E}}{4\pi}}\mu_f
\label{eq::muMSbar}
\end{equation}
is the $\overline{\textrm{MS}}$ renormalisation scale of the full theory, which is introduced such that the renormalised coupling $\kappa$ has vanishing mass dimension for all $d$. Using $\tmu_f$ ensures that all factors of the Euler-Mascheroni constant $\gamma_E$ and $\log(4\pi)$ that stem from momentum integrals systematically cancel. The counterterms $\delta_i$ are related to the renormalisation factors $Z_i$ via
\begin{equation}
    \delta_Z\equiv Z_{\phi}-1\,,\quad \delta_m\equiv(Z_{\phi}-1)m^2+Z_{\phi}\delta m^2\,,\quad \delta_{\kappa}\equiv Z_{\kappa}Z^2_{\phi}-1\,.
    \label{eq:fullTheoryRenormalisation}
\end{equation}
The full-theory mass counterterm was computed at the one-loop order in \cite{Beneke:2023wmt} and can be written as
\begin{equation}
\delta m^2=\frac{\kappa H^2}{8\pi^2}\bigg[-\frac{1}{2\ve}+\delta\hat m^2_{\textrm{fin}}\bigg]\,,
\label{eq::deltamfin}
\end{equation}
where $\delta\hat m^2_{\textrm{fin}}$ denotes the finite part, which defines the mass-renormalisation scheme. 

The above scheme applies to any value of mass $m$. In the following, however, we are interested in particular in the physics of the massless scalar field.
In this case, in $d=4$ dimensions $\nu=\frac{3}{2}$, and the Hankel function reduces to the elementary function
\begin{equation}\label{eq:HankelMassless4d}
    H^{(1)}_{\frac{3}{2}}(-k\eta) =\sqrt{\frac{2}{\pi}}\frac{e^{-ik\eta}(-i+k\eta)}{(-k\eta)^{\frac{3}{2}}}\,,
\end{equation}
but in $d=4-2\varepsilon$ dimensions $\nu=\frac{3}{2}-\varepsilon$. This makes computations quite cumbersome, as the Hankel function for generic $\nu$ is defined in terms of Bessel functions.
Therefore, we include as part of the bare mass in the action \eqref{eq:FullTheoryAction} the evanescent mass term\footnote{Recently, this scheme was used in \cite{Braglia:2025cee} in the context of the EFT of inflation \cite{Cheung:2007st}, and it was suggested that it is inconsistent, due to their finding of spurious logarithms of the conformal-time variable in the renormalised results. This point was addressed and clarified in \cite{Jain:2025maa} using toy models of a shift-symmetric scalar in rigid dS space. It was pointed out that, while the evanescent mass term breaks the shift symmetry, which is present in the EFT of inflation as well, the careful treatment of counterterms leads to consistent renormalised results, which respect unitarity, analyticity, and scale invariance. Since the model we are considering does not possess such a shift symmetry, the use of the evanescent mass should not lead to any subtleties, and we can proceed safely.}~\cite{Melville:2021lst} 
\begin{equation}
    m^2_d=\frac{H^2(d-4)(d+2)}{4}\,,
    \label{eq::md}
\end{equation}
which fixes $\nu=\frac{3}{2}$ in any spacetime dimension $d$, so that one employs~\eqref{eq:HankelMassless4d} for the mode functions.
In this scheme, the $d$-dimensional momentum-space Wightman function \eqref{eq::mphiprop} reduces to
\begin{equation}
\cor{}{\phi(\eta,\vec k)\phi(\eta',-\vec k)}'=(-H\eta)^{\frac{d-4}{2}}(-H\eta')^{\frac{d-4}{2}}H^2\,\frac{e^{ik(\eta'-\eta)}
(-i+k\eta)
(i+k\eta')}{2k^3}\,,
\label{eq::phiprop}
\end{equation}
and its late-time limit \eqref{eq::lateps} simplifies to 
\begin{equation}
    \lim\limits_{-k\eta\rightarrow0}\cor{}{\phi(\eta,\vec k)\phi(\eta,-\vec k)}'=H^{d-2}\bigg[\frac{(-\eta)^{d-4}}{2k^{3}}+\frac{(-\eta)^{d+2}}{18k^{-3}}\bigg]\,.
    \label{eq::m0lateps}
\end{equation}

The choice of a renormalisation scheme in this theory is subtle, as was already remarked in \cite{Beneke:2023wmt}. In fixed-order computations, imposing any renormalisation condition on the correlation functions inevitably leads to IR-divergent finite parts of the counterterms including the mass counterterm.
This is because the massless correlation functions are not well-defined when computed perturbatively, and do not qualify as sensible physical observables, which could be used to define physical renormalisation conditions.
Instead, in a physical scheme, these conditions should be imposed on the non-perturbative (``resummed") correlation functions.
Our strategy will therefore be the following: we only specify the pole parts of the counterterms but leave the finite parts undetermined.
When matching the EFT onto the full theory, the unspecified finite parts contribute through the late-time, small-momentum correlation functions and via the matching coefficients. The EFT then allows us to resum the respective correlation functions, which depend on the yet-unspecified quantities. These resummed correlators are the suitable objects to define physical renormalisation schemes. This in turn fully determines the finite parts of the counterterms such that they are free of IR divergences.

Due to the presence of these divergences, one also has to think about a suitable infrared regulator. While dimensional regularisation, combined with the evanescent mass term \eqref{eq::md}, can also handle these divergences,\footnote{For the strictly massless field in dimensional regularisation, i.e.\ using mode functions with $\nu=\frac{3}{2}-\varepsilon$, the appearing integrals e.g.\ for the one-loop power spectrum are not IR-regulated for any $\varepsilon$, as the $d$-dependence of the integration measure cancels with the one of the mode functions for small comoving momenta. Employing the evanescent mass term $m_d$ is equivalent to employing a separate analytic regulator in the Hankel functions using $\nu=\frac{3}{2}-\varepsilon+\delta$ and setting $\delta\to\varepsilon$.} we instead employ a comoving momentum-cutoff $\Lambda$ as IR regulator. This simplifies the identification of the UV divergences, which manifest as poles in $\ve$, and IR divergences of the correlation functions, which appear in the form of explicit power- and logarithmic divergences in the limit $\Lambda\rightarrow 0$. This IR regulator introduces dS-invariance-violating terms for the correlation functions, but they are of no consequence: since the same IR regulator will be implemented in the full theory and in SdSET, the dS-breaking terms drop out in the matching computation, because the EFT faithfully reproduces the IR structure of the full theory. The IR-divergent full theory correlation functions are simply the convenient quantities to extract the IR-insensitive matching coefficients of SdSET.

We implement the cutoff at the Lagrangian level, by modifying the kinetic term as
\begin{equation}
    S_{\mathrm{kin}} = \int \frac{\der^dx}{(-H\eta)^d}\biggl[
    \frac{(-H\eta)^2}{2}\Bigl((\partial_\eta\phi_0)^2-(\partial_i\phi_0)^2 - \Lambda_0^2\phi^2\Bigr) - \frac{1}{2}m_0^2\phi_0^2
    \biggr]\,,
    \label{eq::Lambdakin}
\end{equation}
where $\Lambda_0$ is the (bare) comoving cutoff. Just as we renormalise the bare mass parameter $m_0^2$ to remove logarithmic divergences proportional to $H^2$, we also renormalise the IR-regularisation parameter $\Lambda^2_0$ multiplicatively as
\begin{equation}
\Lambda^2_0= \Lambda^2(1+\delta\Lambda^2)
\end{equation} 
to remove UV-divergences proportional to powers of $\Lambda$. The counterterm $\delta\Lambda^2$ is assumed to be $\Lo(\kappa)$, and is treated perturbatively. In renormalised perturbation theory, the change of the mode functions~\eqref{eq:HankelMassless4d} and the Wightman function \eqref{eq::phiprop} due to the modification of the kinetic term is accounted for by shifting 
\begin{equation}
k\to k_\Lambda \equiv \sqrt{\vec{k}^2+\Lambda^2}
\label{eq::kLambda}
\end{equation}
in all arguments, as can be seen from \eqref{eq::Lambdakin}. The presence of $\Lambda$ in the mode functions now provides a scale for power-like UV divergences in loop integrals, which are therefore no longer set to zero by dimensional regularisation. 

Unlike the case of the counterterms associated with the renormalisation of the field, mass, and coupling, which contain physically relevant information, the $\Lambda$ regulator in the current scheme has no physical meaning by itself. Therefore, we can define the renormalisation scheme for this quantity simply by the requirement that the counterterm $\delta\Lambda^2$ should subtract both the pole term, as well as the finite terms associated with this pole that come with factors of $\Lambda$. This scheme choice will simplify the matching onto SdSET, and we will further comment on it below when discussing explicit results. 

Let us also note that one can give a loose physical interpretation to this IR-regularisation scheme: adding the term $-\frac{1}{2}(-H\eta)^2\Lambda^2\phi^2$ to the action corresponds to giving the scalar field a ``thermal mass''
\begin{equation}
m^2_{\textrm{th}}(\eta)=\frac{\Lambda^2}{a(\eta)^2} \sim T^2(\eta)\,,
\end{equation}
generated by coupling it to a thermal bath with comoving temperature $T(\eta)$. Such a term was briefly discussed in~\cite{Starobinsky:1982ee}, and in~\cite{Starobinsky:1986fx} it was noticed that it can give an effective IR regularisation of the theory. 

\section{Soft de Sitter Effective Theory}
\label{sec::SdSET}

Soft de Sitter Effective Theory~\cite{Cohen:2020php,Cohen:2021fzf} is the framework describing the late-time limit of equal-time correlation functions in a fixed dS background.
Its degrees of freedom are the superhorizon modes, i.e.\ modes satisfying $k/(a(t)H)\ll1$, and it follows the standard rules of flat-space effective field theories encountered in high-energy physics, such as HQET or SCET.
Most importantly, its construction is based on mode separation, power counting and symmetries. Fundamental to the construction is the small power-counting parameter
\begin{equation}
    \lambda\sim\frac{k}{a(t)H}=-k \eta \ll 1\,,
    \label{eq::pwrcount}
\end{equation}
which is used to expand the correlation functions and to assign appropriate scalings in powers of $\lambda$ to the modes appearing in the theory. This allows one to estimate the scaling of contributions to the correlation functions without an explicit calculation.
An unconventional feature of SdSET is that the power-counting parameter is time dependent in comoving momentum space. However, the EFT is constructed to compute the late-time limit of equal-time correlation functions~\cite{Cohen:2020php}, 
thus the time $\eta$ entering \eqref{eq::pwrcount} is unambiguously the correlation time, which serves as an external, fixed parameter.

In the following, we define and construct the effective theory in $d$ dimensions and explain in detail the regularisation and renormalisation procedure, as well as the matching to the full theory.
In addition to the pioneering works~\cite{Cohen:2020php,Cohen:2021fzf}, we rely heavily on the intuition gained by the method-of-regions analysis of correlation functions~\cite{Beneke:2023wmt} to identify all ingredients of the effective theory. 

\subsection{Free SdSET action}
\label{sec:freematch}

As in the full theory, we use dimensional regularisation for the UV divergences, and hence extend the four-dimensional effective action given in~\cite{Cohen:2020php} to $d$ dimensions. We further construct the kinetic (bilinear) terms to all orders in the power-counting parameter $\lambda$. In the following the scalar field is not required 
to have zero mass. 

As observed in~\cite{Cohen:2020php}, SdSET has a strikingly similar form to HQET, or the theory of a non-relativistic scalar field. In non-relativistic EFT, the full-theory real scalar 
field is represented by two real scalar fields or a complex scalar field in order to implement the symmetry leading to particle-number conservation. To preserve the number of degrees of freedom, the two fields (or the complex field and its complex conjugate) must form a pair of canonically conjugate variables. For SdSET,  one can adapt the construction of the non-relativistic action of~\cite{Namjoo:2017nia}, which uses a field redefinition relating the full-theory field $\phi$ to the effective fields $\varphi_{\pm}$, to obtain an all-order expression for the action. The detailed construction is presented in \appref{app::freered}. The main idea is to construct a canonical transformation relating the pair of canonically conjugated variables $\{\phi,\pi\}$ in the full theory to the pair $\{\vp_+,-2\nu\vp_-\}$ of SdSET fields. This allows one to identify the appropriate field redefinition relating the two pairs of variables, and to construct the free effective action of the SdSET without relying on power-counting arguments. At leading order in the gradient expansion, the field redefinition for $\phi$ in terms of $\vp_{\pm}$ reduces to
\begin{equation}
    \phi(\hat{t},\vec x)=H^{\frac{d}{2}-1}\Bigl[(a(\hat{t})H)^{-\alpha}\vp_+(\hat{t},\vec x)+(a(\hat{t})H)^{-\beta}\vp_-(\hat{t},\vec x)\Bigr] + \mathcal{O}(\partial_i^2\vp_{\pm})\,,
    \label{eq::fullphitoEFTphi}
\end{equation}
where we used the parameters introduced in \eqref{eq::abdef}.
Here and in the following, we always work with the rescaled time variable $\Ht\equiv Ht$, and drop the hat for brevity. Beyond the leading order in gradients, there is some freedom in the choice of the canonical transformation. The definition adopted in \appref{app::freered} is motivated by the resemblance to non-relativistic EFT and leads to the $d$-dimensional free action
\begin{equation}
S=-2\nu\int\der^{d-1}x\,\der t\;\vp_-\left[\dot\vp_+ + \nu\left(1-\sqrt{1+\frac{\p^2_i}{(\nu a(t)H)^2}}\right)\vp_+\right]\,.
\label{eq::fullSkin}
\end{equation}
The square root is defined by its series expansion. Taking $d=4$ and expanding in $\lambda$, the previous two equations agree with the ones of~\cite{Cohen:2020php}, but additionally contain the gradient corrections to all orders.
The effective fields satisfy the canonical, $d$-dimensional equal-time commutation relations
\begin{align}
    [\vp_+(t,\vec x),\vp_-(t,\vec y)]&=-\frac{i}{2\nu}\delta^{(d-1)}(\vec x-\vec y)
    \label{eq::comm1}\,,\\
    [\vp_+(t,\vec x),\vp_+(t,\vec y)]&=[\vp_-(t,\vec x),\vp_-(t,\vec y)]=0\,.
    \label{eq::comm2}
\end{align}
One immediately notices that $\vp_+$ and $-2\nu\vp_-$ are conjugate variables. In turn, the theory originally formulated as a second-order description turned into a first-order one, by trading the scalar field $\phi$ for $\vp_+$ and its conjugate momentum $-2\nu\vp_-$.
This is the natural formulation of a non-relativistic theory, which obeys a Schr\"odinger-like equation of motion.
Indeed, the fields satisfy the free equation of motion 
\begin{equation}
    \dot\vp_{\pm}(t,\vec x)=\mp\nu\left[1-\sqrt{1+\frac{\p_i^2}{(\nu a(t)H)^2}}\right]\vp_{\pm}(t,\vec x)\,,
    \label{eq::effEOM}
\end{equation}
which reduces at leading order in the gradients to
\begin{equation}
    \dot\vp_{\pm}(t,\vec x)=0+\Lo(\lambda^2)\,.
    \label{eq::leadEOM}
\end{equation}
As is commented upon in \appref{app::freered}, in the infinite-mass limit $m\rightarrow\infty$ the mass parameter becomes $\nu\rightarrow im/H$, the field decomposition 
\eqref{eq::fullphitoEFTphi} reduces to the non-relativistic decomposition presented in \cite{Namjoo:2017nia} with $\vp_+\rightarrow\psi^*$, $\vp_-\rightarrow\psi$, up to powers of $H$, and \eqref{eq::effEOM} turns into the free, non-relativistic equation of motion of \cite{Namjoo:2017nia}, with the replacement $\p_i\rightarrow\p_i/a(t)$. This result should be expected, since in this limit the curvature of the spacetime becomes negligible compared to the physical wavelength of the field modes of the EFT. The gradient expansion should therefore coincide with the non-relativistic expansion in flat spacetime. 
In the following we treat the gradient terms as power-suppressed interactions, hence the free theory is \eqref{eq::fullSkin} with the gradient terms set to zero and time-independent solutions to its equation of motion.

Eq.~\eqref{eq::leadEOM} implies that the effective fields are constant in time, as anticipated from the decomposition~\eqref{eq::fullphitoEFTphi}, 
but contains no information about their dependence on position (or momentum in Fourier-space). As shown in~\cite{Cohen:2020php}, bilinear products of creation $(\tilde a_{\vec k})$ and annihilation ($\tilde b_{\vec k}$) operators of the SdSET fields have non-vanishing, time-independent vacuum expectation values $\cor{}{\tilde a_{\vec k}\tilde a_{\vec k'}}$, $\cor{}{\tilde b_{\vec k}\tilde b_{\vec k'}}$. This feature is peculiar to SdSET, and the resulting non-trivial two-point functions of the effective fields $\vp_{\pm}$ must be provided as additional input to the EFT by specifying Gaussian initial conditions. 
In practice, this amounts to matching the free two-point functions of $\vp_+$ and $\vp_-$ at leading power in $\lambda$ to the full-theory two-point functions  in the Bunch-Davies vacuum, which can be done in $d$ dimensions for any value of the mass parameter $\nu$. Using the leading-power field decomposition~\eqref{eq::fullphitoEFTphi} for the Fourier-transformed field $\phi(t,\vec k)$, we find
\begin{align}
    \cor{}{\phi(t,\vec k)\phi(t,-\vec k)}'&=
    H^{d-2}\Big[(a(t)H)^{-2\alpha}\cor{}{\vp_+(t,\vec k)\vp_+(t,-\vec k)}'
    \nonumber\\
    &\phantom{=}+\,(a(t)H)^{-2\beta}\cor{}{\vp_-(t,\vec k)\vp_-(t,-\vec k)}'\nonumber\\
    &\phantom{=}+\,(a(t)H)^{1-d}\Big(\cor{}{\vp_+(t,\vec k)\vp_-(t,-\vec k)}'
    +\cor{}{\vp_-(t,\vec k)\vp_+(t,-\vec k)}'\Big)\Big]\,.
    \label{eq:fulltoEFT2ptmap}
\end{align}
Comparing this result to \eqref{eq::lateps} with $\eta=-\frac{1}{aH}$ determines the power spectra 
\begin{align}
\cor{}{\vp_+(t,\vec k)\vp_+(t,-\vec k)}'&=\frac{C_{\nu}}{k^{2\nu}}\,,\label{eq::EFTps1}\\
\cor{}{\vp_-(t,\vec k)\vp_-(t,-\vec k)}'&=\frac{C_{-\nu}}{k^{-2\nu}}\,,\label{eq::EFTps2}
\end{align}
as well as
\begin{equation}
    \cor{}{\vp_+(t,\vec k)\vp_-(t,-\vec k)}'+\cor{}{\vp_-(t,\vec k)\vp_+(t,-\vec k)}'=-\frac{\cot(\pi\nu)}{2\nu}\,.
    \label{eq::mixedps}
\end{equation}
This equation, together with the equal-time commutation relation \eqref{eq::comm1}, which implies
\begin{equation}
\cor{}{[\vp_+(t,\vec k),\vp_-(t,-\vec k)]}'=-\frac{i}{2\nu}\,,
\end{equation}
fixes the $\vp_+\vp_-$ equal-time two-point functions to be
\begin{equation}
\cor{}{\vp_{\pm}(t,\vec k)\vp_{\mp}(t,-\vec k)}'=-\frac{\cot(\pi\nu)\pm i}{4\nu}\,.
\label{eq::mixed2pt}
\end{equation}
Eqs.~\eqref{eq::EFTps1} and \eqref{eq::EFTps2} define the power spectra of the EFT fields $\vp_{\pm}$ at leading power in the gradient expansion. For $d=4$, they agree with the ones given in \cite{Cohen:2020php}.  We also note that the mixed two-point functions \eqref{eq::mixed2pt} are complex-valued, and therefore always appear in combination 
with each other when matching onto the real-valued full-theory correlators.

From the Fourier transformation
\begin{equation}
    \cor{}{\vp_{\pm}(t,\vec x)\vp_{\pm}(t,\vec y)}=\int\frac{\der^{d-1}k}{(2\pi)^{d-1}}e^{i\vec k\cdot(\vec x-\vec y)}\cor{}{\vp_{\pm}(t,\vec k)\vp_{\pm}(t,-\vec k)}'\,,
\end{equation}
one infers the power counting of the position-space fields~\cite{Cohen:2020php}
\begin{equation}\label{eq::vppower}
    \vp_+(t,\vec x)\sim\lambda^{\alpha}\,,\quad \vp_-(t,\vec x)\sim\lambda^{\beta}\,.
\end{equation}
The field redefinition containing all gradient terms, obtained in \appref{app::freered}, is consistent with this power counting, since all corrections to \eqref{eq::fullphitoEFTphi} are suppressed by powers of gradients, which count as $\lambda$. The power-counting of the position-space EFT fields coincides with their mass dimension
\begin{equation}
[\vp_+(t,\vec x)]=\alpha\,,\quad [\vp_-(t,\vec x)]=\beta.
\label{eq::EFTmassdim}
\end{equation}
With this power-counting of the effective fields, the leading-power term in the free effective action \eqref{eq::fullSkin} is
\begin{equation}
S_{\textrm{kin, lead}}=-2\nu\int\der^{d-1}x\der t\;\vp_-\dot\vp_+\sim\lambda^0\,.
\label{eq::Skinlead}
\end{equation}
This follows from $\vec x\sim\lambda^{-1}$, which in turn follows from \eqref{eq::pwrcount} and the fact that $\vec x$ is the canonically conjugated variable to $\vec k$, that $t\sim\lambda^0$ and $\alpha+\beta=d-1$. This is an important self-consistency property of the framework.

\begin{figure}[t]
\centering
\includegraphics[width=0.46\textwidth]{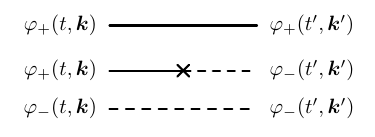}
\caption{Pictorial representation of the three $\vp_+\vp_+$, $\vp_+\vp_-$, $\vp_-\vp_-$ two-point functions used to connect the SdSET fields. The corresponding mathematical expressions for $\nu=\frac{3}{2}$ are given in \eqref{eq::nu32prop1}-\eqref{eq::nu32prop2}.}
\label{fig:props}
\end{figure}

The free theory \eqref{eq::Skinlead} forms the basis for the perturbative expansion of SdSET correlation functions, which is carried out in the Schwinger-Keldysh (SK) / Closed-Time-Path (CTP) formalism. We will use three types of lines to represent the free SdSET two-point functions (of any SK type), which are depicted in \figref{fig:props}. Since all free two-point functions are time-independent at leading power in $\lambda$, we can generalise the commutator \eqref{eq::comm1}, as well as the two-point functions \eqref{eq::EFTps1}, \eqref{eq::EFTps2} and \eqref{eq::mixed2pt} to unequal times:
\begin{align}
\cor{}{[\vp_+(t,\vec k),\vp_-(t',-\vec k)]}'&=-\frac{i}{2\nu}\label{eq::unequaltcomm}\,,\\
\cor{}{\vp_{\pm}(t,\vec k)\vp_{\pm}(t',-\vec k)}'&=\frac{C_{\pm\nu}}{k^{\pm 2\nu}}\,,\label{eq::EFTprop1}\\
\cor{}{\vp_{\pm}(t,\vec k)\vp_{\mp}(t',-\vec k)}'&=-\frac{\cot(\pi\nu)\pm i}{4\nu}\label{eq::EFTprop2}\,.
\end{align}
Each of the four two-point functions of the effective fields has four SK propagators associated to it, see App.~A of \cite{Beneke:2023wmt} for our conventions. However, since the field $\vp_+$ commutes with itself, as does $\vp_-$, all SK propagators associated to their two-point functions reduce to \eqref{eq::EFTprop1}. For the mixed two-point functions \eqref{eq::EFTprop2} the order of the fields matters, and the four SK propagators associated to each are inequivalent. 

While both the form of the kinetic term \eqref{eq::fullSkin} and the two-point functions \eqref{eq::EFTprop1}, \eqref{eq::EFTprop2} are valid for generic values of the mass, in the following we are interested in the physics of massless fields.
We introduce the same evanescent mass \eqref{eq::md} used in the full theory also in the EFT. This mass term only enters through $\nu$, setting it to $\nu=\frac{3}{2}$ for any dimension $d$. The two-point function of $\vp_+$ simplifies for any value of $d$ to
\begin{flalign}
    \cor{}{\vp_+(t,\vec k)\vp_+(t',-\vec k)}'&=\frac{1}{2k^3}\,,\label{eq::nu32prop1}\\
    \cor{}{\vp_{\pm}(t,\vec k)\vp_{\mp}(t',-\vec k)}'&=\mp\frac{i}{6}\,,
    \label{eq:nu32comm}\\
    \cor{}{\vp_-(t,\vec k)\vp_-(t',-\vec k)}'&=\frac{k^3}{18}\,.\label{eq::nu32prop2}
\end{flalign}
matching the leading term in \eqref{eq::m0lateps}. Furthermore, this choice of $\nu$ also fixes
\begin{equation}
\alpha=\frac{d-4}{2}\,,\quad \beta=\frac{d+2}{2}\,.
\label{eq::ouralphabeta}
\end{equation} 

\subsection{Symmetries of the effective fields}
\label{sec::SdSETsymm}

To characterise the effective fields $\vp_{\pm}$ and the field theory that describes their dynamics, we first work out the implications of the full-theory symmetries. The field $\phi$ is a scalar under the action of the isometries of dS space, which are spatial translations and rotations (T\&R), dilatations (D) and special conformal transformations (SCT), more details in \appref{app::dSiso}. This means that $\phi$ transforms as
\begin{equation}
\phi'(t',\vec x')=\phi(t,\vec x)
\label{eq::phiscalar}
\end{equation}
under their action. From this equation, the field decomposition \eqref{eq::fullphitoEFTphi}, and the transformation of $a=(-H\eta)^{-1}$, which can be inferred from the one of $\eta$ discussed in \appref{app::dSiso}, the transformation  of the effective fields $\vp_{\pm}$ under the dS isometries follows. Their active transformation was discussed in \cite{Cohen:2020php}, and a derivation of it can be found in \appref{app::dSiso}. For our purposes we will find the simultaneous transformation of the fields and their coordinate arguments useful, which reads
\begin{flalign}
\textrm{T\&R:}\quad \vp_{\pm}(t,\vec x)\rightarrow&\;\vp'_{\pm}(t',\vec x')=\vp_{\pm}(t,\vec x)\,,\label{eq::vptrafo1}\\
\textrm{D:}\quad \vp_{\pm}(t,\vec x)\rightarrow&\;\vp'_{\pm}(t',\vec x')=\bigg[1+\bigg(\frac{d-1}{2}\mp\nu\bigg)\delta\bigg]\vp_{\pm}(t,\vec x)\,,\label{eq::vptrafo2}\\
\textrm{SCT:}\quad\vp_{\pm}(t,\vec x)\rightarrow&\;\vp'_{\pm}(t',\vec x')=\Big[1-(d-1\mp2\nu)\vec b\cdot\vec x\Big]\vp_{\pm}(t,\vec x)\,,\label{eq::vptrafo3}
\end{flalign}
where $\delta$ and $\vec b$ are the infinitesimal parameters of dilatations and special conformal transformations, respectively.

Another important symmetry of the effective-theory fields is the field ``reparametrisation invariance" (RPI)  under transformations of the form \cite{Cohen:2020php}
\begin{equation}
\vp_+\rightarrow\vp_++\iota(a(t)H)^{\alpha-\beta}\vp_-\,,\quad \vp_-\rightarrow(1-\iota)\vp_-\,,
\label{eq::RPI}
\end{equation}
which follows from the field decomposition \eqref{eq::fullphitoEFTphi}. 
Here $\iota$ is an infinitesimal parameter. This transformation leaves the decomposition of $\phi$ in terms of $\vp_{\pm}$ invariant, even though it mixes the effective fields among each other. Correlation functions are also left invariant. As SdSET is constructed to reproduce these correlators, this transformation must be a symmetry of the EFT. Requiring that SdSET is reparametrisation-invariant enforces non-trivial relations among the terms of the SdSET action, linking them at different orders in the power-counting parameter $\lambda$.\footnote{RPI is broken by the canonical transformation that brings the action into the form of \eqref{eq::fullSkin}, since any specific choice of the representation of the full-theory canonical momentum  $\pi$ in \eqref{eq:pidef} is not invariant under \eqref{eq::RPI}. However, since the full-theory interactions are independent of $\pi$, the RPI relations hold for the SdSET interaction Lagrangian \eqref{eq::EFTint} below.}

\subsection{SdSET interaction terms and power-counting}
\label{sec::SdSETints}

For the minimally coupled scalar field, derivative interactions 
in SdSET are always of sub-leading power in $\lambda$. 
We therefore focus on the leading non-derivative interactions 
in the following. 
The bare, regularised EFT action containing all local interaction terms which are left invariant by the isometries \eqref{eq::vptrafo1}-\eqref{eq::vptrafo3} reads
\begin{equation}
S_{\textrm{int}}=-\int\der^{d-1}x\der t\;\sum_{n=1}^{\infty}\sum_{m=0}^{2n}a(t)^{2(n-1)\ve-3(m-1)}\frac{c^0_{2n-m,m}}{(2n-m)!m!}\,\vp^{2n-m}_{+,0}(t,\vec x)\vp^m_{-,0}(t,\vec x)\,,
\label{eq::EFTint}
\end{equation}
where $c^0_{2n-m,m}$ denote the bare couplings, $\vp_{\pm,0}$ the bare effective fields, and the values of $\alpha$ and $\beta$ for the massless scalar given in \eqref{eq::ouralphabeta} have been used. We restrict ourselves to interaction terms involving an even number of fields, since SdSET inherits the $\Z_2$-symmetry $\phi\rightarrow-\phi$ of the full theory.
The RPI transformation of the effective fields relates the effective couplings $c^0_{2n-m,m}$ with fixed $n$ to each other. Using \eqref{eq::RPI} with \eqref{eq::ouralphabeta}, and requiring that the interaction terms stay invariant under this infinitesimal transformation, we find 
\begin{equation}
c^0_{2n-m+1,m-1}=H^3c^0_{2n-m,m}\,.
\label{eq::RPIcouplings}
\end{equation}
This implies that for fixed $n$, all $2n$ effective couplings can be reduced to a single one.

Eq.~\eqref{eq::EFTint} includes interactions involving only $\vp_{+,0}$-fields.
These terms are problematic, since the (bare) equation of motion for $\vp_{-,0}$ at leading power in $\lambda$ is then of the form
\begin{equation}
    \dot\vp_{-,0}=\sum_{n=1}^{\infty}\,\vp^{2n-1}_{+,0}\bigg[a(t)^{3+2(n-1)\ve}\frac{c^0_{2n,0}}{(2n-1)!}+\Lo(\lambda^2)\bigg]\,.
\end{equation}
The left-hand side of this equation is subleading in the power-counting parameter $\lambda$ relative to the right-hand side, which naively leads to the conclusion that the theory is trivial,
\begin{equation}
    \vp_{\pm,0}=0\,.
\end{equation}
Another way to see this is that the interaction $\vp^n_{+,0}$ is super-leading with respect to the kinetic term of the action \eqref{eq::Skinlead}. 
It is therefore tempting to just exclude these interactions.
However, they do contribute to the matching computations and cannot be neglected.
Note that due to the commutation relations \eqref{eq::comm2}, $\vp^n_{+,0}$ interactions only contribute to correlation functions containing at least one $\vp_{-,0}$ field, which can be most easily seen using the multiple-commutator formula \cite{Weinberg:2005vy}. Hence, this interaction never results in a super-leading contribution when inserted into a correlation function. Its power-enhancement is precisely countered by the suppression of $\vp_{-,0}$.
Therefore, if such terms are included in the action, they do not render the theory trivial. This feature is not specific to SdSET, but related to the fact that we consider a theory where the field and its conjugate momentum differ in power counting. 

As a consequence, the naive power counting that is assigned by the rules \eqref{eq::vppower} using the free theory does not reflect the power-counting of correlators containing $\vp_{-,0}$-fields in the interacting theory.
For instance, correlation functions of $\vp_{+,0}$-fields and a single $\vp_{-,0}$-field involving an insertion of a $\vp^n_{+,0}$ interaction term contribute leading-power terms in $\lambda$, and are thus just as important as correlation functions involving only $\vp_{+,0}$, schematically  \begin{equation}
\cor{}{\vp_{+1}...\vp_{+m} \,[c_{n-1,1}\vp^{n-1}_+\vp_-]}\sim\cor{}{\vp_{+1}...\vp_{+ m-1}\vp_{-m}\,[c_{n,0}\vp^n_+]}
\sim\lambda^{\alpha(m+n-1)+\beta}\,.
\end{equation}
In order to be able to apply the power-counting rules \eqref{eq::vppower} also to correlation functions in the interacting theory, the interactions $\vp_{+,0}^n$ should be removed from the action. This can be achieved by a field redefinition of $\vp_{-,0}$~\cite{Cohen:2020php}, which in $d$ dimensions takes the form
\begin{equation}
\vp_{-,0}\rightarrow\vp_{-,0}+\sum_{n=0}^{\infty}a(t)^{3+2n\ve}\,r_n\vp^{2n+1}_{+,0}\,,
\label{eq::vpmredef}
\end{equation}
where $r_n$ are some constants still to be determined. 
The use of this field redefinition seems problematic at first glance, since it violates the power-counting rules~\eqref{eq::vppower}. However, the construction of the free EFT action performed above does not rely on power-counting arguments, as it is exact to all orders in the gradient terms. Therefore, it is consistent to use field redefinitions of the form \eqref{eq::vpmredef} to derive an action for the interacting theory which is free of the problematic terms.
From the perspective of the correlation functions, the redefinition~\eqref{eq::vpmredef} moves the insertions of the super-leading interactions in a $\varphi_-$ correlator into correlation function consisting only of $\vp_+$ fields.

After performing the redefinition, one obtains the appropriate basis of effective \textit{interacting} fields in which the EFT-correlators manifestly respect the desired power-counting \eqref{eq::vppower}.
However, the in-in correlators are \emph{not} invariant under field redefinitions.
Therefore, one must keep track of the field redefinitions when computing correlation functions. The derivation of the explicit field redefinition in $d$ dimensions is presented in \appref{app::intredef}. Up to terms proportional to $\vp^5_{+,0}$ we find
\begin{flalign}
\vp_{-,0}\rightarrow\;&\vp_{-,0}+(a(t)H)^3\,\Bigg\{\frac{c^0_{1,1}}{9}\vp_{+,0}+\frac{c^0_{3,1}a(t)^{2\ve}}{18(3+2\ve)}\vp^3_{+,0}+\frac{2a(t)^{4\ve}}{3+4\ve}\bigg[\frac{c^0_{5,1}}{6!}+\frac{(c^0_{3,1})^2}{108(3+2\ve)}\bigg]\vp^5_{+,0}\Bigg\}\nonumber\\
&+\Lo(\vp^7_{+,0})\,.
\label{eq::vpredefexplicit}
\end{flalign}
As explained in \appref{app::intredef}, the above expression contains all terms  which will be relevant for us in this paper.\footnote{We further  remark that the field redefinition that eliminates the superficially super-leading interactions $\vp_+^{2n}$ fixes the field basis, and hence the reparametrisation symmetry, in the sense that the coefficients of the $\vp_+^{2n-m}\vp_-^m$ Lagrangian terms after the field redefinition no longer obey the relations \eqref{eq::RPIcouplings}. Instead, we apply these relations {\em before} the field redefinition, so that all $c_{2n-m,m}^{0}$ are expressed in terms of $c_{2n-1,1}^{0}$, and {\em then} eliminate the 
$\vp_+^{2n}$ interaction.} The $d$-dimensional, interacting full-theory and bare EFT fields are then related as
\begin{flalign}
\phi(t,\vec x)=\;&H^{\frac{d}{2}-1}(a(t)H)^{\ve}\,\Bigg\{\bigg[1+\frac{c^0_{1,1}}{9}\bigg]\vp_{+,0}(t,\vec x)+\frac{c^0_{3,1}a(t)^{2\ve}}{18(3+2\ve)}\vp^3_{+,0}(t,\vec x)\nonumber\\
&+\frac{2a(t)^{4\ve}}{3+4\ve}\bigg[\frac{c^0_{5,1}}{6!}+\frac{(c^0_{3,1})^2}{108(3+2\ve)}\bigg]\vp^5_{+,0}(t,\vec x)+(a(t)H)^{-3}\vp_{-,0}(t,\vec x)\Bigg\}\,.
\label{eq::nonlinearphi}
\end{flalign}
This relation must be taken into account when matching full-theory correlation functions, as noted in~\cite{Cohen:2021fzf}. 
In particular, a correlator of single fields $\phi$ in the full theory is matched to a sum of different EFT-correlators including powers of the EFT field $\vp_{+,0}$. This is consistent with the power-counting rules \eqref{eq::vppower}, since $\vp_{+,0}$ has 
scaling dimension $-\varepsilon$, so any power of the field $\vp_{+,0}$ is relevant as $\varepsilon\to 0$. 

\begin{figure}[t]
\centering
\includegraphics[width=0.5\textwidth]{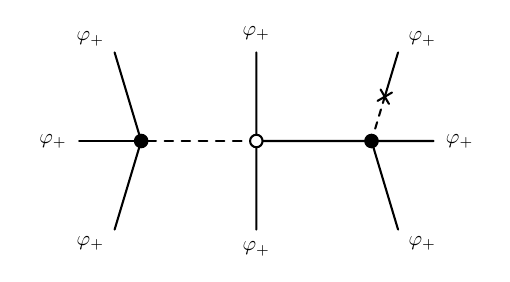}
\caption{Sample tree-level $\vp_+$ eight-point function diagram with three insertions of the $\vp^3_+\vp_-$ vertex of both Schwinger-Keldysh types containing the various propagators shown in \figref{fig:props}.}
\label{fig:SKvertices}
\end{figure}

When computing SdSET in-in correlation functions in the SK formalism, the action \eqref{eq::EFTint} is doubled, and appears in the weight of the SdSET path integral in the form
\begin{equation}
\sum_{\sigma=\pm}\sigma S_{\textrm{int}}[\vp^{\sigma}_{\pm}]=S_{\textrm{int}}[\vp^+_{\pm}]-S_{\textrm{int}}[\vp^-_{\pm}]\,,
\label{eq::SKint}
\end{equation}
where $\sigma$ is the CTP index. In the diagrammatic representation of the perturbative expansion of in-in correlators, the $(+)$- and $(-)$-type SK vertices are represented by full and empty dots, respectively. The fields at these vertices are connected to each other and to the external fields by means of the appropriate SK propagators discussed in \secref{sec:freematch} following the standard rules summarised in App.~A of \cite{Beneke:2023wmt}. Further, we use a diagrammatic representation that is closer to flat-space Feynman diagrams. Namely, we do not place the ends of all the external legs on a horizontal line at the top, as is done for Witten diagrams. This is exemplified in \figref{fig:SKvertices}. 

\subsection{Non-Gaussian initial conditions}
\label{sec::NGICs}

To correctly match the full-theory correlation functions to SdSET, non-Gaussian initial conditions (ICs) must be included~\cite{Cohen:2020php}. The reason is that the EFT is 
valid only at late times, but the full-theory time evolution from 
the Bunch-Davies vacuum at past infinity generates non-Gaussian correlations until 
the time when the matching is performed. Since at early times the theory is 
weakly coupled, the matching is perturbative. 

To implement initial conditions into the SdSET path integral, we adapt the formalism presented in \cite{Garny:2009ni} to our setting. The initial time slice for SdSET is located at some finite time $t_*$. It plays the role of a time factorisation scale, in that it separates the evolution of the fluctuations captured in the IC from the ones described by the EFT, i.e.\ it separates early-time from late-time effects when thinking in terms of the method of regions \cite{Beneke:2023wmt}.
Physically,  $t_*$ may be thought of as being close to the time of horizon crossing of the momentum modes $k$ described by the EFT fields, $t_*\sim t_H(k)$. This is where the initial-condition functional  should be localised. We add the factor $e^{i\mathcal{F}}$ to the SdSET path integral and make the following ansatz for $\mathcal{F}$:
\begin{flalign}
\mathcal{F}[\vp_{\pm}]= &\sum_{n=1}^{\infty}\sum_{\sigma_1,...,\sigma_{2n}=\pm}\int\bigg[\prod_{j=1}^{2n}\der^{d-1}x_j\bigg]\;\sum_{m=0}^{2n}\;\frac{a^{n(6-2\ve)-3m}_*}{(2n-m)!m!}\,\Xi^{\sigma_1,...,\sigma_{2n};0}_{2n-m,m}(\vec x_1,...,\vec x_{2n})
\nonumber\\
&\times\prod_{k=m+1}^{2n}\vp^{\sigma_k}_{+,0}(t_*,\vec x_k)\,\prod_{l=1}^m\vp^{\sigma_l}_{-,0}(t_*,\vec x_l)\,.
\label{eq::ICfunctional}
\end{flalign}
Similarly to the case of the EFT couplings, we added the superscript ``0" to the definition of the bare initial-condition functions $\Xi^{\sigma_1,...,\sigma_{n};0}_{n-m,m}$, since we anticipate that they will need to be renormalised. We also implemented the restrictions implied by the $\Z_2$-symmetry under $\phi\rightarrow-\phi$ of the full theory. The functional $\mathcal{F}$ receives a series of corrections in {\em positive} powers of $\eta$ (more precisely, $\eta$ times momenta in momentum space), with the same field content as in \eqref{eq::ICfunctional}, from higher-order terms in the expansion in the correlation time $\eta$ in the early-time region, as discussed around Eq.~(4.7) in~\cite{Beneke:2023wmt}. Since these terms are power-suppressed relative to the corresponding leading terms, and disappear when $\eta\to 0$, we do not write them here.

The $\Xi^{\sigma_1,...,\sigma_n;0}_{n-m,m}$ parameterise a Hermitian density matrix describing the initial state of the theory, 
so they must satisfy \cite{Garny:2009ni}
\begin{equation}
i\Xi^{\sigma_1,...,\sigma_n;0}_{n-m,m}(\vec x_1,...,\vec x_n)=\Big[i\Xi^{-\sigma_1,...,-\sigma_n;0}_{n-m,m}(\vec x_1,...,\vec x_n)\Big]^*\,.
\label{eq::IChermitean}
\end{equation}
In principle, all possible combinations of the CTP indices $\sigma_i$ are allowed. 
However, for the purpose of calculating equal-(late-)time correlation functions at given correlation time, we find for all examples considered that it is sufficient to match the underlying $\phi^4$ theory to an effective theory involving only interactions with all-plus or all-minus fields. Therefore, it will be enough to restrict the IC functions (and Lagrangian interactions as was assumed in \eqref{eq::SKint}) to the ones involving all-plus and all-minus CTP indices only. It is plausible that this holds generally, since the effective Lagrangian and IC functional by construction account for weakly coupled large momentum and early time correlations generated by the unitary and conservative dynamics of the underlying theory. The IC function can therefore be written in terms of a single index $\sigma=\pm$ as
\begin{equation}
\Xi^{\sigma_1,...,\sigma_n;0}_{n-m,m}(\vec x_1,...,\vec x_n)=\bigg[\prod_{i=1}^n\delta_{\sigma_i\sigma}\bigg]\,\Xi^{\sigma;0}_{n-m,m}(\vec x_1,...,\vec x_n)\,.
\end{equation}
Since they should be real-valued functions of the position arguments, \eqref{eq::IChermitean} implies that they flip their sign with $\sigma$,
so we can express both $\Xi^{\pm;0}_{n-m,m}$ through a single, bare initial-condition function\footnote{The following equation holds under the stated assumption that the IC functions are real-valued as is the case for the present theory. However, reality is not enforced by \eqref{eq::IChermitean}, and  may be too restrictive in other settings, see the discussion in \cite{Fiore:2026ivs}.} $\Xi^0_{n-m,m}$
\begin{equation}
\Xi^{\sigma;0}_{n-m,m}(\vec x_1,...,\vec x_n)=\sigma\,\Xi^0_{n-m,m}(\vec x_1,...,\vec x_n)\,.
\end{equation}

For practical calculations, we represent the IC functions $\Xi^0_{n-m,m}$ in momentum space. Using translation invariance, one finds 
\begin{flalign}
\Xi^0_{n-m,m}(\vec x_1,...,\vec x_n)&=\int\bigg[\prod_{i=1}^n\frac{\der^{d-1}k_i}{(2\pi)^{d-1}}\bigg]\,(2\pi)^{d-1}\delta^{(d-1)}\bigg(\sum_{l=1}^n\vec k_l\bigg)\exp\bigg[i\sum_{j=1}^n\vec k_j\cdot\vec x_j\bigg]
\nonumber\\[0.2cm] 
&\phantom{=}\times\, a^{-(n-1)(3-2\ve)}_*\,\Xi^0_{n-m,m}(\vec k_1,...,\vec k_n)\,.
\end{flalign}
Therefore, introducing the momentum-space functional
\begin{flalign}
F[\vp_{\pm}]\equiv&\sum_{n=1}^{\infty}\int\bigg[\prod_{j=1}^{2n}\frac{\der^{d-1}k_j}{(2\pi)^{d-1}}\bigg]\,(2\pi)^{d-1}\delta^{(d-1)}\bigg(\sum_{l=1}^{2n}\vec k_l\bigg)
\label{eq::ICF}\\
&\times \sum_{m=0}^{2n}\;\frac{a^{2(n-1)\ve+3(1-m)}_*}{(2n-m)!m!}\,\Xi^{0}_{2n-m,m}(\vec k_1,...,\vec k_{2n})
\,\prod_{l=m+1}^{2n}\vp_{+,0}(t_*,\vec k_l)
\prod_{q=1}^m\vp_{-,0}(t_*,\vec k_q)\,,
\nonumber
\end{flalign}
the IC functional \eqref{eq::ICfunctional} to be added to the SdSET path integral may be reduced to
\begin{equation}
\mathcal{F}[\vp_{\pm}]\rightarrow\sum_{\sigma=\pm}\sigma F[\vp^{\sigma}_{\pm}]=F[\vp^+_{\pm}]-F[\vp^-_{\pm}]\,,
\vspace*{-0.2cm}
\end{equation}
in full analogy to \eqref{eq::SKint}.
From this expression we see that the non-Gaussian initial conditions can be treated as non-local, momentum-conserving vertices. Moreover, we see that 
the SK formalism with two types of vertices (filled and empty) works exactly the same as for the Lagrangian interactions. Following \cite{Garny:2009ni}, the insertion of any $(+)$- and $(-)$-type IC $\Xi^0_{n-m,m}$ in SdSET correlation functions is depicted diagrammatically by means of full and empty polygons, respectively, with $n$ corners.

Eq.~\eqref{eq::ICfunctional} depends explicitly on $t_*$, the parameter that defines the time slice at which the initial conditions for the time evolution of SdSET correlation functions should be matched to the full theory, but since the bare IC functional $\mathcal{F}$ does not depend on the precise value of $t_*$,  this dependence on the time factorisation scale cancels in the product and sum of all terms on the right-hand side, including the IC functions, for which the $t_*$ argument has been omitted for simplicity. 
The factors of $a_*$ in \eqref{eq::ICfunctional} are motivated in analogy to the factors of $a(t)$ present in the vertices in Sec.~\ref{sec::SdSETints}, since the exact (all-order) IC functional should be invariant under the  dS isometries. We already observed above that the combinations $a(t)^{-\frac{3}{2}\pm\nu}\vp_{\pm}(t,\vec x)$ are dS-invariant by themselves. The factors of $a_*$ then compensate for the transformation of all other objects appearing in \eqref{eq::ICfunctional} under dS dilatations, while ``dS boost'' invariance implies relations among IC functions with different power-like $a_*$ dependence. 

There is an explicit power-like dependence on $a_*$ even in $d=4$ for $m\not= 1$ due to the factor $a_*^{3(1-m)}$ in \eqref{eq::ICF}. Since the full IC functional is $a_*$-independent, this implies an implicit power-like dependence on $a_*$ of the bare IC functions $\Xi^0_{2n-m,m}$, since the fields $\vp_{\pm}$ depend on $t_*$ only through power-suppressed gradient corrections. This complication can be avoided by putting $t_*\to t$ in the IC functional \eqref{eq::ICF} except for the factor $a_*^{2 (n-1)\varepsilon}$. In this case $a_*$ appears only in logarithms and the systematics of matching can be performed as usual in dimensional regularisation. While this choice defies the physical interpretation of the IC functional being set at horizon crossing, it comes with manifest power counting and close relation to the method-of-region computation \cite{Beneke:2023wmt}, where in fact the early-time region corresponds to expanding the correlator around the singular value at future infinity, $\eta=0$. However, as explained in the following paragraph, unless one is interested in power-suppressed effects that vanish in the late-time limit, only IC functions with $m=1$ are relevant and there is no difference between the above two prescriptions.

As was the case for the Lagrangian interactions, the IC functional contains the superficially super-leading terms $\Xi_{2n,0}\,\vp_+^{2n}$. Again, the enhancement is spurious, since these terms contribute to correlators only through a commutator with a $\vp_-$-field, which implies $\lambda^3$-suppression. However, unlike the $c_{2n,0}$ couplings, the $\Xi_{2n,0}$ IC functions cannot be removed by a local field redefinition, since this freedom has already been fully exploited. Instead we find that full-theory correlation functions can be consistently matched to SdSET by setting $\Xi_{2n,0}$ to zero. To make this point, we note that for the tree-level four-point correlation function, which will be explicitly computed below, one obtains from the insertion of the IC functional \eqref{eq::ICF}
\begin{equation}
\frac{(a_*\tmu)^{2\ve}}{24k_1^3\dots k_4^3}\sum_{i=1}^4k_i^3\,\biggl(\Xi^0_{3,1}(\vec{k}_1,\dots,\vec{k}_4) - \frac{1}{H^3}\biggl(\frac{a_*}{a(t)}\biggr)^{\!3}\,\Xi^0_{4,0}(\vec{k}_1,\dots,\vec{k}_4)\biggr)+\ldots\,.
\label{eq:Xi3140}
\end{equation}
The explicit factor $a_*^3$ in front of the second term is cancelled by an 
implicit $1/a_*^3$ dependence of $\Xi^0_{4,0}$. Time-independence of $\Xi^0_{4,0}$ and dS scaling then imply that the second term in brackets is suppressed by the factor $(k/(a(t) H))^3\sim \lambda^3$. 
The additional power suppression arises from the assumption that the IC functional \eqref{eq::ICfunctional} does not contain terms with negative powers of $\eta$. Matching at leading power therefore determines  $\Xi^0_{3,1}$, while $\Xi^0_{4,0}$ is degenerate with irrelevant power corrections and can be dropped. On the other hand, the scheme that employs $a_*^{2(n-1)\ve} a(t)^{3(1-m)}$ in \eqref{eq::ICF} introduces a term proportional to $\eta^{-3}$ into \eqref{eq::ICfunctional}, which is technically not excluded, since it contributes to large-time correlation functions only logarithmically in 
$(-\eta)$. In this case, the round bracket in \eqref{eq:Xi3140} reads 
$\Xi^0_{3,1}-\frac{1}{H^3}\,\Xi^0_{4,0}$
and one can show that $\Xi^0_{3,1}$ and $\Xi^0_{4,0}$ {\em always} appear in this combination. While one is now in principle free to 
assign any value to the two IC functions as long as the above difference remains the same, a consistent and convenient choice is $\Xi_{4,0}=0$, since in this case the IC functional does not contain naively power-counting violating terms. Thus independent of the two schemes for the IC functional, one can set $\Xi_{2n,0}=0$, and this will be assumed in the following 
for all $\Xi_{2n,0}$. IC functions with $m>1$ are not relevant in the late-time limit and are not of interest here. This discussion shows that there is a redundancy in the IC functional, which can be resolved only when considering power corrections and remains to be explored.

In general, SdSET diagrams involve a combination of vertices and IC functions. Since all IC terms are localised in time at $t_*$, which is earlier than any time after horizon crossing, the (anti-)time ordering appearing in the propagators can always be resolved uniquely. Propagators connecting ICs to each other involve (anti-)time-ordered two-point functions of $\vp_+$ and $\vp_-$ evaluated at equal time $t_*$, which are ambiguous, since the two fields do not commute. Adhering to the definition of the (anti-)time-ordering operations in App.~A of \cite{Beneke:2023wmt}, and assigning the value $\theta(0)=\frac{1}{2}$, together with \eqref{eq::mixedps}, this leads to
\begin{equation}
\cor{}{T\{\vp_{\pm}(t,\vec k)\vp_{\mp}(t,-\vec k)\}}'=\cor{}{\overline T\{\vp_{\pm}(t,\vec k)\vp_{\mp}(t,-\vec k)\}}'=-\frac{\cot(\pi\nu)}{4\nu}\stackrel{\nu=\frac{3}{2}}{\,=\,\,}0\,.
\label{eq::mixedequalt}
\end{equation}
Different definitions of the (anti-)time-ordering operation can be used as well, and amount to a change in the matched values of the IC functions, ensuring that the complete SdSET correlation function always reproduces its full-theory counterpart.

\subsection{Regularisation, renormalisation, and matching}
\label{sec:SdSETreg}

\subsubsection{Regularisation scheme}

We already set up the theory in $d=4-2\varepsilon$ dimensions, supplemented by the same evanescent mass term \eqref{eq::md} that we use in the full theory, to be able to regularise the UV divergences using dimensional regularisation.
In SdSET, the time integrals associated with vertices generate UV divergences when inserted in diagrams, which is not the case in the full theory. This is essentially due to the fact that the two-point functions of EFT fields are time-independent, and it 
implies that there is renormalisation already at tree level. Loop integrals lead, in general, to UV- and IR-divergent momentum integrals. 

At leading power in $\lambda$, the SdSET interaction terms \eqref{eq::EFTint} reduce to
\begin{equation}
S_{\textrm{int}}=-\int\der^{d-1}x\der t\;\sum_{n=0}^{\infty}a(t)^{2n\ve}\frac{c^0_{2n+1,1}}{(2n+1)!}\,\vp^{2n+1}_{+,0}(t,\vec x)\vp_{-,0}(t,\vec x)\,.
\label{eq::Sintleadill}
\end{equation} 
All summands, except for the $n=0$ one, feature $a(t)$ with a dimension-dependent exponent which regularises the divergent time integrals, since  \cite{Cohen:2020php}
\begin{equation}
\int_{-\infty}^t\der t'\;a(t')^{p+\ve}\equiv\frac{a(t)^{p+\ve}}{p+\ve}
\label{eq::EFTtints}
\end{equation}
for any value of $p$, as long as the dimensional-regularisation parameter $\ve$ does not drop out of the exponent. Since the summand with $n=0$ in \eqref{eq::Sintleadill} has no factor of $a(t)$, the associated time integral is not regularised. This is the same accidental cancellation of the dimension-dependent terms in the mass counterterm insertion diagrams that was observed in \cite{Beneke:2023wmt}, which can be identified as a peculiarity of the regularisation scheme. We therefore introduce an additional analytic regulator
\begin{equation}
\bigg(\frac{\nu}{a(t)H}\bigg)^{\!-2\delta}
\label{eq::EFTanareg}
\end{equation}
in the bilinear interaction term appearing in \eqref{eq::Sintlead}, such that $S_{\textrm{int}}$ becomes
\begin{flalign}
S_{\textrm{int}}=-\int\der^{d-1}x\der t\;\bigg[&c^0_{1,1}\bigg(\frac{\nu}{a(t)H}\bigg)^{\!-2\delta}\vp_{+,0}(t,\vec x)\vp_{-,0}(t,\vec x)\nonumber\\
&+\sum_{n=1}^{\infty}a(t)^{2n\ve}\frac{c^0_{2n+1,1}}{(2n+1)!}\,\vp^{2n+1}_{+,0}(t,\vec x)\vp_{-,0}(t,\vec x)\bigg]\,.
\label{eq::Sintlead}
\end{flalign} 
The explicit powers of $a(t)$ regularise the time integrals, while the $d$-dependence of the spatial integration measure regularises the loop integrals.

Similarly, written in terms of momentum-space fields, the leading-power IC functional \eqref{eq::ICF} reads
\begin{flalign}
F[\vp_{\pm}]&=\sum_{n=0}^{\infty}\int\bigg[\prod_{j=1}^{2n+2}\frac{\der^{d-1}k_j}{(2\pi)^{d-1}}\bigg]\;(2\pi)^{d-1}\delta^{(d-1)}\bigg(\sum_{l=1}^{2n+2}\vec k_l\bigg)
\frac{a^{2n\ve}_*}{(2n+1)!}\,\Xi^0_{2n+1,1}(\vec k_1,\dots,\vec k_{2n+2})
\nonumber\\[0.1cm]
&\phantom{=}\times\, \bigg[\prod_{l=1}^{2n+1}\vp_{+,0}(t_*,\vec k_l)\bigg]\,\vp_{-,0}(t_*,\vec k_{2n+2})\,,
\end{flalign}
and the $d$-dependence of the momentum-integration measure regularises the loop integrals into which the IC functions are inserted. It will prove convenient to mirror the addition of the analytic regulator \eqref{eq::EFTanareg} in the bilinear term of the IC functional, by multiplying it with $(\frac{\nu}{a_*H})^{-2\delta}$, so we use the following modified form of $F[\vp_{\pm}]$ for the explicit computations below
\begin{flalign}
F[\vp_{\pm}]&=\int\frac{\der^{d-1}k}{(2\pi)^{d-1}}\;\bigg(\frac{\nu}{a_*H}\bigg)^{\!-2\delta}\Xi^0_{1,1}(\vec k,-\vec k)\,\vp_{+,0}(t_*,\vec k)\vp_{-,0}(t_*,-\vec k)\nonumber\\
&\phantom{=}+\,\sum_{n=1}^{\infty}\int\bigg[\prod_{j=1}^{2n+2}\frac{\der^{d-1}k_j}{(2\pi)^{d-1}}\bigg]\;(2\pi)^{d-1}\delta^{(d-1)}\bigg(\sum_{l=1}^{2n+2}\vec k_l\bigg)
\frac{a^{2n\ve}_*}{(2n+1)!}\,\Xi^0_{2n+1,1}(\vec k_1,...,\vec k_{2n+2})
\nonumber\\
&\phantom{=}\quad\times\,\bigg[\prod_{l=1}^{2n+1}\vp_{+,0}(t_*,\vec k_l)\bigg]\,\vp_{-,0}(t_*,\vec k_{2n+2})\,.
\label{eq::ICFmom}
\end{flalign}
We effectively implemented the same UV-regularisation scheme in SdSET as in the full theory, even though this is not a requirement,
since the difference between the two schemes used in the full and effective theories can always be compensated by the matching coefficients. 

For the IR regularisation, however, it is crucial that the same scheme is used in the full and the effective theory. SdSET is constructed to faithfully reproduce the infrared structure of the full theory, and this includes all of the IR-divergent terms in correlation functions of the massless, minimally coupled scalar field when they are computed perturbatively. Since the divergent terms appear on both sides of the matching equations between full-theory and EFT correlators, they will cancel in the matching, provided the same IR regulator is employed. 
This guarantees that the matching coefficients are IR-insensitive and can be computed 
from IR-divergent correlation functions.

Therefore, we need to repeat the matching procedure onto the free SdSET of  \secref{sec:freematch}, now for the free full theory with the modified action \eqref{eq::Lambdakin}. The presence of $\Lambda$ in the full theory amounts to the global replacement $-\p^2_i\rightarrow-\p^2_i+\Lambda^2$. This feature persists throughout the matching procedure, so we find the modified free SdSET action
\begin{equation}
S_{\Lambda}=-3\int\der^{d-1}x\,\der t\;\vp_-\bigg[\dot\vp_+ + \frac{3}{2}\bigg(1-\sqrt{1+\frac{\p^2_i-\Lambda^2}{(\nu a(t)H)^2}}\bigg)\vp_+\biggr]\,,
\end{equation}
where we used that $\nu=\frac{3}{2}$ for any $d$. The matching of the $\vp_{\pm}$ power spectra leads to the same expressions as in \eqref{eq::nu32prop1}-\eqref{eq::nu32prop2} with the replacement
\begin{equation}
k\rightarrow k_{\Lambda}
\end{equation}
with $k_\Lambda$ defined in \eqref{eq::kLambda}. In particular, the two-point function \eqref{eq::nu32prop1} of $\vp_+$ for $\nu=\frac{3}{2}$ now reads 
\begin{equation}
\cor{}{\vp_+(t,\vec k)\vp_+(t,\vec k')}=(2\pi)^{d-1}\delta^{(d-1)}(\vec k+\vec k')\,\frac{1}{2k_{\Lambda}^3}\,.
\end{equation}
The two-point function \eqref{eq::nu32prop2} of $\vp_-$ is modified analogously.
An important feature of this IR-regularisation scheme is that the canonical commutation relations of the SdSET fields \eqref{eq::comm1}, \eqref{eq::comm2} are left unchanged. The proof is given in \appref{app:comm}. Therefore, the mixed two-point functions \eqref{eq:nu32comm} are unchanged as well. 

\subsubsection{Renormalisation of fields and couplings}

The treatment of UV divergences follows the standard approach in quantum field theory.
One renormalises the bare fields as
\begin{equation}
    \vp_{\pm,0} = \sqrt{Z_{\vp}}\vp_{\pm}\,,
\end{equation}
where the wave-function renormalisation factors $Z_{\vp}$ for the two effective fields $\vp_\pm$ must be equal due to RPI~\eqref{eq::RPI}. 
Regarding the renormalisation of the effective couplings, we first note that given the  dimension of the effective fields \eqref{eq::EFTmassdim} the bare, $d$-dimensional couplings have mass dimension
\begin{equation}
[c^0_{2n-m,m}]=2(n-1)\ve+3(1-m)\,.
\end{equation}
For $\ve=0$, the $n$-dependence drops out since in $d=4$, $\vp_+$ has vanishing mass dimension, so all couplings $c_{2n-m,m}$ with fixed $m$ have equal mass dimension. Therefore, in general, these couplings all mix under renormalisation. Since all effective couplings are related by \eqref{eq::RPIcouplings}, it suffices to consider the renormalisation of a single one of the $2n$ couplings for fixed $n$.
We choose the couplings $c^0_{2n+1,1}$, with $n\geq 0$, since these are the only marginal interactions in $d=4$. 
The bare and renormalised couplings are related via
\begin{equation}
c^0_{2n+1,1}=\tmu^{2n\ve}\hat c_{2n+1,1}
=\tmu^{2n\ve}\,\Big[c_{2n+1,1}+
\delta c_{2n+1,1}(\{c_{2 l+1,1}\},\varepsilon)\Big]\,,
\label{eq:couplingren}
\end{equation}
where we introduced the SdSET $\overline{\rm MS}$ renormalisation scale $\tmu$ defined as in \eqref{eq::muMSbar}, and the dimensionless, but still divergent, couplings $\hat c_{2n+1,1}$. 
The counterterms $\delta c_{2n+1,1}$ are polynomial in all relevant renormalised couplings. 
Even though there are infinitely many counterterms, when matching to a UV theory with a finite number of relevant couplings, which is weakly coupled at the matching scale, only finitely many counterterms contribute at a given order in the perturbative expansion of the correlation functions. This makes the problem tractable. The counterterms $\delta c_{2n+1,1}$ can be computed order-by-order in perturbation theory in powers of the renormalised effective couplings $c_{2l+1,1}$ as in flat-space EFTs. We adopt the $\overline{\rm MS}$ scheme and choose $\delta c_{2n+1,1}$ to cancel only the poles in $\ve$. 
We note that when \eqref{eq:couplingren} is inserted into \eqref{eq::Sintlead} every factor of 
$\Tilde{\mu}$ introduced through the renormalisation of the couplings will be accompanied by $a(t)$ and hence 
appear only in the product $(a(t)\tilde{\mu})^{2n\epsilon}$.

In addition one needs in principle a field renormalisation counterterm defined as $\delta_{\vp}\equiv Z_{\vp}-1$.
We can, however, make the following simplifying observation: when working at leading power in $\lambda$, that is, when neglecting the gradient terms in the SdSET action, the kinetic term, expressed in terms of renormalised fields, reduces to
\begin{equation}
S_{\textrm{kin}}=-3\int\der^{d-1}x\der t\,Z_{\vp}\,\dot\vp_+\vp_-\,.
\end{equation}
Here, we assumed that $Z_{\vp}$ is time-independent and commutes with the time derivatives. The counterterm $\delta_{\vp}$ is treated perturbatively, i.e.\ one considers $\delta_{\vp}\dot\vp_+\vp_-$ as a vertex to be inserted in diagrams. 
The corresponding Feynman rule involves a time derivative acting on the free two-point functions that connect to the counterterm vertex. However, at leading power in $\lambda$, the free SdSET fields are time-independent, hence the insertion of such a vertex leads to a vanishing result. This situation changes only at $\Lo(\lambda^2)$, since the time dependence of the non-interacting effective fields enters through powers of the gradient terms $\p^2_i/(a(t)H)^2$. We can thus conclude
\begin{equation}
\delta_{\vp}=\Lo(\lambda^2)\,,
\end{equation}
which means that the renormalisation of the effective fields is trivial at leading power in $\lambda$.\footnote{This situation is analogous to the case of the wave-function renormalisation of the non-relativistic quark field in NRQCD, which also starts at $\Lo(v^2)$, see e.g. \cite{Beneke:1997jm,Bodwin:1994jh}. }  Therefore, we set $\vp_{\pm,0}=\vp_{\pm}$ in the following. 

\subsubsection{Renormalisation of the non-Gaussian initial conditions}

The appearance of divergent time integrals, even at tree level, is a novel and unusual feature of SdSET. The origin of these divergences can be understood from the method of regions applied to cosmological correlators \cite{Beneke:2023wmt}. Each vertex in the full theory is associated with a finite time integral. When decomposing the time integral into its two regions, the early-time and late-time contributions develop divergences, which cancel in the sum over all regions. The contributions from the early-time, soft-momentum regions of correlation functions enter in SdSET as (bare) non-Gaussian initial conditions, while the late-time, soft-momentum regions are reproduced by the insertion of EFT vertices into SdSET correlation functions. Therefore, the divergent time integrals generated by the EFT vertices mirror the ones found in the late-time, soft-momentum regions of the full-theory correlators. The resulting divergences must be subtracted, resulting in renormalised initial conditions. 

\begin{figure}[t]
\centering
\begin{subfigure}{0.3\textwidth}
\centering
\includegraphics[width=0.8\textwidth]{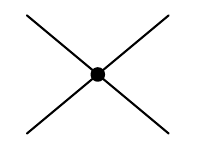}
\caption{}
\end{subfigure}
\begin{subfigure}{0.3\textwidth}
\centering
\includegraphics[width=0.8\textwidth]{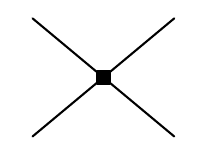}
\caption{}
\end{subfigure}
\begin{subfigure}{0.3\textwidth}
\centering
\includegraphics[width=0.8\textwidth]{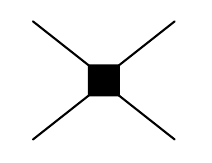}
\caption{}
\end{subfigure}
\caption{Examples of the three vertex objects: Lagrangian interaction vertex (a), IC counterterm (b), and renormalised non-Gaussian IC function (c). Each diagram appears twice, once with a full and once with an empty vertex, due to the doubling of each vertex type in the Schwinger-Keldysh formalism. We omitted these copies here.}
\label{fig::SKexamples}
\end{figure}

In momentum space, the bare initial-condition functions $\Xi^0_{2n+1,1}$ have the same mass dimension as the effective couplings $c_{2n+1,1}$,
\begin{equation}
\Big[\Xi^0_{2n+1,1}(\vec k_1,...,\vec k_{2n+2})\Big]=2\ve n\,.
\end{equation}
We can therefore proceed analogously and define a dimensionless version  $\hat \Xi^0_{2n+1,1}$ of the bare momentum-space initial-condition functions,
which is then split into an initial-condition counterterm $\xi_{2n+1,1}$ and the renormalised initial-condition function $\Xi_{2n+1,1}$:
\begin{flalign}
\Xi^0_{2n+1,1}(\vec k_1,...,\vec k_{2n+2})&=\tmu^{2n\ve}\,\hat \Xi^0_{2n+1,1}(\vec k_1,...,\vec k_{2n+2})\nonumber\\
&=\tmu^{2n\ve}\Big[\xi_{2n+1,1}(\vec k_1,...,\vec k_{2n+2})+\Xi_{2n+1,1}(\vec k_1,...,\vec k_{2n+2})\Big]\,.
\label{eq::renICs}
\end{flalign}
When inserted into \eqref{eq::ICFmom}, the dependence on $\Tilde{\mu}$ introduced in this way will appear in the combination $a_*\Tilde{\mu}$ to yield the products $(a_*\tilde{\mu})^{2n\epsilon}$.
The IC counterterms can, in general, be momentum-dependent as well, and may appear already at tree level to remove the UV poles generated by the time integrals of EFT tree diagrams. At loop level, the $\xi_{2n+1,1}$ will also be used to remove UV divergences generated by momentum integrations. 
The insertions of both $\xi_{2n+1,1}$ and $\Xi_{2n+1,1}$ in SdSET correlation functions are represented diagrammatically as described in \secref{sec::NGICs}, with the former and latter being represented as small and large polygons, respectively. Together with the Lagrangian interaction vertices, this results in three 
vertex objects, as illustrated in \figref{fig::SKexamples}. 

To the best of our knowledge, there is presently no general theory of perturbative renormalisation of the non-Gaussian initial-condition functions that we can build on, so we will refrain from making general statements about their all-order renormalisation or mixing structure beyond what has been stated above.\footnote{However, for a study in the context of super-renormalisable theories, see \cite{Garny:2015oza}.} For the purposes of this work, \eqref{eq::renICs} suffices.

\subsubsection{Matching procedure}

Once the SdSET vertices and non-Gaussian initial conditions have been renormalised, this still leaves the renormalised couplings and initial-condition functions undetermined. They are fixed by matching the renormalised SdSET correlation functions onto the corresponding renormalised functions in the full theory. The effective couplings can be determined by comparing the time-dependent parts of the two expressions, 
while the initial-condition functions can compensate for the mismatch of any time-independent and spatially non-local terms. This procedure will be exemplified using several examples below.


\section{Tree-level trispectrum}
\label{sec:trispectrum}

In this section we match the full-theory tree-level trispectrum onto SdSET. This  calculation demonstrates a few key features of the EFT, namely the existence of UV poles already at tree level, the necessity of the IC counterterm $\xi_{3,1}$, and the origin of the corresponding secular logarithm. The first step is the computation of the tree-level trispectrum in SdSET. After performing the calculation and renormalising the correlator, we match it to the full theory to determine the effective coupling $c_{3,1}$ and the renormalised IC function $\Xi_{3,1}$ at tree level. Since we compute a tree-level correlation function, the IR-regulator $\Lambda$ can be set to zero from the start, and hence we drop the subscript $\Lambda$ on the absolute values of momenta.

\subsection{Tree-level SdSET trispectrum}

The aim is to evaluate the correlation function
\begin{equation}
\cor{}{\vp_+(t,\vec k_1)\vp_+(t,\vec k_2)\vp_+(t,\vec k_3)\vp_+(t,\vec k_4)}' \equiv\cor{}{\vp_{+,1}\vp_{+,2}\vp_{+,3}\vp_{+,4}}'\,.
\end{equation}
It receives contributions from the quartic EFT vertex $c_{3,1}$, the IC counterterm $\xi_{3,1}$ and the renormalised IC $\Xi_{3,1}$. We will denote these contributions by corresponding subscripts. 

\begin{figure}[t]
\hskip0.5cm
\begin{subfigure}{0.45\textwidth}
\centering
\includegraphics[width=0.9\textwidth]{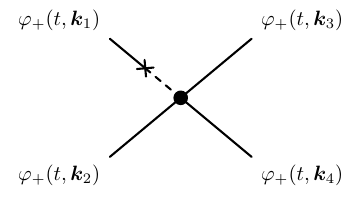}
\caption{}
\end{subfigure}%
\begin{subfigure}{0.45\textwidth}
\centering
\includegraphics[width=0.9\textwidth]{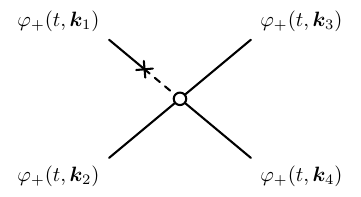}
\caption{}
\end{subfigure}\\
\hspace*{0.5cm}
\begin{subfigure}{0.45\textwidth}
\centering
\includegraphics[width=0.9\textwidth]{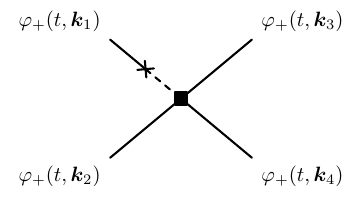}
\caption{}
\end{subfigure}%
\begin{subfigure}{0.45\textwidth}
\centering
\includegraphics[width=0.9\textwidth]{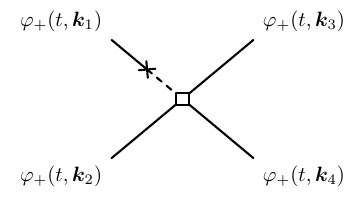}
\caption{}
\end{subfigure}
\caption{Upper line: 
The two diagrams for the insertion of the quartic vertex with coupling 
$c_{3,1}$ into the trispectrum. The remaining permutations where the $\vp_+\vp_-$ line attaches to the other external legs also contribute but are not shown.  
Lower line: The corresponding diagrams from the insertion of the IC counterterm $\xi_{3,1}$.}
\label{fig:trivertex}
\end{figure}

The first contribution arises from the insertion of the leading-power quartic vertex through diagrams as depicted in \figref{fig:trivertex}. Using the diagrammatic rules explained in \secref{sec::SdSET} and summing over all diagrams, we find 
\begin{flalign}
&\cor{}{\vp_{+,1}\vp_{+,2}\vp_{+,3}\vp_{+,4}}'_{|\,c_{3,1}}\nonumber\\
&=-ic_{3,1}\int_{-\infty}^t\der t'\;(a(t')\tmu)^{2\ve}\sum_{i=1}^4\bigg[\cor{}{T\{\vp_+(t,\vec k_i)\vp_-(t',-\vec k_i)\}}'\prod_{\substack{j=1 \\ j\neq i}}^4\,\cor{}{T\{\vp_+(t,\vec k_j)\vp_+(t',-\vec k_j)\}}'\nonumber\\
&\quad-\cor{}{\overline T\{\vp_+(t,\vec k_i)\vp_-(t',-\vec k_i)\}}'\prod_{\substack{j=1 \\ j\neq i}}^4\,\cor{}{\overline T\{\vp_+(t,\vec k_j)\vp_+(t',-\vec k_j\}}'\bigg]\label{eq::triplrelim1}\\
&=-ic_{3,1}\tmu^{2\ve}\int_{-\infty}^t\der t'\;a(t')^{2\ve}\sum_{i=1}^4\cor{}{[\vp_+(t,\vec k_i),\vp_-(t',-\vec k_i)]}'\prod_{\substack{j=1 \\ j\neq i}}^4\,\cor{}{\vp_+(t',\vec k_j)\vp_+(t,\vec k_j)}'\,,\label{eq::triplrelim2}
\end{flalign}
where we stripped off the overall momentum-conserving delta distribution by employing the primed notation~\eqref{eq::corprime}.
The relative minus sign in \eqref{eq::triplrelim1} is due to the two types of SK vertices inserted into the trispectrum. Eq.~\eqref{eq::triplrelim2} was obtained by resolving the (anti-)time-ordered products and taking into account that the $\vp_+$ commute among themselves. Using the commutation relation~\eqref{eq::unequaltcomm} and the two-point function~\eqref{eq::EFTprop1}, one obtains
\begin{equation}
\cor{}{\vp_{+,1}\vp_{+,2}\vp_{+,3}\vp_{+,4}}'_{|\,c_{3,1}}=-\frac{c_{3,1}\tmu^{2\ve}}{8(k_1k_2k_3k_4)^3}\sum_{j=1}^4\frac{k^3_j}{3}\int_{-\infty}^t\der t'\;a(t')^{2\ve}\,.
\end{equation}
The time integral is evaluated using \eqref{eq::EFTtints}, leading to
\begin{flalign}
\cor{}{\vp_{+,1}\vp_{+,2}\vp_{+,3}\vp_{+,4}}'_{|\,c_{3,1}}&=-\frac{c_{3,1}}{8(k_1k_2k_3k_4)^3}\frac{(a(t)\tmu)^{2\ve}}{2\ve}\sum_{j=1}^4\frac{k^3_j}{3}\,.
\label{eq::trispecc31}
\end{flalign}
We see that the SdSET correlation function features a pole in $\varepsilon$ already at tree level, which originates from the lower boundary $t=-\infty$ of the time integral,  and corresponds to a time-UV pole. The $\ve$-expansion of the above expression then generates a secular logarithm $\ln(a(t))$.
This pole must be removed using the initial-condition counterterm $\xi_{3,1}$ to yield a finite correlation function. This is always possible, since the pole is time-independent due to its origin from $t\to-\infty$. Its insertion into the trispectrum leads to diagrams as shown in the lower line of \figref{fig:trivertex}. These diagrams are very similar to the ones corresponding to the vertex 
insertion. Employing \eqref{eq::ICFmom}, their expression is the same 
as the previous one with $-i c_{3,1}\int _{-\infty}^t\der t' \;a(t')^{2\ve}$ 
replaced by $i \xi_{3,1} a_*^{2\ve}$. Thus
\begin{flalign}
&\cor{}{\vp_{+,1}\vp_{+,2}\vp_{+,3}\vp_{+,4}}'_{|\,\xi_{3,1}}\nonumber\\[0.1cm]
&=i(a_*\tmu)^{2\ve}\xi_{3,1}\sum_{i=1}^4\cor{}{[\vp_+(t,\vec k_i),\vp_-(t_*,-\vec k_i)]}'\prod_{\substack{j=1 \\ j\neq i}}^4\cor{}{\vp_+(t_*,\vec k_j)\vp_+(t,\vec k_j)}'\nonumber\\[-0.3cm]
&=\frac{\xi_{3,1}}{8(k_1k_2k_3k_4)^3}(a_*\tmu)^{2\ve}\sum_{j=1}^4\frac{k^3_j}{3}\,.
\end{flalign}
Comparing this result to~\eqref{eq::trispecc31}, we see that the momentum structure of the two results matches.
We minimally subtract the pole by choosing
\begin{equation}
\xi_{3,1}=\frac{c_{3,1}}{2\ve}\,.
\label{eq::xi31}
\end{equation}
Note that the counterterm is fully determined within SdSET, without the need of any input from matching, and thus one can compute finite, renormalised correlation functions in the EFT without knowledge of a full theory. 
In addition, this pole term matches the late-time pole that was generated by the time integral appearing in the early-time region of the full-theory trispectrum~\cite{Beneke:2023wmt}. 

Expanding the resulting contribution to the trispectrum in $\ve$ and letting $\ve\rightarrow0$ after subtracting the pole, we find the renormalised result
\begin{equation}
\cor{}{\vp_{+,1}\vp_{+,2}\vp_{+,3}\vp_{+,4}}'_{|\,c_{3,1}+\xi_{3,1}}=\frac{c_{3,1}}{8(k_1k_2k_3k_4)^3}\log\bigg(\frac{a_*}{a(t)}\bigg)\sum_{j=1}^4 \frac{k^3_j}{3}\,.
\label{eq::trispectrumEFTfinite}
\end{equation}
This combination is now finite, but depends on the reference scale factor $a_*$, which is similar to the factorisation scale $\mu$ in ordinary flat-space calculations. 
Here the scale factor $a_*$ tracks the time-dependence of the tree-level trispectrum.

\begin{figure}[t]
\centering
\begin{subfigure}{0.45\textwidth}
\centering
\includegraphics[width=0.9\textwidth]{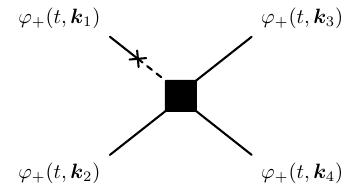}
\caption{}
\end{subfigure}%
\begin{subfigure}{0.45\textwidth}
\centering
\includegraphics[width=0.9\textwidth]{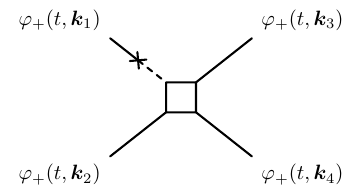}
\caption{}
\end{subfigure}
\caption{The two diagrams with insertion of the renormalised IC $\Xi_{3,1}$.}
\label{fig:triIC}
\end{figure}

Finally, we compute the insertion of the finite part of the four-point initial condition $\Xi_{3,1}$, which represents the finite piece of the early-time region of the full-theory trispectrum~\cite{Beneke:2023wmt}. 
The corresponding diagrams are depicted in \figref{fig:triIC}.
The computation is completely analogous to the initial-condition counterterm $\xi_{3,1}$, resulting in
\begin{equation}
\cor{}{\vp_{+,1}\vp_{+,2}\vp_{+,3}\vp_{+,4}}'_{|\,\Xi_{3,1}}=\frac{(a_*\tmu)^{2\ve}}{24(k_1k_2k_3k_4)^3}\,\Xi_{3,1}(\vec k_1,\vec k_2,\vec k_3,\vec k_4)\sum_{j=1}^4 k^3_j\,.
\label{eq:EFTtrispectrumIC}
\end{equation}
Since $\Xi_{3,1}(\vec k_1,\vec k_2,\vec k_3,\vec k_4)$ is finite by definition, one can set $\ve\rightarrow0$.
Summing all contributions, the renormalised trispectrum computed in SdSET reads
\begin{flalign}
&\cor{}{\vp_{+,1}\vp_{+,2}\vp_{+,3}\vp_{+,4}}'_{|\,c_{3,1}+\xi_{3,1}+\Xi_{3,1}}\nonumber\\
&=\frac{1}{24(k_1k_2k_3k_4)^3}\bigg\{c_{3,1}\log\bigg(\frac{a_*}{a(t)}\bigg)+\Xi_{3,1}(\vec k_1,\vec k_2,\vec k_3,\vec k_4)\bigg\}\sum_{j=1}^4 k^3_j\,.
\label{eq::EFTtrispectrumComplete}
\end{flalign}
As anticipated, this result contains two undetermined quantities, the coupling $c_{3,1}$ and the initial-condition function $\Xi_{3,1}$, both of which need to be determined by matching onto the tree-level full-theory result.

\subsection{Matching the trispectrum}

The late-time limit of the full-theory tree-level trispectrum reads in four dimensions \cite{Beneke:2023wmt}
\begin{flalign}
&\lim\limits_{-k_i\eta\rightarrow0}\cor{}{\phi(t,\vec k_1)\phi(t,\vec k_2)\phi(t,\vec k_3)\phi(t,\vec k_4)}'\nonumber\\
&=\frac{\kappa H^4}{8(k_1k_2k_3k_4)^3}\Biggl\{\biggl[\frac{1}{3}+\ln\bigg(\frac{e^{\gamma_E}k_t}{a(t)H}\bigg)\biggr]\sum_{j=1}^4\frac{k^3_j}{3}-\frac{k_1k_2k_3k_4}{k_t}+k_t\biggl(\sum_{j<l}k_jk_l-\frac{4}{9}k^2_t\biggr)\Biggr\}\,,
\label{eq::fulltri2}
\end{flalign}
where the total momentum is defined as $k_t \equiv k_1+k_2+k_3+k_4$.

To match this result in SdSET, one has to keep track of the relation between the full-theory field $\phi$ and the effective field $\vp_+$, since the correlation functions are not invariant under field redefinitions. The relevant terms in \eqref{eq::nonlinearphi} are
\begin{align}
\phi(t,\vec x)&=H^{1-\varepsilon}\bigg\{(a(t)H)^{\varepsilon}\bigg[\vp_+(t,\vec x)+\frac{c^0_{3,1}a(t)^{2\varepsilon}}{18(3+2\varepsilon)}\vp^3_+(t,\vec x)\bigg]
\nonumber\\
&\quad
+(a(t)H)^{-3+\varepsilon}\vp_-(t,\vec x)\bigg\}\,,
\label{eq::phivpcubic}
\end{align}
which anticipates that in $\kappa\phi^4$ theory $c_{3,1}$ is non-vanishing at $\mathcal{O}(\kappa)$, while $c_{2n+1,1}$ 
are at least $\mathcal{O}(\kappa^2)$ when $n>1$.\footnote{Eqs.~\eqref{eq:c31match}, \eqref{eq::c11} below and discussion in \appref{app::intredef}.}
It is convenient to introduce the spatial (inverse) Fourier-transform of the local composite operator $\vp^n_+(t,\vec x)$ in $d$ dimensions as\footnote{The notation is ambiguous. We therefore emphasise that the left-hand side is not the $n$th power of the Fourier transform of 
$\vp_+(t,\vec x)$, which will never appear in this paper.}
\begin{equation}
\vp^n_+(t,\vec k)\equiv\int\der^{d-1}x\;e^{-i\vec k\cdot\vec x}\vp^n_+(t,\vec x)\,.
\label{eq::vpnFT}
\end{equation}
Then the following combination of renormalised EFT correlators, taking $d\to 4$, yields the desired relation
\begin{flalign}
\cor{}{\phi(t,\vec k_1)&\phi(t,\vec k_2)\phi(t,\vec k_3)\phi(t,\vec k_4)}'_{|\,\Lo(\kappa)}
=H^4\,\bigg\{\cor{}{\vp_{+,1}\vp_{+,2}\vp_{+,3}\vp_{+,4}}'_{|\,c_{3,1}+\xi_{3,1}+\Xi_{3,1}}\nn\\
&\hspace{1.6cm}+\frac{c_{3,1}}{54}\Big[\cor{}{\vp^3_+(t,\vec k_1)\vp_+(t,\vec k_2)\vp_+(t,\vec k_3)\vp_+(t,\vec k_4)}'_{|\,\mathrm{free}} + \mathrm{perms.}\Bigr]\bigg\}\,,
\label{eq::trimatch}
\end{flalign}
where ``$\mathrm{perms.}$'' denotes all other permutations of the cubic insertion. The subscripts of the various EFT correlation functions denote at which order in the perturbative expansion in effective couplings and initial conditions they should be evaluated. The correlators involving only single fields have been computed in the previous subsection and are given in~\eqref{eq::EFTtrispectrumComplete}.
The correlators involving the composite operator $\vp^3_+$ are multiplied by $c_{3,1}$, so one just needs to evaluate their connected component in the Gaussian approximation, without inserting any additional interaction terms. Exemplarily, we find
\begin{equation}
\cor{}{\vp^3_+(t,\vec k_1)\vp_+(t,\vec k_2)\vp_+(t,\vec k_3)\vp_+(t,\vec k_4)}'_{|\,\mathrm{free}}=\frac{3}{4(k_2k_3k_4)^3}\,.
\end{equation}
The result for the remaining three contributions can be obtained by permuting the labels of the momentum vectors. 
The result for the field-redefinition contribution in the second line of 
\eqref{eq::trimatch} then reads 
\begin{flalign}
\cor{}{\vp_{+,1}\vp_{+,2}\vp_{+,3}\vp_{+,4}}'_{|\,\textrm{redef.}}
&=\frac{c_{3,1}}{24(k_1k_2k_3k_4)^3}\sum_{j=1}^4\frac{k^3_j}{3}\,.
\end{flalign}

Adding up the previous results yields the renormalised combination of SdSET correlators 
\begin{flalign}
&\cor{}{\vp_{+,1}\vp_{+,2}\vp_{+,3}\vp_{+,4}}'_{|\,c_{3,1}+\xi_{3,1}+\Xi_{3,1}+\textrm{redef.}}\nonumber\\
&=\frac{1}{8(k_1k_2k_3k_4)^3}\,\bigg\{c_{3,1}\bigg[\log\bigg(\frac{a_*}{a(t)}\bigg)+\frac{1}{3}\bigg]+\Xi_{3,1}(\vec k_1,\vec k_2,\vec k_3,\vec k_4)\bigg\}\sum_{j=1}^4\frac{k^3_j}{3}\,,
\end{flalign}
which can now be matched onto the full-theory result \eqref{eq::fulltri2} by means of  \eqref{eq::trimatch}. The effective coupling $c_{3,1}$ is uniquely determined by comparing the coefficients of the time-dependent logarithm, and at this order in the perturbative expansion we find simply
\begin{equation}
c_{3,1}=\kappa\,.
\label{eq:c31match}
\end{equation}
This result would receive higher-order corrections in the full-theory coupling $\kappa$, if the matching of the trispectrum were pushed to loop level. The IC function $\Xi_{3,1}(\vec k_1,\vec k_2,\vec k_3,\vec k_4)$ then needs to reproduce the remaining, time-independent terms, and compensate the dependence of the SdSET expression on $a_*$. 
At tree level, from the above we find 
\begin{flalign}
\Xi_{3,1}(\vec k_1,\vec k_2,\vec k_3,\vec k_4)&=3\kappa\,\Bigg\{\frac{1}{3}\log\bigg(\frac{e^{\gamma_E}k_t}{a_*H}\bigg)\nonumber\\
&\phantom{=}+\bigg(\sum_{j=1}^4k^3_j\bigg)^{-1}\bigg[-\frac{k_1k_2k_3k_4}{k_t}+k_t\bigg(\sum_{j<l}k_jk_l-\frac{4}{9}k_t^2\bigg)\bigg]\Bigg\}\,.
\label{eq::Xi31tree}
\end{flalign}
Comparing the explicit result~\eqref{eq::Xi31tree} to the early-time region of the full-theory trispectrum given in (4.19) in~\cite{Beneke:2023wmt}, one sees that after accounting for the prefactor and subtracting the pole, both expressions agree. This is to be expected, since no momentum integration and only a single time integral is involved.
The logarithm $\ln(k_t/(a_* H))$ in~\eqref{eq::Xi31tree} has a clear physical interpretation. Its natural scale, defined as the scale where the logarithm is not parametrically large, is $a_*\sim k_t/H$, identifying $t_*$ as the time of horizon crossing of the soft modes.
This reinforces the interpretation of $a_*$ as a time-factorisation scale separating the early-time subhorizon dynamics from the late-time superhorizon evolution captured by the EFT.
Moreover, the logarithm can already be inferred prior to the explicit matching calculation from the single-pole structure of the SdSET correlation function alone and the $\ln(a_*)$ dependence associated with this pole, together with the requirement that the full correlation function must be independent of $a_*$. 
The remaining polynomial momentum-dependent terms are fixed by the matching, but as said here they coincide exactly with the early-time region of the full result. 


\section{Tree-level six-point function}
\label{sec:penta}

To illustrate the matching procedure for correlation functions involving more than one vertex we consider the late-time limit of the tree-level six-point function (or ``pentaspectrum"). The computation of the complete full-theory result is presented in \appref{app::fullpenta}. We follow the same logic as in the previous section, and start by computing the renormalised tree-level SdSET six-point function. We then proceed to  matching, which requires us to consider tree-level SdSET correlation functions involving the composite operators $\vp^3_+$ and $\vp^5_+$. 
The main result of this section is the determination of the effective coupling $c_{5,1}$ and of the IC function $\Xi_{5,1}$ at $\Lo(\kappa^2)$.

\subsection{SdSET six-point function}

We start by computing the tree-level SdSET six-point function
\begin{equation}
\cor{}{\vp_+(t,\vec k_1)\vp_+(t,\vec k_2)\vp_+(t,\vec k_3)\vp_+(t,\vec k_4)\vp_+(t,\vec k_5)\vp_+(t,\vec k_6)}'\,.
\end{equation}
We consider the diagrams with external fields with momenta $\{\vec k_1,\vec k_2,\vec k_3\}$ and $\{\vec k_4,\vec k_5,\vec k_6\}$ connected to the same vertex, respectively. These terms may be thought of as the analogue to the $s$-channel contribution to a $3\rightarrow 3$ scattering process in flat spacetime. We will denote their contribution to the six-point function by 
\begin{equation}
P(t;\vec k_1,...,\vec k_6)\,,
\end{equation}
and the various contributions to $P$ using subscripts. The remaining contributions to the six-point function can then be obtained by permutations of the external momenta,
\begin{flalign}
&\cor{}{\vp_+(t,\vec k_1)\vp_+(t,\vec k_2)\vp_+(t,\vec k_3)\vp_+(t,\vec k_4)\vp_+(t,\vec k_5)\vp_+(t,\vec k_6)}'\nonumber\\
&=P(t;\vec k_1,...,\vec k_6)+(\vec k_1\leftrightarrow\vec k_4)+(\vec k_1\leftrightarrow\vec k_5)+(\vec k_1\leftrightarrow\vec k_6)\nonumber\\
&\phantom{=}+(\vec k_2\leftrightarrow\vec k_4)+(\vec k_2\leftrightarrow\vec k_5)+ (\vec k_2\leftrightarrow\vec k_6)+\{\vec k_1,\vec k_2\}\leftrightarrow\{\vec k_4,\vec k_5\}\nonumber\\
&\phantom{=}+\{\vec k_1,\vec k_2\}\leftrightarrow\{\vec k_4,\vec k_6\}+\{\vec k_1,\vec k_2\}\leftrightarrow\{\vec k_5,\vec k_6\}\,,
\label{eq::pentaperms}
\end{flalign}
resulting in a total of ten summands. 
Because of the large number of diagrams involved, we shall only draw a representative diagram for each topology, namely the all-$(+)$ version of the occurring vertices.

\subsubsection{Contribution from SdSET vertices}

We start by computing the contributions to $P$ from the insertion of SdSET vertices. At this order in the perturbative expansion we need to consider the following interaction terms: 
\begin{equation}
S_{\textrm{int}}\supset-\int\der^{d-1}x\der t\;\bigg[\frac{c_{3,1}}{3!}(a(t)\tmu)^{2\ve}\vp^3_+\vp_-+\bigg(\frac{c_{5,1}}{5!}+\frac{c^2_{3,1}}{36(3+2\ve)}\bigg)(a(t)\tmu)^{4\ve}\vp^5_+\vp_-\bigg]\,.
\label{eq::pentaints}
\end{equation}
The quartic interaction term was already considered in the previous section, in which we determined its tree-level value $c_{3,1}=\kappa$. We now need to consider for the first time the six-point interaction $\vp_+^5\vp_-$. The term  involving $c^2_{3,1}$ is generated by the field redefinition of $\vp_-$ that  removed the $\vp^4_+$ interaction, see \appref{app::intredef}. 

\begin{figure}[t]
\centering
\hskip0.5cm
\begin{subfigure}{0.45\textwidth}
\centering
\includegraphics[width=0.9\textwidth]{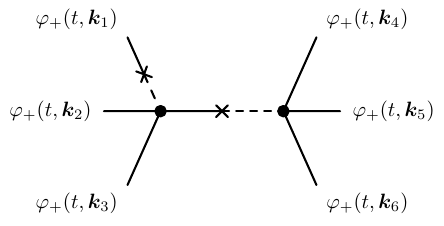}
\caption{}
\end{subfigure}%
\begin{subfigure}{0.45\textwidth}
\centering
\includegraphics[width=0.9\textwidth]{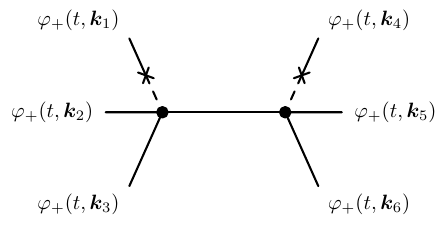}
\caption{}
\end{subfigure}\\
\hspace*{0.5cm}
\begin{subfigure}{0.45\textwidth}
\centering
\includegraphics[width=0.9\textwidth]{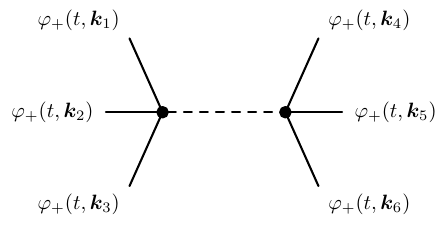}
\caption{}
\end{subfigure}%
\begin{subfigure}{0.45\textwidth}
\centering
\includegraphics[width=0.95\textwidth]{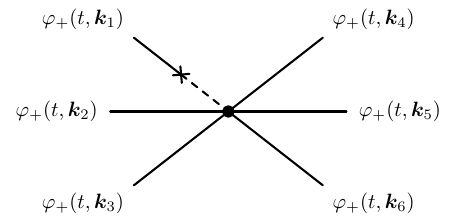}
\caption{}
\end{subfigure}
\caption{Upper line: the two non-vanishing diagram classes contributing to $P(t;\vec k_1,...,\vec k_6)$ featuring two quartic vertices. Lower line: vanishing diagram featuring two quartic vertices (c) and diagram with insertion of the six-point Lagrangian interaction vertex (d).}
\label{fig::penta1}
\end{figure}

Since we match onto an expression which is $\Lo(\kappa^2)$, we must consider diagrams involving two insertions of the quartic interaction and diagrams involving a single insertion of the six-point interaction. We start with the former. Summing over all four SK diagrams with topology as shown in the panel (a) of \figref{fig::penta1} we find 
\begin{flalign}
&\bigg(\frac{ic_{3,1}}{3!}\bigg)^2\frac{(3!)^2}{2^5(k_2k_3k_4k_5k_6)^3}\int_{-\infty}^t\der t_1\;(a(t_1)\tmu)^{2\ve}\int_{-\infty}^t\der t_2\;(a(t_2)\tmu)^{2\ve}\Big[\cor{}{T\{\vp_+(t,\vec k_1)\vp_-(t_1,\vec k_1)\}}'\nonumber\\
&\phantom{=}\times\Big(\cor{}{T\{\vp_+(t_1,\vec K)\vp_-(t_2,\vec K)\}}'-\cor{}{\vp_-(t_2,\vec K)\vp_+(t_1,\vec K)}'\Big)-\cor{}{\overline T\{\vp_+(t,\vec k_1)\vp_-(t_1,\vec k_1)\}}'\nonumber\\
&\phantom{=}\times\Big(\cor{}{\vp_+(t_1,\vec K)\vp_-(t_2,\vec K)}'-\cor{}{\overline T\{\vp_+(t_1,\vec K)\vp_-(t_2,\vec K)\}}'\Big)\Big]\nonumber\\
&=-\frac{c_{3,1}^2\tmu^{4\ve}}{32(k_2k_3k_4k_5k_6)^3}\int_{-\infty}^t\der t_1\;a(t_1)^{2\ve}\int_{-\infty}^{t_1}\der t_2\;a(t_2)^{2\ve}\cor{}{[\vp_+(t,\vec k_1),\vp_-(t_1,\vec k_1)]}'\nonumber\\
&\quad\times\cor{}{[\vp_+(t_1,\vec K),\vp_-(t_2,\vec K)]}'\label{eq::pentaintermediate}\\[0.1cm]
&=\frac{c_{3,1}^2(a(t)\tmu)^{4\ve}}{288(k_2k_3k_4k_5k_6)^3}\frac{1}{8\ve^2}\,,
\end{flalign}
using the abbreviation 
\begin{equation}
\vec K\equiv\sum_{j=1}^3\vec k_j=-\sum_{j=4}^6\vec k_j\,.
\end{equation} 
To obtain \eqref{eq::pentaintermediate} we resolved the (anti-)time-ordered two-point functions, which results in nested time integrals, and the integrand reduces to a product of commutators. Evaluating them using \eqref{eq::unequaltcomm} and computing the time integrals using \eqref{eq::EFTtints} yields the last equality. Summing over the other five ways of connecting an external field to the $\vp_-$ in the vertex in the diagram in \figref{fig::penta1} (a) gives
\begin{equation}
\frac{c_{3,1}^2(a(t)\tmu)^{4\ve}}{288}\frac{1}{8\ve^2}\frac{1}{(k_1k_2k_3k_4k_5k_6)^3}\sum_{j=1}^6k^3_j\,.
\end{equation}
Following the same steps, the sum over the four SK diagrams with topology shown in the panel (b) of \figref{fig::penta1} yields 
\begin{flalign}
&\bigg(\frac{ic_{3,1}}{3!}\bigg)^2\frac{(3!)^2}{2^5(k_2k_3k_5k_6K)^3}\int_{-\infty}^t\der t_1\;(a(t_1)\tmu)^{2\ve}\int_{-\infty}^t\der t_2\;(a(t_2)\tmu)^{2\ve}\Big[\cor{}{T\{\vp_+(t,\vec k_1)\vp_-(t_1,\vec k_1)\}}'\nonumber\\
&\phantom{=}\times\Big(\cor{}{T\{\vp_+(t,\vec k_4)\vp_-(t_2,\vec k_4)\}}'-\cor{}{\overline T\{\vp_+(t,\vec k_4)\vp_-(t_2,\vec k_4)\}}'\Big)-\cor{}{\overline T\{\vp_+(t,\vec k_1)\vp_-(t_1,\vec k_1)\}}'\nonumber\\
&\phantom{=}\times\Big(T\cor{}{\vp_+(t,\vec k_4)\vp_-(t_2,\vec k_4)}'-\cor{}{\overline T\{\vp_+(t,\vec k_4)\vp_-(t_2,\vec k_4)\}}'\Big)\Big]\nonumber\\
&=\frac{c_{3,1}^2(a(t)\tmu)^{4\ve}}{288(k_2k_3k_5k_6K)^3}\frac{1}{4\ve^2}\,,
\end{flalign}
where $K\equiv|\vec K|$. Summing over the remaining possibilities of connecting two external $\vp_+$ to the $\vp_-$ fields in the vertices gives
\begin{equation}
\frac{c_{3,1}^2(a(t)\tmu)^{4\ve}}{288}\frac{1}{4\ve^2}\frac{1}{(k_1k_2k_3k_4k_5k_6K)^3}\sum_{j=1}^3k^3_j\,\sum_{l=4}^6k^3_l\,.
\end{equation}

We can also draw the diagrams of the type shown in the panel (c) of \figref{fig::penta1}, but they vanish, since  they only involve $\cor{}{\vp_+\vp_+}$ and $\cor{}{\vp_-\vp_-}$ two-point functions. Then the four SK diagrams resulting from this diagram only differ in sign (two diagrams having positive and the other two negative sign) and in the ordering of the fields in the two-point functions. Since the $\vp_+$-fields commute with each other, as do the $\vp_-$-fields, one adds and subtracts two copies of the same expression, resulting in zero. Therefore, the overall contribution to $P$ from the insertion of two quartic vertices into the SdSET six-point function reads
\begin{equation}
P(t;\vec k_1,...,\vec k_6)_{|\,\vp^3_+\vp_-\times \vp^3_+\vp_-}=\frac{c^2_{3,1}(a(t)\tmu)^{4\ve}}{576\,(k_1k_2k_3k_4k_5k_6K)^3}\frac{1}{4\ve^2}\bigg[K^3\,\sum_{j=1}^6k^3_j+2\sum_{j=1}^3k^3_j\,\sum_{l=4}^6k^3_l\bigg]\,.
\end{equation} 

Finally, we consider the diagrams generated by the insertion of the six-point interaction $\vp^5_+\vp_-$ terms in \eqref{eq::pentaints}, which correspond to diagrams as shown in the panel (d) of \figref{fig::penta1}. This diagram yields the expression 
\begin{flalign}
&5!\times i\,\bigg[\frac{c_{5,1}}{5!}+\frac{c^2_{3,1}}{36(3+2\ve)}\bigg]\frac{1}{2^5(k_2k_3k_4k_5k_6)^3}\int_{-\infty}^t\der t'\;(a(t')\tmu)^{4\ve}\cor{}{[\vp_-(t',\vec k_1),\vp_+(t,\vec k_1)]}'\nonumber\\
&=-\frac{5(a(t)\tmu)^{4\ve}}{864\,(k_2k_3k_4k_5k_6)^3}\bigg[\frac{9c_{5,1}}{20\ve}+c^2_{3,1}\bigg(\frac{1}{2\ve}-\frac{1}{3}\bigg)\bigg]\,.
\end{flalign}
To obtain the second line, we expanded the square bracket to $\mathcal{O}(\ve)$, but did not expand the prefactor for conciseness.
This will be done in all following results unless stated otherwise. 
Summing over all possibilities of connecting the external $\vp_+$ with the $\vp_-$ in the vertex we find 
\begin{equation}
-\frac{5(a(t)\tmu)^{4\ve}}{864(k_1k_2k_3k_4k_5k_6K)^3}\,\bigg[\frac{9c_{5,1}}{20\ve}+c^2_{3,1}\bigg(\frac{1}{2\ve}-\frac{1}{3}\bigg)\bigg]K^3\sum_{j=1}^6k^3_j\,.
\label{psd1}
\end{equation}
This expression already contains all possible permutations of the external momenta, and it should therefore be thought of as the contribution to all ten summands on the right-hand side of \eqref{eq::pentaperms}. Since it is totally symmetric under exchange of the momenta, for the contribution to $P$ alone we need to divide by ten
\begin{equation}
P(t;\vec k_1,...,\vec k_6)_{|\,\vp^5_+\vp_-}=-\frac{(a(t)\tmu)^{4\ve}}{1728\,(k_1k_2k_3k_4k_5k_6K)^3}\,\bigg[\frac{9c_{5,1}}{20\ve}+c^2_{3,1}\bigg(\frac{1}{2\ve}-\frac{1}{3}\bigg)\bigg]K^3\sum_{j=1}^6k^3_j\,.
\end{equation}
As in the case of the trispectrum, 
these results exhibit time-integral divergences, and the poles must be removed by a counterterm, which is supplied by the IC counterterm $\xi_{5,1}$. 

\subsubsection{Contribution from the non-Gaussian initial conditions}

At this order in the perturbative expansion we have to consider insertions of the IC $\Xi_{3,1}$, and the associated counterterm, which are already known at tree level from the previous section, and $\Xi_{5,1}$ and the associated counterterm.
The finite IC function $\Xi_{5,1}$ will be matched at the end of this section. Since $\Xi_{3,1}\sim \kappa$, there are mixed contributions, involving the insertion of one vertex and one quartic IC, and double insertions of two $\Xi_{3,1}$. On the other hand $\Xi_{5,1}$ is of order $\kappa^2$, and we therefore only need its single insertion.

\begin{figure}[t]
\centering
\hskip0.5cm
\begin{subfigure}{0.45\textwidth}
\centering
\includegraphics[width=0.9\textwidth]{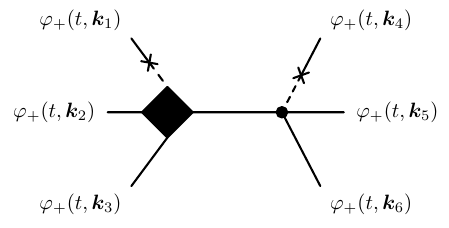}
\caption{}
\end{subfigure}%
\begin{subfigure}{0.45\textwidth}
\centering
\includegraphics[width=0.9\textwidth]{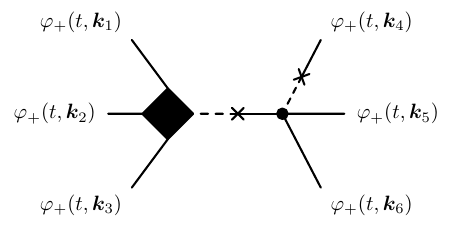}
\caption{}
\end{subfigure}\\
\hspace*{0.45cm}
\begin{subfigure}{0.45\textwidth}
\centering
\includegraphics[width=0.9\textwidth]{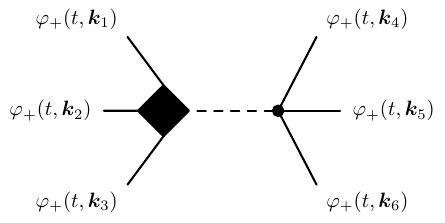}
\caption{}
\end{subfigure}%
\begin{subfigure}{0.45\textwidth}
\centering
\includegraphics[width=0.9\textwidth]{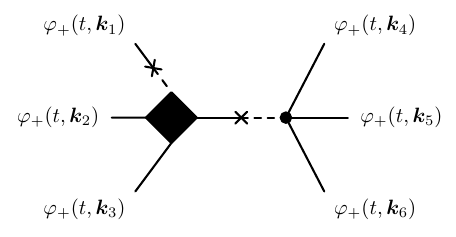}
\caption{}
\end{subfigure}
\caption{Diagrams featuring the insertion of the IC $\Xi_{3,1}$ and the quartic vertex. Upper line: the two non-vanishing diagrams. Lower line: the two vanishing diagrams.}
\label{fig::penta2}
\end{figure}

The non-vanishing mixed IC-vertex contributions are generated by diagrams as shown in the upper two panels of \figref{fig::penta2}. Further, we need to consider the corresponding diagrams with the insertion of the initial-condition counterterm $\xi_{3,1}$ instead of $\Xi_{3,1}$.
For brevity, we refrain from drawing these diagrams here and in the following. The sum over all diagrams of the form shown in the upper line of \figref{fig::penta2}, where the external fields with momenta $\{\vec k_1,\vec k_2,\vec k_3\}$ are connected to the IC insertion, and the remaining external fields to the vertex insertion, yields
\begin{flalign}
&-\frac{i(a_*\tmu)^{2\ve}}{3!}\frac{1}{2^5}\frac{1}{(k_1k_2k_3k_4k_5k_6K)^3}(3!)^2\bigg(-\frac{i}{3}\bigg)^2\bigg[K^3+\sum_{l=1}^3k^3_l\bigg]\Big[\xi_{3,1}+\Xi_{3,1}(\vec K,\vec k_1,\vec k_2,\vec k_3)\Big]\nonumber\\
&\times\frac{ic_{3,1}\tmu^{2\ve}}{6}\sum_{j=4}^6k^3_j\int_{-\infty}^t\der t'\;(a(t')\tmu)^{2\ve}\nonumber\\
&=-\frac{c_{3,1}(a_*a(t)\tmu^2)^{2\ve}}{576(k_1k_2k_3k_4k_5k_6K)^3}\frac{1}{\ve}\bigg[K^3+\sum_{l=1}^3k^3_l\bigg]\Big[\xi_{3,1}+\Xi_{3,1}(\vec K,\vec k_1,\vec k_2,\vec k_3)\Big]\sum_{j=4}^6k^3_j\,.
\end{flalign}
Exchanging the role of the vertex and initial-condition insertion in the diagrams in \figref{fig::penta2} amounts to letting $\{\vec k_1,\vec k_2,\vec k_3\}\leftrightarrow\{\vec k_4,\vec k_5,\vec k_6\}$ in the above expression. The diagrams of the type shown in the lower line of \figref{fig::penta2} vanish. The diagrams in panel (c) have the same propagator structure as those of \figref{fig::penta1}~(c), and therefore vanish for the same reason. The diagram shown in \figref{fig::penta2}~(d) yields 
\begin{flalign}
&-\frac{i^2c_{3,1}(a_*\tmu)^{2\ve}}{2^5(k_2k_3k_4k_5k_6)^3}\Big[\xi_{3,1}+\Xi_{3,1}(\vec K,\vec k_1,\vec k_2,\vec k_3)\Big]\hspace{-2.3pt}\int_{-\infty}^t\der t'\;(a(t')\tmu)^{2\ve}\Big[\cor{}{T\{\vp_+(t,\vec k_1)\vp_-(t_*,\vec k_1)\}}'\nonumber\\
&\phantom{=}\times\Big(\cor{}{T\{\vp_+(t_*,\vec K)\vp_-(t',\vec K)\}}'-\cor{}{\vp_-(t',\vec K)\vp_+(t_*,\vec K)}'\Big)-\cor{}{\overline T\{\vp_+(t,\vec k_1)\vp_-(t_*,\vec k_1)\}}'\nonumber\\
&\phantom{=}\times\Big(\cor{}{\vp_+(t_*,\vec K)\vp_-(t',\vec K)}'-\cor{}{\overline T\{\vp_+(t_*,\vec K)\vp_-(t',\vec K)\}}'\Big)\Big]=0\,,
\end{flalign}
and the same cancellation pattern occurs for all other diagrams of this type. Thus, the contribution to $P$ from diagrams featuring the insertion of one IC and one vertex reads
\begin{flalign}
&P(t;\vec k_1,...,\vec k_6)_{|\,\vp^3_+\vp_-\times (\xi_{3,1}+\Xi_{3,1})}\nonumber\\
&=-\frac{c_{3,1}(a_*a(t)\tmu^2)^{2\ve}}{576(k_1k_2k_3k_4k_5k_6K)^3}\frac{1}{\ve}\bigg[K^3+\sum_{l=1}^3k^3_l\bigg]\Big[\xi_{3,1}+\Xi_{3,1}(\vec K,\vec k_1,\vec k_2,\vec k_3)\Big]\sum_{j=4}^6k^3_j\nonumber\\
&\phantom{=}+\{\vec k_1,\vec k_2,\vec k_3\}\leftrightarrow\{\vec k_4,\vec k_5,\vec k_6\}\,.
\label{eq::EFTpentamixed}
\end{flalign}

\begin{figure}[t]
\centering
\hskip0.5cm
\begin{subfigure}{0.45\textwidth}
\centering
\includegraphics[width=0.9\textwidth]{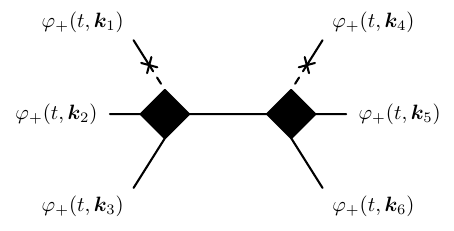}
\caption{}
\end{subfigure}%
\begin{subfigure}{0.45\textwidth}
\centering
\includegraphics[width=0.9\textwidth]{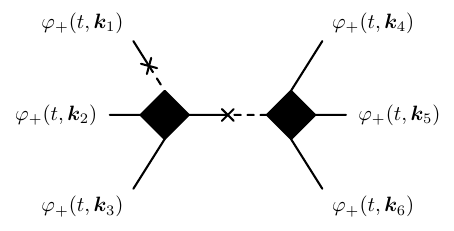}
\caption{}
\end{subfigure}\\
\hspace*{0.5cm}
\begin{subfigure}{0.45\textwidth}
\centering
\includegraphics[width=0.9\textwidth]{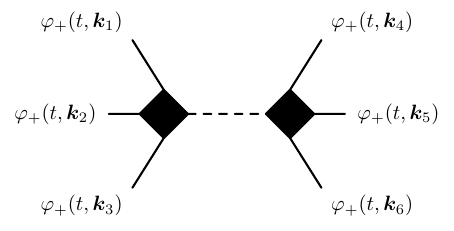}
\caption{}
\end{subfigure}%
\begin{subfigure}{0.45\textwidth}
\centering
\includegraphics[width=0.9\textwidth]{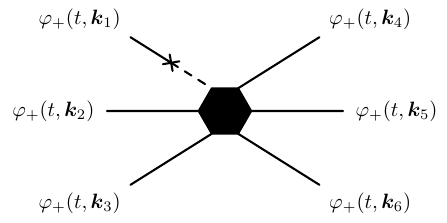}
\caption{}
\end{subfigure}
\caption{Diagrams featuring only IC insertions. Upper line: non-vanishing diagrams with double insertions of $\Xi_{3,1}$. Lower line: vanishing diagram from the double insertion of $\Xi_{3,1}$~(c) and from the insertion of $\Xi_{5,1}$~(d).}
\label{fig::penta3}
\end{figure}

We next turn to the double insertion of the four-point IC. The first two diagram classes are depicted in the upper line of \figref{fig::penta3}. The third, shown in panel (c) of the same figure, again gives a vanishing result. Summing over all possible attachments of the external fields to the two IC insertions, we obtain for diagram (a)
\begin{equation}
\frac{(a_*\tmu)^{4\ve}}{288(k_1k_2k_3k_4k_5k_6K)^3}\Big[\xi_{3,1}+\Xi_{3,1}(\vec K,\vec k_1,\vec k_2,\vec k_3)\Big]\Big[\xi_{3,1}+\Xi_{3,1}(\vec K,\vec k_4,\vec k_5,\vec k_6)\Big]\sum_{j=1}^3k^3_j\sum_{l=4}^6k^3_l\,.
\end{equation}
Diagram (b) requires us to consider for the first time (anti-)time-ordered two-point functions of $\vp_+$ and $\vp_-$ evaluated at equal times:
\begin{flalign}
&-\frac{(a_*\tmu)^{4\ve}k^3_1K^3}{32(k_1k_2k_3k_4k_5k_6K)^3}\Big[\xi_{3,1}+\Xi_{3,1}(\vec K,\vec k_1,\vec k_2,\vec k_3)\Big]\Big[\xi_{3,1}+\Xi_{3,1}(\vec K,\vec k_4,\vec k_5,\vec k_6)\Big]\nonumber\\
&\times\bigg\{\cor{}{T\{\vp_+(t,\vec k_1)\vp_-(t_*,\vec k_1)\}}'\Big[\cor{}{T\{\vp_+(t_*,\vec K)\vp_-(t_*,\vec K)\}}'-\cor{}{\vp_-(t_*,\vec K)\vp_+(t_*,\vec K)}'\Big]\nonumber\\
&-\cor{}{\overline T\{\vp_+(t,\vec k_1)\vp_-(t_*,\vec k_1)\}}'\Big[\cor{}{\vp_+(t_*,\vec K)\vp_-(t_*,\vec K)}'-\cor{}{\overline T\{\vp_+(t_*,\vec K)\vp_-(t_*,\vec K)\}}'\Big]\bigg\}\,.
\label{eq::pentaequaltstar}
\end{flalign}
Using \eqref{eq::mixedequalt} 
the above reduces to 
\begin{equation}
\frac{(a_*\tmu)^{4\ve}k^3_1K^3}{288(k_1k_2k_3k_4k_5k_6K)^3}\,\frac{1}{2}\Big[\xi_{3,1}+\Xi_{3,1}(\vec K,\vec k_1,\vec k_2,\vec k_3)\Big]\Big[\xi_{3,1}+\Xi_{3,1}(\vec K,\vec k_4,\vec k_5,\vec k_6)\Big]\,.
\end{equation}
Hence, the sum of both diagram classes reads 
\begin{flalign}
&P(t;\vec k_1,...,\vec k_6)_{|\,(\xi_{3,1}+\Xi_{3,1})\times(\xi_{3,1}+\Xi_{3,1})}=\frac{(a_*\tmu)^{4\ve}}{288(k_1k_2k_3k_4k_5k_6K)^3}\Big[\xi_{3,1}+\Xi_{3,1}(\vec K,\vec k_1,\vec k_2,\vec k_3)\Big]\nonumber\\
&\times\,\Big[\xi_{3,1}+\Xi_{3,1}(\vec K,\vec k_4,\vec k_5,\vec k_6)\Big]\bigg[\sum_{j=1}^3k^3_j\sum_{l=4}^6k^3_l+\frac{1}{2}K^3\sum_{j=1}^6k^3_j\bigg]\,.
\end{flalign} 

Finally, we consider the insertion of $\Xi_{5,1}$, shown in panel (d) of \figref{fig::penta3}, as well as the insertion of $\xi_{5,1}$. For the following computations it is convenient to decompose the initial-condition counterterm and -function into ten terms that mirror the symmetries of the EFT diagrams with an intermediate propagator, as:
\begin{flalign}
\xi_{5,1}(\vec k_1,...,\vec k_6)&=\hat\xi_{5,1}(\vec k_1,\vec k_2,\vec k_3|\vec k_4,\vec k_5,\vec k_6)+\textrm{permutations as in \eqref{eq::pentaperms}}\,,\\
\Xi_{5,1}(\vec k_1,...,\vec k_6)&=\hat\Xi_{5,1}(\vec k_1,\vec k_2,\vec k_3|\vec k_4,\vec k_5,\vec k_6)+\textrm{permutations as in \eqref{eq::pentaperms}}\,.\label{eq::hatXi51}
\end{flalign}
The hatted quantities are defined to be totally symmetric in their first and last three momentum arguments, as well as under the exchange of these two groups of arguments. Adding  the remaining nine permutations of their arguments, as in \eqref{eq::pentaperms}, we recover the original, totally symmetric $\xi_{5,1}$ and $\Xi_{5,1}$. 
The hatted functions can then be conveniently associated to each configuration of external momenta, e.g. to the ``$s$-channel'' contribution $P(t;\vec k_1,...,\vec k_6)$ considered in this section
\begin{flalign}
P(t;\vec k_1,...,\vec k_6)_{|\,\hat\xi_{5,1}+\hat\Xi_{5,1}}&=\frac{(a_*\tmu)^{4\ve}}{96(k_1k_2k_3k_4k_5k_6K)^3}\Big[\hat\xi_{5,1}(\vec k_1,\vec k_2,\vec k_3|\vec k_4,\vec k_5,\vec k_6)\nonumber\\
&\phantom{=}+\hat\Xi_{5,1}(\vec k_1,\vec k_2,\vec k_3|\vec k_4,\vec k_5,\vec k_6)\Big]K^3\sum_{j=1}^6k^3_j\,.
\end{flalign}
We will only consider $\hat\xi_{5,1}$ and $\hat\Xi_{5,1}$ with arguments as given above from now on, so in the following we will abbreviate them in intermediate steps.

\subsubsection{Renormalised SdSET six-point function}

Summing the contributions to $P$, and introducing the abbreviations
\begin{equation}
(k^n_{i...j})^m\equiv \bigg(\sum_{l=i}^jk^n_l\bigg)^{\!m}\,,\quad (k^n_{Ki...j})^m\equiv \bigg(K^n+\sum_{l=i}^jk^n_l\bigg)^{\!m}\,,
\label{eq:kns}
\end{equation}
we find 
\begin{flalign}
&P(t;\vec k_1,...,\vec k_6)_{|\,\textrm{6-pt.}}\nonumber\\[0.2cm]
&\equiv P(t;\vec k_1,...,\vec k_6)_{|\,\vp^3_+\vp_-\times\vp^3_+\vp_-+\vp^5_+\vp_-+\vp^3_+\vp_-\times(\xi_{3,1}+\Xi_{3,1})+(\xi_{3,1}+\Xi_{3,1})\times(\xi_{3,1}+\Xi_{3,1})+\hat\xi_{5,1}+\hat\Xi_{5,1}}\nonumber\\
&=\frac{(a(t)\tmu)^{4\ve}k^3_{123}}{576(k_1k_2k_3k_4k_5k_6K)^3}\,\Bigg\{\frac{K^3}{6\ve}\bigg[-c^2_{3,1}-\frac{9c_{5,1}}{10}+3c_{3,1}\Big[\Xi_{3,1}(\vec K,\vec k_1,\vec k_2,\vec k_3)\nonumber\\
&\phantom{=}-\Xi_{3,1}(\vec K,\vec k_4,\vec k_5,\vec k_6)\Big]\bigg]+c_{3,1}k^3_{K456}\log\bigg(\frac{a_*}{a(t)}\bigg)\bigg[c_{3,1}\log\bigg(\frac{a_*}{a(t)}\bigg)+2\,\Xi_{3,1}(\vec K,\vec k_1,\vec k_2,\vec k_3)\bigg]\nonumber\\
&\phantom{=}+\frac{c^2_{3,1}K^3}{9}+k^3_{K456}\,\Xi_{3,1}(\vec K,\vec k_1,\vec k_2,\vec k_3)\,\Xi_{3,1}(\vec K,\vec k_4,\vec k_5,\vec k_6)
\label{eq::regP}\\
&\phantom{=}+6\,\bigg(\frac{a_*}{a(t)}\bigg)^{\!4\ve}K^3\,\Big[\hat\xi_{5,1}(\vec k_1,...,\vec k_6)+\hat\Xi_{5,1}(\vec k_1,...,\vec k_6)\Big]\Bigg\}+\{\vec k_1,\vec k_2,\vec k_3\}\leftrightarrow\{\vec k_4,\vec k_5,\vec k_6\}\,.
\nonumber
\end{flalign}
To obtain this result, $(a(t)\tmu)^{4\ve}$ has been factored out, $\xi_{3,1}$ from \eqref{eq::xi31} has been used, and all terms in the curly brackets, except for the last line of \eqref{eq::regP}, have been expanded in $\ve$. The counterterm $\xi_{5,1}$ subtracts all pole terms appearing in the SdSET six-point function, including those of the other nine summands in \eqref{eq::pentaperms}. We then find 
\begin{flalign}
\hat\xi_{5,1}(\vec k_1,\vec k_2,\vec k_3|\vec k_4,\vec k_5,\vec k_6)&=\frac{1}{2\ve}\bigg\{\frac{c^2_{3,1}}{18}+\frac{c_{5,1}}{20}-\frac{c_{3,1}(k^3_{123}-k^3_{456})}{6k^3_{123456}}\bigg[\Xi_{3,1}(\vec K,\vec k_1,\vec k_2,\vec k_3)\nonumber\\
&\phantom{=}-\Xi_{3,1}(\vec K,\vec k_4,\vec k_5,\vec k_6)\bigg]\bigg\}\,.
\end{flalign}
Returning to $P$, this implies the renormalised expression 
\begin{flalign}
P(t;\vec k_1,...,\vec k_6)_{|\,\textrm{6-pt.}}&=\frac{k^3_{123}}{576(k_1k_2k_3k_4k_5k_6K)^3}\Bigg\{c^2_{3,1}k^3_{K456}\log^2\bigg(\frac{a_*}{a(t)}\bigg)\nonumber\\
&\phantom{=}+\bigg[\bigg(\frac{2c^2_{3,1}}{3}+\frac{3c_{5,1}}{5}\bigg)K^3+2c_{3,1}k^3_{K456}\,\Xi_{3,1}(\vec K,\vec k_4,\vec k_5,\vec k_6)\bigg]\log\bigg(\frac{a_*}{a(t)}\bigg)\nonumber\\
&\phantom{=}+\frac{c^2_{3,1}K^3}{9}+k^3_{K456}\Xi_{3,1}(\vec K,\vec k_1,\vec k_2,\vec k_3)\,\Xi_{3,1}(\vec K,\vec k_4,\vec k_5,\vec k_6)\nonumber\\
&\phantom{=}+6K^3\,\hat\Xi_{5,1}(\vec k_1,...,\vec k_6)\Bigg\}+\{\vec k_1,\vec k_2,\vec k_3\}\leftrightarrow\{\vec k_4,\vec k_5,\vec k_6\}\,.
\label{eq::renP}
\end{flalign}

\subsection{Matching the six-point function}

\begin{figure}[t]
\centering
\begin{subfigure}{0.45\textwidth}
\centering
\includegraphics[width=0.9\textwidth]{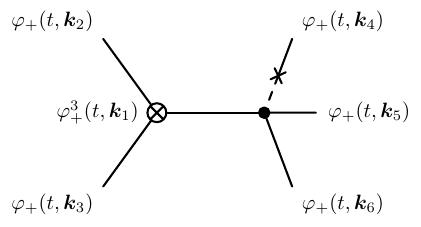}
\caption{}
\end{subfigure}%
\begin{subfigure}{0.45\textwidth}
\centering
\includegraphics[width=0.9\textwidth]{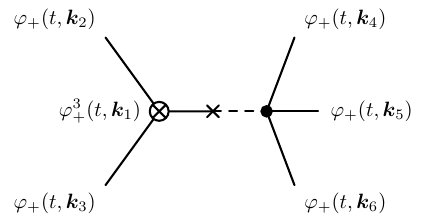}
\caption{}
\end{subfigure}\\
\begin{subfigure}{0.45\textwidth}
\centering
\includegraphics[width=0.9\textwidth]{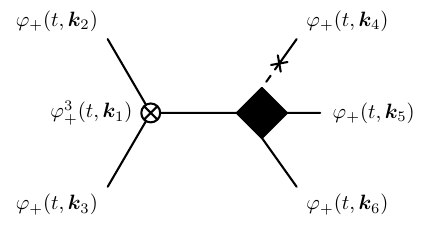}
\caption{}
\end{subfigure}%
\begin{subfigure}{0.45\textwidth}
\centering
\includegraphics[width=0.9\textwidth]{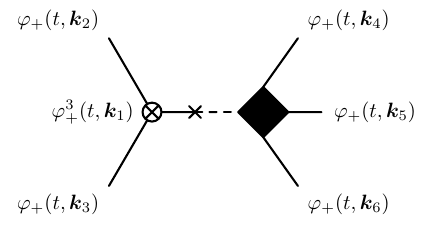}
\caption{}
\end{subfigure}
\caption{Diagrammatic representation of composite-operator correlation functions involving $\vp^3_+$. Upper line: sample diagrams from the additional insertion of a quartic vertex. Lower line: sample of the diagrams resulting from the additional insertion of  $\Xi_{3,1}$.}
\label{fig::penta4}
\end{figure}

The matching equation between the full-theory and SdSET six-point functions needs to account for the non-linear leading-power relation  \eqref{eq::nonlinearphi} between the full-theory and effective fields. This implies 
\begin{flalign}
&\cor{}{\phi(t,\vec k_1)\phi(t,\vec k_2)\phi(t,\vec k_3)\phi(t,\vec k_4)\phi(t,\vec k_5)\phi(t,\vec k_6)}'\nonumber\\[0.2cm]
&=H^6\Bigg[\cor{}{\vp_+(t,\vec k_1)\vp_+(t,\vec k_2)\vp_+(t,\vec k_3)\vp_+(t,\vec k_4)\vp_+(t,\vec k_5)\vp_+(t,\vec k_6)}'\nonumber\\
&\phantom{=}+\frac{c_{3,1}(a(t)\tmu)^{2\ve}}{18(3+2\ve)}\cor{}{\vp^3_+(t,\vec k_1)\vp_+(t,\vec k_2)\vp_+(t,\vec k_3)\vp_+(t,\vec k_4)\vp_+(t,\vec k_5)\vp_+(t,\vec k_6)}'\nonumber\\
&\phantom{=}+\frac{c^2_{3,1}(a(t)\tmu)^{4\ve}}{324(3+2\ve)^2}\cor{}{\vp^3_+(t,\vec k_1)\vp^3_+(t,\vec k_2)\vp_+(t,\vec k_3)\vp_+(t,\vec k_4)\vp_+(t,\vec k_5)\vp_+(t,\vec k_6)}'\nonumber\\
&\phantom{=}+\frac{(a(t)\tmu)^{4\ve}}{54(3+4\ve)}\bigg[\frac{3c_{5,1}}{20}+\frac{c^2_{3,1}}{3+2\ve}\bigg]\cor{}{\vp^5_+(t,\vec k_1)\vp_+(t,\vec k_2)\vp_+(t,\vec k_3)\vp_+(t,\vec k_4)\vp_+(t,\vec k_5)\vp_+(t,\vec k_6)}'\nonumber\\
&\phantom{=}+\textrm{perms.}\Bigg]
\label{eq::pentamatch}
\end{flalign}
for the six-point function. ``Perms." refers to the composite-operator correlation functions obtained from the ones given by taking permutations of the momentum labels. Since the correlation function in the third line of \eqref{eq::pentamatch} is multiplied by $c_{3,1}\sim\kappa$, to order $\kappa^2$, we can insert a quartic EFT vertex or IC into it. On the other hand, since the correlation functions appearing in the fourth and fifth line of \eqref{eq::pentamatch} are multiplied by $c_{5,1}\sim c^2_{3,1}\sim\kappa^2$, 
we evaluate them in the Gaussian approximation. 
We show representatives of the relevant contractions 
in \figref{fig::penta4} and \figref{fig::penta5}, 
and use the symbol $\otimes$ to denote the insertion of a composite operator. 
As before we limit ourselves to computing explicitly the 
contractions which combine with the renormalised result \eqref{eq::renP} for $P$ to reproduce the corresponding full-theory expression. 

\subsubsection{Composite-operator correlation functions}

We begin with the correlation function in the third line of \eqref{eq::pentamatch}. The leading connected contribution  involves the insertion of the quartic SdSET vertex $c_{3,1}$ or IC $\Xi_{3,1}$. For the vertex-insertion contributions we consider 
the following contractions: composite operators $\vp^3_+$ with momentum argument belonging to the set $\{\vec k_1,\vec k_2,\vec k_3\}$ are connected to remaining single fields with momentum arguments belonging to the same set, while the single fields with momenta belonging to the set $\{\vec k_4,\vec k_5,\vec k_6\}$ are connected to the vertex, and vice-versa. These are the contractions that yield the correct momentum structure to be grouped with the remaining contributions to $P$. Sample diagrams corresponding to such contractions are shown in \figref{fig::penta4} (a) and (b). All remaining contributions from other possible contractions are taken into account by summing over momentum permutations as in \eqref{eq::pentaperms}. Summing over all such contractions leads to the expression
\begin{equation}
P(t;\vec k_1,...,\vec k_6)_{|\,\vp^3_+\times\vp^3_+\vp_-}=-\frac{c_{3,1}^2(a(t)\tmu)^{4\ve}}{864(k_1k_2k_3k_4k_5k_6K)^3}\,\bigg[\frac{1}{2\ve}-\frac{1}{3}\bigg]\bigg[K^3\sum_{j=1}^6k^3_j+2\sum_{j=1}^3k^3_j\,\sum_{l=4}^6k^3_l\bigg]\,.
\label{psd5}
\end{equation}
Similarly, the insertion of the IC, and the associated counterterm, corresponding to the diagrams with topologies as shown in the second line of \figref{fig::penta4}, lead to the expression
\begin{flalign}
&P(t;\vec k_1,...,\vec k_6)_{|\,\vp^3_+\times(\xi_{3,1}+\Xi_{3,1})}\nonumber\\
&=\frac{c_{3,1}(a(t)a_*\tmu^2)^{2\ve}}{288(k_1k_2k_3k_4k_5k_6K)^3}\frac{1}{(3+2\ve)}\bigg[k^3_{K123}\,k^3_{456}\Big[\xi_{3,1}+\Xi_{3,1}(\vec K,\vec k_1,\vec k_2,\vec k_3)\Big]\nonumber\\
&\phantom{=}+\{\vec k_1,\vec k_2,\vec k_3\}\leftrightarrow\{\vec k_4,\vec k_5,\vec k_6\}\bigg]\,,
\end{flalign}
where we kept the full $\varepsilon$ dependence as $\xi_{3,1}$ contains a pole.

\begin{figure}[t]
\centering
\begin{subfigure}{0.45\textwidth}
\centering
\includegraphics[width=0.8\textwidth]{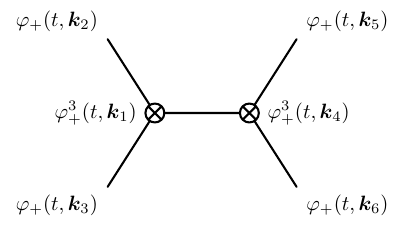}
\caption{}
\end{subfigure}%
\begin{subfigure}{0.45\textwidth}
\centering
\raisebox{0.3cm}[0pt][0pt]{%
\includegraphics[width=\textwidth]{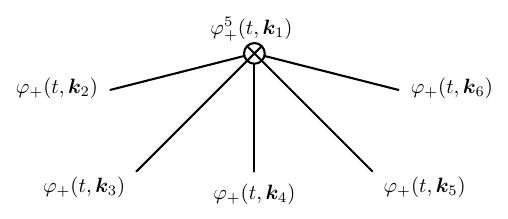}}
\caption{}
\end{subfigure}
\caption{Sample diagrams representing the correlation functions involving two composite operators $\vp^3_+$ (a) and the composite operator $\vp^5_+$ (b), evaluated in the Gaussian approximation.}
\label{fig::penta5}
\end{figure}

Next, we consider the correlation function in the fourth line of \eqref{eq::pentamatch} involving two composite operators $\vp^3_+$, which we evaluate in the Gaussian approximation. Similarly to the correlators involving a single $\vp^3_+$ operator, we identify and group into $P$ the contributions with the appropriate symmetry of the momentum labels under permutations. This corresponds to 
the subset of all possible contractions where one $\vp^3_+$ carries a momentum belonging to the set $\{\vec k_1,\vec k_2,\vec k_3\}$, while the other $\vp^3_+$ has an argument belonging to $\{\vec k_4,\vec k_5,\vec k_6\}$, respectively. The single fields are then connected to the $\vp^3_+$ associated with their respective momentum set. A sample diagram corresponding to such a contribution is shown in \figref{fig::penta5} (a). All remaining contributions to the correlation function in question are recovered when summing over permutations of the momenta. Summing over all such terms gives 
\begin{equation}
P(t;\vec k_1,...,\vec k_6)_{|\,\vp^3_+\times\vp^3_+}=\frac{c_{3,1}^2(a(t)\tmu)^{4\ve}}{2592(k_1k_2k_3k_4k_5k_6K)^3}\sum_{j=1}^3k^3_j\sum_{l=4}^6k^3_l\,.
\label{psredef3}
\end{equation}
Finally, from the correlation function in the fifth line of \eqref{eq::pentamatch} and its permutations, which can be evaluated in the free-theory approximation and represented diagrammatically as in panel (b) of \figref{fig::penta5}, we get 
\begin{equation}
P(t;\vec k_1,...,\vec k_6)_{|\,\vp^5_+}=\frac{(a(t)\tmu)^{4\ve}}{1296(k_1k_2k_3k_4k_5k_6K)^3}\bigg[c^2_{3,1}+\frac{9c_{5,1}}{20}\bigg]K^3\sum_{j=1}^6k^3_j\,.
\end{equation}
As for the previous insertions of six-point vertices, we divided the result obtained from the correlation functions involving the operator $\vp^5_+$ by ten to obtain the  contribution to $P$.

Summing the various contributions stemming from the field redefinition, and fully expanding the resulting expression in $\ve$, we find the result
\begin{flalign}
&P(t;\vec k_1,...,\vec k_6)_{|\,\textrm{redef.}}\equiv P(t;\vec k_1,...,\vec k_6)_{|\,\vp^3_+\times\vp^3_+\vp_-+\vp^3_+\times(\xi_{3,1}+\Xi_{3,1})+\vp^3_+\times\vp^3_++\vp^5_+}\\[0.15cm]
&=\frac{c_{3,1}k^3_{123}}{576(k_1k_2k_3k_4k_5k_6K)^3}\,\Bigg\{\frac{2c_{3,1}}{3}\,k^3_{K456}\,\log\bigg(\frac{a_*}{a(t)}\bigg)+\bigg[\frac{c_{3,1}}{9}\bigg(4K^3+k^3_{456}\bigg)+\frac{c_{5,1}}{5}K^3\nonumber\\
&\phantom{=}+\frac{2}{3}k^3_{K456}\,\Xi_{3,1}(\vec K,\vec k_4,\vec k_5,\vec k_6)\bigg]\Bigg\}+\{\vec k_1,\vec k_2,\vec k_3\}\leftrightarrow\{\vec k_4,\vec k_5,\vec k_6\}\,.
\label{eq:compop}
\end{flalign}
Notice that this expression is finite on its own, even though its parts were divergent. This must be the case to be able to successfully match onto the finite full-theory result, since the SdSET six-point function has already been renormalised and is finite on its own.

We now have computed all of the pieces contributing to $P$. Summing \eqref{eq::renP} and \eqref{eq:compop}, we find 
\begin{flalign}
&P_{\textrm{SdSET}}(t;\vec k_1,...,\vec k_6)\equiv P(t;\vec k_1,...,\vec k_6)_{|\,\textrm{6-pt.}}+P(t;\vec k_1,...,\vec k_6)_{|\,\textrm{redef.}}\\[0.15cm]
&=\frac{k^3_{123}}{576(k_1k_2k_3k_4k_5k_6K)^3}\,\Bigg[c^2_{3,1}k^3_{K456}\log^2\bigg(\frac{a_*}{a(t)}\bigg)+\bigg[\frac{2c^2_{3,1}}{3}\Big(2K^3+k^3_{456}\Big)+\frac{3c_{5,1}}{5}K^3\nonumber\\
&\phantom{=}+2c_{3,1}k^3_{K456}\,\Xi_{3,1}(\vec K,\vec k_4,\vec k_5,\vec k_6)\bigg]\log\bigg(\frac{a_*}{a(t)}\bigg)+\frac{c^2_{3,1}}{9}\Big(5K^3+k^3_{456}\Big)+\frac{c_{5,1}}{5}K^3\nonumber\\
&\phantom{=}+\frac{2c_{3,1}}{3}k^3_{K456}\,\Xi_{3,1}(\vec K,\vec k_4,\vec k_5,\vec k_6)+k^3_{K456}\,\Xi_{3,1}(\vec K,\vec k_1,\vec k_2,\vec k_3)\,\Xi_{3,1}(\vec K,\vec k_4,\vec k_5,\vec k_6)\nonumber\\
&\phantom{=}+6K^3\,\hat\Xi_{5,1}(\vec k_1,...,\vec k_6)\Bigg]+\{\vec k_1,\vec k_2,\vec k_3\}\leftrightarrow\{\vec k_4,\vec k_5,\vec k_6\}\,.
\label{eq:SdSET6ptall}
\end{flalign}

\subsubsection{Six-point function in the full theory}

The six-point function in the full theory in the late-time limit is computed in \appref{app::fullpenta} using the method-of-region approach \cite{Beneke:1997zp ,Beneke:2023wmt} and reads
\begin{flalign}
&P_{\textrm{full}}(\eta;\vec k_1,...,\vec k_6)\nonumber\\
&=\frac{\kappa^2H^6}{576(k_1k_2k_3k_4k_5k_6K)^3}\Bigg\{-\frac{K^3}{2}\Big[k^3_{123}-k^3_{456}\Big]\log^2(-e^{\gamma_E}k_{K123}\eta)+k^3_{K123}k^3_{456}\log(-e^{\gamma_E}k_{K123}\eta)\nonumber\\
&\phantom{=}\times\log(-e^{\gamma_E}k_{K456}\eta)+\bigg[K^3k^3_{456}+k^3_{123}\bigg(\frac{K^3}{3}+\frac{2}{3}k^3_{456}+6\bigg(-\frac{Kk_4k_5k_6}{k_{K456}}\nonumber\\
&\phantom{=}+k_{K456}\bigg(Kk_{456}+k_4k_{56}+k_5k_6-\frac{4}{9}(k_{K456})^2\bigg)\bigg)\bigg)\bigg]\log(-e^{\gamma_E}k_{K123}\eta)\nonumber\\
&\phantom{=}+k^3_{123}\bigg[\frac{1}{9}K^3-\frac{1}{3}k^3_{456}-\frac{2Kk_4k_5k_6}{k_{K456}}-\frac{2}{3}\Big[5k^2_{K456}k_{K456}-2\Big(K[k_4k_{56}+k_5k_6]+k_4k_5k_6\Big)\Big]\nonumber\\
&\phantom{=}+\frac{\pi^2}{12}k^3_{K456}\bigg]+f(\vec k_1,...,\vec k_6)\Bigg\}+\{\vec k_1,\vec k_2,\vec k_3\}\leftrightarrow\{\vec k_4,\vec k_5,\vec k_6\}\,,
\label{eq:full6pt}
\end{flalign}
where $f(\vec k_1,...,\vec k_6)$ is given in \eqref{eq:fdef}.

\subsubsection{Matching of $c_{5,1}$ and $\Xi_{5,1}$}

From the previously matched value $c_{3,1}=\kappa$, and matching the time-dependent terms in \eqref{eq:SdSET6ptall} and \eqref{eq:full6pt}, we conclude 
\begin{equation}
c_{5,1}=0+\Lo(\kappa^3)\,.
\end{equation}
This result is expected from the method of regions. The effective couplings $c_{2n+1,1}$ with $n>1$, which possess no counterparts in the full theory, contain the physics of the late-time, hard-momentum regions of full-theory loop diagrams. For $c_{5,1}$, the lowest such diagram is $\Lo(\kappa^3)$.\footnote{There are no tree-diagram contributions at $\mathcal{O}(\kappa^2)$, since there is no expansion of the intermediate propagators in small momentum. We expect this to hold for the tree-level matching of generic couplings $c_{2n+1,1}$ which have no counterpart in the full theory.\label{ft:coupling}} 

The renormalised component $\hat\Xi_{5,1}$ of the initial-condition function is then determined by 
equating \eqref{eq:full6pt} to \eqref{eq:SdSET6ptall}, using $c_{3,1}=\kappa$, $c_{5,1}=0$, and solving for 
\begin{flalign}
&\hat\Xi_{5,1}(\vec k_1,\vec k_2,\vec k_3|\vec k_4,\vec k_5,\vec k_6)\nonumber\\
&=\frac{96(k_1k_2k_3k_4k_5k_6)^3}{k^3_{123456}}\bigg[H^{-6}P_{\textrm{full}}(t;\vec k_1,...,\vec k_6)-P_{\textrm{SdSET}}(t;\vec k_1,...,\vec k_6)|_{\textrm{no }\hat\Xi_{5,1}}\bigg]\nonumber\\
&=\frac{\kappa^2}{6K^3k^3_{123456}}\,\Bigg\{-\frac{K^3}{2}\Big[k^3_{123}-k^3_{456}\Big]\log^2\bigg(\frac{e^{\gamma_E}k_{K123}}{a_*H}\bigg)+k^3_{K123}k^3_{456}\log\bigg(\frac{e^{\gamma_E}k_{K123}}{a_*H}\bigg)\nonumber\\
&\phantom{=}\times\log\bigg(\frac{e^{\gamma_E}k_{K456}}{a_*H}\bigg)+\Bigg[K^3k^3_{456}+k^3_{123}\bigg[\frac{K^3}{3}+\frac{2}{3}k^3_{456}+6\,\bigg(-\frac{Kk_4k_5k_6}{k_{K456}}\nonumber\\
&\phantom{=}+k_{K456}\bigg(Kk_{456}+k_4k_{56}+k_5k_6-\frac{4}{9}(k_{K456})^2\bigg)\bigg)\bigg]\Bigg]\log\bigg(\frac{e^{\gamma_E}k_{K123}}{a_*H}\bigg)+k^3_{123}\Bigg[-\frac{4}{9}k^3_{K456}\nonumber\\
&\phantom{=}-\frac{2}{3\kappa}k^3_{K456}\,\Xi_{3,1}(\vec K,\vec k_4,\vec k_5,\vec k_6)-\frac{k^3_{K456}}{\kappa^2}\,\Xi_{3,1}(\vec K,\vec k_1,\vec k_2,\vec k_3)\,\Xi_{3,1}(\vec K,\vec k_4,\vec k_5,\vec k_6)\nonumber\\
&\phantom{=}-\frac{2Kk_4k_5k_6}{k_{K456}}-\frac{2}{3}\Big[5k^2_{K456}k_{K456}-2\Big(K[k_4k_{56}+k_5k_6]+k_4k_5k_6\Big)\Big]+\frac{\pi^2}{12}\,k^3_{K456}\Bigg]\nonumber\\
&\phantom{=}+f(\vec k_1,...,\vec k_6)\Bigg\}+\{\vec k_1,\vec k_2,\vec k_3\}\leftrightarrow\{\vec k_4,\vec k_5,\vec k_6\}\,,
\end{flalign}
and the full $\Xi_{5,1}$ is recovered by substituting $\hat\Xi_{5,1}$ in \eqref{eq::hatXi51}. The explicit expression for $\Xi_{5,1}$ exhibits a number of features that closely mirror, while extending, those encountered for the trispectrum.
Most prominently, the appearance of double logarithms in $a_*$ directly reflects the presence of double poles in the SdSET correlator, due to the two-fold time integral.
As in the four-point case, these logarithms have a direct interpretation in terms of the early-time region of both time integrals, with their natural scale at horizon crossing for the respective soft momenta.
Likewise, the absence of $\mu$-dependence is a consequence of the fact that matching at this order only involves time but no momentum integrals.
In direct comparison to the full-theory early-early time region, one notices that the momentum dependence of $\Xi_{5,1}$ is similar but slightly different, as $\Xi_{5,1}$ must compensate additional terms due to single- and the double-$\Xi_{3,1}$ insertion contributions to \eqref{eq::pentaperms}.

While the structure of the matching calculation is conceptually identical to that of the trispectrum, the technical complexity increases substantially due to the number of fields in the correlation function, and consequently the proliferation of contributing SdSET correlation functions, as well as the intricate non-local momentum dependence of the initial-condition vertices.
One can therefore expect that when pushing the matching procedure to higher orders in the perturbative expansion and to more complicated correlation functions, the technical complexity of the involved computation will keep increasing, however, no additional conceptual difficulties should arise.
Nevertheless, the fact that SdSET reproduces all time-dependent structures of the full-theory six-point function, which contains two time integrals, provides a non-trivial consistency check of the effective framework and supports its applicability to higher-order corrections, as the six-point function is an ingredient for the computation of NNLO corrections to the Fokker-Planck equation \cite{Cohen:2021fzf}.


\section{One-loop power spectrum}
\label{sec:pwr}

Matching the one-loop power spectrum constitutes the simplest example of a one-loop matching computation in SdSET. We start by computing the renormalised one-loop power spectrum of the effective field $\vp_+$, which depends on the finite part of the EFT mass counterterm, and on the perturbative correction to the Gaussian initial-condition function $\Xi_{1,1}$. We then set up the matching equation to the full-theory one-loop power spectrum, and determine these two quantities. Since the power spectrum is both UV- and IR-divergent, we employ dimensional regularisation and the $\Lambda$ IR-regulator. 

\subsection{One-loop SdSET power spectrum}

At the one-loop order the correlation function
\begin{equation}
\cor{}{\vp_+(t,\vec k)\vp_+(t,-\vec k)}'_{|\,\Lo(\kappa)}
\end{equation}
receives contributions from the insertions of the quartic vertex $c_{3,1}$, the IC function $\Xi_{3,1}$ and its  counterterm $\xi_{3,1}$, which have all been determined to the required tree-level accuracy in Sec.~\ref{sec:trispectrum}.
In addition, the insertions of the vertex $c_{1,1}$ and IC functions $\xi_{1,1},\,\Xi_{1,1}$ contribute. 

\begin{figure}[t]
\centering
\begin{subfigure}{0.5\textwidth}
\centering
\includegraphics[width=\textwidth]{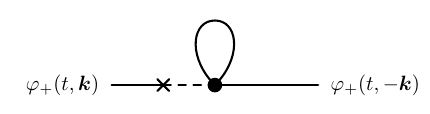}
\caption{}
\end{subfigure}%
\begin{subfigure}{0.5\textwidth}
\centering
\includegraphics[width=\textwidth]{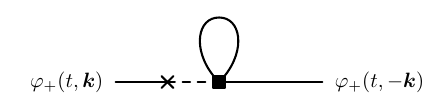}
\caption{}
\end{subfigure}
\caption{The two diagram classes resulting from the insertion of the quartic SdSET vertex and four-point initial-condition counterterm into the $\vp_+$-power spectrum. Only the diagrams with $(+)$-type Schwinger-Keldysh vertices are shown.}
\label{fig:pwrrelevant}
\end{figure}

Before performing the calculation, it proves useful to explore the structure of the contributions and their UV divergences to organise the calculation.
The insertion of the quartic vertex $c_{3,1}$ and IC counterterm $\xi_{3,1}$, shown in Fig.~\ref{fig:pwrrelevant}, result in tadpole diagrams. 
The vertex diagram contains a divergent time integral, whose pole is cancelled by $\xi_{3,1}$. The momentum integrals also generate UV poles, which must be subtracted by the pole part of the mass counterterm $c_{1,1}$ for the vertex diagram\footnote{Recall that for mass renormalisation, we do not use the $\overline{\mathrm{MS}}$ scheme but allow for a finite part in the mass counterterm, as explained below~\eqref{eq::m0lateps}.} 
and $\xi_{1,1}$ for the $\xi_{3,1}$ counterterm diagram, as will become evident from the time-dependence of the respective prefactors.

In addition, the two-point vertex requires the analytic regulator $(\nu/a(t)H)^{-2\delta}$ specified in~\eqref{eq::Sintlead}, and the poles in $\delta$ must cancel between $c_{1,1}$ and $\xi_{1,1}$.
The insertion of the finite IC function $\Xi_{3,1}$ results in another divergent momentum integral, and its $\varepsilon$-poles must be absorbed in $\xi_{1,1}$ as they are time-independent. 
This checks the consistency of the renormalisation of SdSET.
Finally, the IC $\Xi_{1,1}$ is finite by construction and is then obtained by matching, together with the finite part of the mass counterterm $\hat{c}_{1,1}$.

This leads us to structure the calculation as follows: First, we determine the contribution of the sum $c_{3,1}$ and $\xi_{3,1}$, and subtract the resulting UV pole from the momentum integral via $c_{1,1}$, determining its pole part.
The result then contains left-over poles in $\delta$.
Next, we compute the IC insertion $\Xi_{3,1}$, collect all remaining poles and determine the IC counterterm $\xi_{1,1}$. The final result must then be finite and not contain any poles in either $\varepsilon$ or $\delta$.
In the end, we compute the insertion of the IC function $\Xi_{1,1}$.
We proceed to match the correlator onto the full theory, keeping in mind additional contributions due to field redefinitions. This matching completely determines the finite pieces of $c_{1,1}$ and $\Xi_{1,1}$.

\subsubsection{Insertions of the leading-power quartic vertex and $\xi_{3,1}$}

\begin{figure}[t]
\centering
\begin{subfigure}{0.5\textwidth}
\centering
\includegraphics[width=\textwidth]{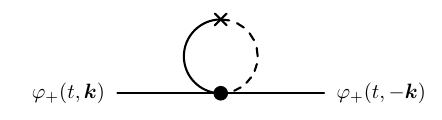}
\caption{}
\end{subfigure}%
\begin{subfigure}{0.5\textwidth}
\centering
\includegraphics[width=\textwidth]{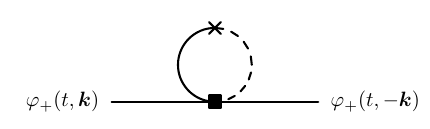}
\caption{}
\end{subfigure}
\caption{The two vanishing diagram classes resulting from the insertion of the quartic SdSET vertex and four-point initial-condition counterterm into the $\vp_+$-power spectrum. Only the diagrams with $(+)$-type Schwinger-Keldysh vertices are shown.}
\label{fig:pwrvanish}
\end{figure}

We start with the contributions to the one-loop power spectrum from the insertions of the SdSET interaction vertex $c_{3,1}\vp_+^3\vp_-$, and IC counterterm $\xi_{3,1}\vp_+^3\vp_-$. The non-vanishing contributions are shown in Fig.~\ref{fig:pwrrelevant}. The diagrams of the type shown in Fig.~\ref{fig:pwrvanish} involve an (anti-)time-ordered $\cor{}{\vp_+\vp_-}$ two-point function evaluated at the same time, which vanishes, see \eqref{eq::mixedequalt}. These diagrams therefore vanish identically. 
This is a general observation: diagrams containing a $\cor{}{\vp_+\vp_-}$-line attaching to the same vertex vanish. We therefore ignore them in the following. 

Moreover, for the external momentum $k$, one can set $k_{\Lambda}\rightarrow k$ after computing the loop integral, since the result does not involve any inverse powers of $\Lambda$. Thus, the expansion of the result as $\Lambda\rightarrow0$ is equivalent to dropping the subscript $\Lambda$ for the external momentum.

The sum of all contributing topologies to the vertex and its IC counterterm then yields
\begin{flalign}
\cor{}{\vp_+(t,\vec k)\vp_+(t,-\vec k)}'_{|\,c_{3,1}+\xi_{3,1}}&=-\frac{1}{12k^3}
\biggl( 
c_{3,1}\int_{-\infty}^t\der t'\;a(t')^{2\ve}
-a_*^{2\varepsilon}\xi_{3,1}\biggr)
\,\tmu^{2\ve}\!\int\frac{\der^{d-1}l}{(2\pi)^{d-1}}\frac{1}{l^3_{\Lambda}}\nonumber\\
&=-\frac{c_{3,1}}{12k^3}\frac{a(t)^{2\ve}-a^{2\ve}_*}{2\ve}\bigg(\frac{\Lambda}{ a(t)\mu}\bigg)^{\!-2\ve}e^{\varepsilon\gamma_E}\Gamma(\ve)\nn\\
&=\frac{c_{3,1}}{48\pi^2k^3}\bigg(\frac{\Lambda}{a(t)\mu}\bigg)^{\!-2\ve}\log\bigg(\frac{a_*}{a(t)}\bigg)\,\bigg[\frac{1}{\ve}+\log\bigg(\frac{a_*}{a(t)}\bigg)\bigg]\,,
\label{eq::c31pwrunexpand}
\end{flalign}
where we used $\xi_{3,1}=\frac{c_{3,1}}{2\ve}$. As anticipated, the result only contains a single pole in $\ve$, due to the cancellation of the time-integral divergence between the two contributions in the combination $[a^{2\ve}_*-a(t)^{2\ve}]/(2\ve)$. The remaining  pole is a UV pole stemming from the momentum integral. It must be subtracted by an appropriate counterterm contained in one of the SdSET vertices, since it needs to come with a time-dependent factor. The vertex $c_{1,1}\vp_+\vp_-$ fulfils this requirement.

\subsubsection{Insertion of the bilinear vertex $c_{1,1}$}

We next turn to the insertion of the bilinear vertex
\begin{equation}
-\int\der^{d-1}x\der t\;\bigg(\frac{\nu}{a(t)H}\bigg)^{-2\delta}c^0_{1,1}\vp_+(t,\vec x)\vp_-(t,\vec x)\,.
\end{equation}
The coupling $c^0_{1,1}$ plays the role of a bare mass term in SdSET.

\begin{figure}[t]
\centering
\begin{subfigure}{0.5\textwidth}
\centering
\raisebox{-0.4cm}[0pt][0pt]{%
\includegraphics[width=0.95\textwidth]{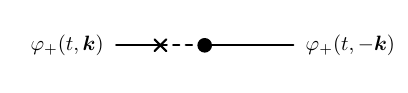}}
\caption{}
\end{subfigure}%
\begin{subfigure}{0.5\textwidth}
\centering
\includegraphics[width=\textwidth]{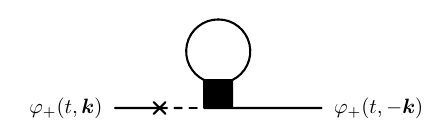}
\caption{}
\end{subfigure}
\caption{The two types of diagrams resulting from the insertion of the SdSET two-point Lagrangian interaction $c^0_{1,1}$ (left panel) and renormalised IC function $\Xi_{3,1}$ (right panel) into the $\vp_+$-power spectrum. Only the diagrams with $(+)$-type Schwinger-Keldysh vertices are shown.}
\label{fig:pwrmassct}
\end{figure}

When inserted into the power spectrum it leads to diagrams as shown in \figref{fig:pwrmassct}~(a), corresponding to the expression
\begin{flalign}
\cor{}{\vp_+(t,\vec k)\vp_+(t,-\vec k)}'_{|\,c_{1,1}}&=2ic^0_{1,1}\frac{i}{3}\frac{1}{2k_{\Lambda}^3}\bigg(\frac{\nu}{H}\bigg)^{-2\delta}\int_{-\infty}^t\der t'\;a(t')^{2\delta}\nonumber\\
&=-\frac{c^0_{1,1}}{3k^3}\bigg(\frac{\nu}{a_*H}\bigg)^{-2\delta}\bigg[\frac{1}{2\delta}-\log\bigg(\frac{a_*}{a(t)}\bigg)\bigg]\,.
\label{eq:c11insunexp}
\end{flalign}
In the above equation we employ the bare mass counterterm $c^0_{1,1}$ as a shorthand to denote the sum of its divergent and finite parts. The $\delta$-pole, which is a time-integral UV pole, demonstrates the necessity of introducing the analytic regulator $(\nu/a(t)H)^{-2\delta}$ in the SdSET action. 
To write the result expanded in $\delta$ as above, we multiplied and divided it by a factor $a^{2\delta}_*$, since this proves to be the most convenient representation in the following. 

Summing \eqref{eq::c31pwrunexpand} and \eqref{eq:c11insunexp}, we get
\begin{flalign}
\cor{}{\vp_+(t,\vec k)\vp_+(t,-\vec k)}'_{|\,c_{3,1}+\xi_{3,1}+c_{1,1}}
&=\frac{c_{3,1}}{48\pi^2k^3}\bigg(\frac{\Lambda}{a(t)\mu}\bigg)^{-2\ve}\log\bigg(\frac{a_*}{a(t)}\bigg)\bigg[\frac{1}{\ve}+\log\bigg(\frac{a_*}{a(t)}\bigg)\bigg]\nonumber\\
&\phantom{=}-\frac{c^0_{1,1}}{3k^3}\bigg(\frac{\nu}{a_*H}\bigg)^{-2\delta}\bigg[\frac{1}{2\delta}-\log\bigg(\frac{a_*}{a(t)}\bigg)\bigg]\,.
\label{eq::pwrcomb1}
\end{flalign}
The pole in $\ve$ can now be subtracted consistently by means of $c^0_{1,1}$, since it appears multiplied by the correct time-dependent factor, by choosing
\begin{flalign}
c^0_{1,1}=-\frac{c_{3,1}}{16\pi^2\ve}+c_{1,1}
\equiv\frac{c_{3,1}}{8\pi^2}\bigg[-\frac{1}{2\ve}+\hat c_{1,1}\bigg]
\label{eq::c11}\,.
\end{flalign}
In the above expressions $c_{1,1}$ denotes the still-undetermined finite part of the bare coefficient $c^0_{1,1}$, which will be determined below by matching onto the full-theory result. For later convenience, we have also introduced its rescaled version $\hat c_{1,1}$. 
With this choice, \eqref{eq::pwrcomb1} reads
\begin{flalign}
&\cor{}{\vp_+(t,\vec k)\vp_+(t,-\vec k)}'_{|\,c_{3,1}+\xi_{3,1}+c_{1,1}}\nonumber\\
&=\frac{c_{3,1}}{24\pi^2k^3}\,\Bigg\{\frac{1}{2}\log\bigg(\frac{a_*}{a(t)}\bigg)\bigg[\log\bigg(\frac{a_*}{a(t)}\bigg)-2\log\bigg(\frac{\Lambda}{a(t)\mu}\bigg)+2\hat c_{1,1}\bigg]\nonumber\\
&\phantom{=}+\bigg(\frac{\nu}{a_*H}\bigg)^{-2\delta}\frac{1}{2\delta}\bigg(\frac{1}{2\ve}-\hat c_{1,1}\bigg)\Bigg\}\,,
\label{eq:1lpower1}
\end{flalign}
but is still not UV-finite since $\delta$- and $\delta\times\ve$-poles are left over. However, some contributions to the SdSET power spectrum are still missing, so we postpone the treatment of these left-over poles.

\subsubsection{Insertion of the initial-condition function $\Xi_{3,1}$}

The technically most challenging part of the present computation is the insertion of the renormalised IC function $\Xi_{3,1}$ into the power spectrum. One of the relevant SK diagrams is shown in \figref{fig:pwrmassct}~(b). In \secref{sec:trispectrum} we found 
\begin{flalign}
\Xi_{3,1}(\vec k_1,\vec k_2,\vec k_3,\vec k_4)&=3\kappa\,\Bigg\{\frac{1}{3}\log\bigg(\frac{e^{\gamma_E}k_t}{a_*H}\bigg)\nonumber\\
&\phantom{=}+\bigg(\sum_{j=1}^4k^3_j\bigg)^{-1}\bigg[-\frac{k_1k_2k_3k_4}{k_t}+k_t\bigg(\sum_{j<l}k_jk_l-\frac{4}{9}k_t^2\bigg)\bigg]\Bigg\}
\end{flalign}
for $\Xi_{3,1}$. For the following, we  have to replace 
\begin{equation}
k_i\rightarrow k_{i\Lambda}\,,\quad i=1,...,4\,
\end{equation}
to implement the IR regularisation and compute 
\begin{flalign}
\cor{}{\vp_+(t,\vec k)\vp_+(t,-\vec k)}'_{|\,\Xi_{3,1}}&=-\frac{i(a_*\tmu)^{2\ve}}{4k^3_{\Lambda}}\frac{i}{3}\int\frac{\der^{d-1}l}{(2\pi)^{d-1}}\;\frac{1}{l^3_{\Lambda}}\,\Xi_{3,1}(\vec k,-\vec k,\vec l,-\vec l)\nonumber\\
&=\frac{\kappa(a_*\tmu)^{2\ve}}{8k_{\Lambda}^3}\int\frac{\der^{d-1}l}{(2\pi)^{d-1}}\;\frac{1}{l^3_{\Lambda}}\,\Bigg\{\frac{2}{3}\log\bigg(\frac{2e^{\gamma_E}(k_{\Lambda}+l_{\Lambda})}{a_*H}\bigg)\label{eq:1lpowerXi31}\\
&\phantom{=}-\frac{1}{k_{\Lambda}^3+l_{\Lambda}^3}\bigg[\frac{k_{\Lambda}^2l_{\Lambda}^2}{2(k_{\Lambda}+l_{\Lambda})}+\frac{2}{9}(7k_{\Lambda}^3+3k_{\Lambda}^2l_{\Lambda}+3k_{\Lambda}l_{\Lambda}^2+7l_{\Lambda}^3)\bigg]\Bigg\}\nonumber\,.
\end{flalign} 
The rational part of the integrand can be simplified to 
\begin{flalign}
&\frac{1}{k_{\Lambda}^3+l_{\Lambda}^3}\bigg[\frac{k_{\Lambda}^2l_{\Lambda}^2}{2(k_{\Lambda}+l_{\Lambda})}+\frac{2}{9}(7k_{\Lambda}^3+3k_{\Lambda}^2l_{\Lambda}+3k_{\Lambda}l_{\Lambda}^2+7l_{\Lambda}^3)\bigg]\nonumber\\
&=\frac{14}{9}+\frac{k_{\Lambda}^2}{6(k_{\Lambda}+l_{\Lambda})^2}+\frac{k_{\Lambda}(-k_{\Lambda}^2+6k_{\Lambda}l_{\Lambda}+4l_{\Lambda}^2)}{6(k_{\Lambda}^3+l_{\Lambda}^3)}\,,
\label{eq:1lXi31rational}
\end{flalign}
which facilitates the integrations considerably, leaving three types of integrals: one involving a logarithm of $k_{\Lambda}+l_{\Lambda}$, one the denominator $(k_{\Lambda}+l_{\Lambda})^2$ and another involving the denominator $k_{\Lambda}^3+l_{\Lambda}^3$. All of these integrals can be evaluated by making use of the Mellin-Barnes identity 
\begin{equation}
\frac{1}{(X+Y)^{\lambda}}=\frac{1}{\Gamma(\lambda)}\frac{1}{2\pi i}\int_{-i\infty}^{i\infty}\der z\;\Gamma(\lambda+z)\Gamma(-z)\frac{X^z}{Y^{\lambda+z}}\,.
\label{eq::MB}
\end{equation}
The details of the computation can be found in \appref{app::pwrloops}.

Evaluating the loop integrals, and keeping only terms which do not vanish as $\Lambda\rightarrow0$, we find 
\begin{align}
\cor{}{\vp_+(t,\vec k)\vp_+(t,-\vec k)}'_{|\,\Xi_{3,1}}&=\frac{\kappa}{96\pi^2k^3}\,\Bigg\{\,\frac{1}{\ve^2}-\frac{2}{\ve}\bigg[\frac{4}{3}-\log\bigg(\frac{e^{\gamma_E}\mu}{H}\bigg)\bigg]+2\log^2\bigg(\frac{2k}{\Lambda}\bigg)\nonumber\\
&\hspace*{-4.1cm}-4\log\bigg(\frac{2k}{\Lambda}\bigg)-2\log\bigg(\frac{\Lambda}{a_*\mu}\bigg)\bigg[\log\bigg(\frac{\Lambda}{a_*\mu}\bigg)+2\log\bigg(\frac{e^{\gamma_E}\mu}{H}\bigg)-\frac{8}{3}\bigg]+5-\frac{20\pi}{3\sqrt{3}}+\frac{5\pi^2}{12}\,\Bigg\}\,.
\end{align}
The result features both double- and single-UV poles, the former being generated by the logarithmic term appearing in the integrand. They are local and independent of $\Lambda$, and therefore removable by means of a local counterterm. 

\subsubsection{Insertion of $\xi_{1,1}$, renormalisation of the remaining UV divergences}

The remaining contribution to the SdSET power spectrum comes from the bare initial-condition function $\Xi^0_{1,1}$, which we split into a counterterm and a finite part.
The counterterm enters the IC functional $F[\vp_{\pm}]$ \eqref{eq::ICF} via the term
\begin{equation}
(a_*\nu)^{2\delta}\xi_{1,1}\int\frac{\der^{d-1}k}{(2\pi)^{d-1}}\;\vp_+(t_*,\vec k)\vp_-(t_*,-\vec k)\,,
\label{eq::xi11vert}
\end{equation}
which, when inserted into the EFT power spectrum, yields
\begin{flalign}
\cor{}{\vp_+(t,\vec k)\vp_+(t,-\vec k)}'_{|\,\xi_{1,1}}&=-\frac{i}{2k^3}\frac{i}{3}\bigg(\frac{\nu}{a_*H}\bigg)^{\!-2\delta}\xi_{1,1}\,.
\end{flalign}
In the above result we have already let $\Lambda\rightarrow0$. Since this is the last counterterm at our disposal for the present computation, we must require the sum
\begin{flalign}
&\cor{}{\vp_+(t,\vec k)\vp_+(t,-\vec k)}'_{|\,c_{3,1}+\xi_{3,1}+\Xi_{3,1}+c_{1,1}+\xi_{1,1}}\nonumber\\
&=\frac{c_{3,1}}{24\pi^2k^3}\,\Bigg\{\,\frac{1}{2}\log\bigg(\frac{a_*}{a(t)}\bigg)\bigg[\log\bigg(\frac{a_*}{a(t)}\bigg)-2\log\bigg(\frac{\Lambda}{a(t)\mu}\bigg)+2\hat c_{1,1}\bigg]+\frac{(a_*\nu)^{2\delta}}{2\delta}\bigg(\frac{1}{2\ve}-\hat c_{1,1}\bigg)\Bigg\}\nonumber\\
&\phantom{=}+\frac{\kappa}{96\pi^2k^3}\,\Bigg\{\frac{1}{\ve^2}-\frac{2}{\ve}\,\bigg[\frac{4}{3}-\log\bigg(\frac{e^{\gamma_E}\mu}{H}\bigg)\bigg]-2\log^2\bigg(\frac{\Lambda}{a_*\mu}\bigg)+4\,\bigg[\frac{4}{3}-\log\bigg(\frac{e^{\gamma_E}\mu}{H}\bigg)\bigg]\log\bigg(\frac{\Lambda}{a_*\mu}\bigg)\nonumber\\
&\phantom{=}\hskip2cm+2\log^2\bigg(\frac{2k}{\Lambda}\bigg)-4\log\bigg(\frac{2k}{\Lambda}\bigg)+5-\frac{20\pi}{3\sqrt{3}}+\frac{5\pi^2}{12}\Bigg\}+\bigg(\frac{\nu}{a_*H}\bigg)^{-2\delta}\frac{\xi_{1,1}}{6k^3}
\label{eq::regSdSETpwr}
\end{flalign}
to be finite. This can be achieved by choosing
\begin{equation}
\xi_{1,1}=-\frac{\kappa}{16\pi^2}\bigg\{\frac{1}{\ve^2}-\frac{2}{\ve}\bigg[\frac{4}{3}-\log\bigg(\frac{e^{\gamma_E}\mu}{H}\bigg)\bigg]\bigg\}-\frac{c_{3,1}}{8\pi^2\delta}\bigg[\frac{1}{2\ve}-\hat c_{1,1}\bigg]\,.
\label{eq::x11}
\end{equation}
We can then send $\delta\rightarrow0$ without generating any additional terms from the expansion of the coefficients $(\nu/a_*H)^{-2\delta}$. 
One may wonder why the full-theory coupling $\kappa$ appears in the previous expression for $\xi_{1,1}$ even though it refers to quantity that is defined in SdSET without reference to a UV theory. The reason is that in order to obtain the result in a simple form we made use of the explicit 
expression \eqref{eq::Xi31tree} for $\Xi_{3,1}$ obtained from tree-level matching. From a purely SdSET 
perspective, the counterterm $\xi_{1,1}$ would instead involve the UV pole of the integral in the first line of \eqref{eq:1lpowerXi31} with $\Xi_{3,1}$ the general, unmatched non-local quartic coupling in the SdSET IC functional. 

For future use of \eqref{eq::x11}, it is important to distinguish the poles multiplying the effective coupling $c_{3,1}$ from the ones arising from the integral over $\Xi_{3,1}$, which in the above are proportional to the full-theory coupling $\kappa$. However, for the matching computation this distinction is immaterial, since the two couplings have the same tree-level value. We therefore set $c_{3,1}=\kappa$ in the renormalised result
\begin{flalign}
&\cor{}{\vp_+(t,\vec k)\vp_+(t,-\vec k)}'_{|\,c_{3,1}+\xi_{3,1}+\Xi_{3,1}+c_{1,1}+\xi_{1,1}}\nonumber\\
&=\frac{\kappa}{24\pi^2k^3}\,\Bigg\{\bigg[\frac{7}{3}-\log\bigg(\frac{2e^{\gamma_E}k}{a(t)H}\bigg)\bigg]\log\bigg(\frac{\Lambda}{a(t)\mu}\bigg)+\log\bigg(\frac{2e^{\gamma_E}k}{a(t)H}\bigg)\,\bigg[\frac{1}{2}\log\bigg(\frac{2e^{\gamma_E}k}{a(t)H}\bigg)\nonumber\\
&\phantom{=}-\log\bigg(\frac{e^{\gamma_E}\mu}{H}\bigg)-1\bigg]-\bigg[\frac{4}{3}-\log\bigg(\frac{e^{\gamma_E}\mu}{H}\bigg)-\hat c_{1,1}\bigg]\log\bigg(\frac{a_*}{a(t)}\bigg)\nonumber\\
&\phantom{=}+\bigg[\frac{1}{2}\log\bigg(\frac{e^{\gamma_E}\mu}{H}\bigg)+1\bigg]\log\bigg(\frac{e^{\gamma_E}\mu}{H}\bigg)+\frac{5}{4}-\frac{5\pi}{3\sqrt{3}}+\frac{5\pi^2}{48}\,\Bigg\}\,.
\end{flalign}

\subsubsection{Insertion of $\Xi_{1,1}$}

Finally, we compute the insertion of $\Xi_{1,1}$, which enters $F[\vp_{\pm}]$ through the term
\begin{equation}
\bigg(\frac{\nu}{a_*H}\bigg)^{\!-2\delta}\int\frac{\der^{d-1}k}{(2\pi)^{d-1}}\;\Xi_{1,1}(\vec k)\vp_+(t_*,\vec k)\vp_-(t_*,-\vec k)\,,
\label{eq::Xi11vert}
\end{equation}
and, in full analogy to the case of $\xi_{1,1}$, it leads to the expression
\begin{flalign}
\cor{}{\vp_+(t,\vec k)\vp_+(t,-\vec k)}'_{|\,\Xi_{1,1}}=\bigg(\frac{\nu}{a_*H}\bigg)^{\!-2\delta}\,\frac{\Xi_{1,1}(\vec k)}{6k^3}\,.
\end{flalign}
We can set $\delta\rightarrow0$, since the function $\Xi_{1,1}(\vec k)$ is finite. The full expression for the renormalised SdSET power spectrum, at order $\kappa$ in the perturbative expansion reads
\begin{flalign}
&\cor{}{\vp_+(t,\vec k)\vp_+(t,-\vec k)}'_{|\,\Lo(\kappa)}=\cor{}{\vp_+(t,\vec k)\vp_+(t,-\vec k)}'_{|\,c_{3,1}+\xi_{3,1}+\Xi_{3,1}+c_{1,1}+\xi_{1,1}+\Xi_{1,1}}\nonumber\\[0.1cm]
&=\frac{\kappa}{24\pi^2k^3}\,\Bigg\{\bigg[\frac{7}{3}-\log\bigg(\frac{2e^{\gamma_E}k}{a(t)H}\bigg)\bigg]\log\bigg(\frac{\Lambda}{a(t)\mu}\bigg)+\log\bigg(\frac{2e^{\gamma_E}k}{a(t)H}\bigg)\bigg[\frac{1}{2}\log\bigg(\frac{2e^{\gamma_E}k}{a(t)H}\bigg)\nonumber\\
&\phantom{=}-\log\bigg(\frac{e^{\gamma_E}\mu}{H}\bigg)-1\bigg]-\bigg[\frac{4}{3}-\log\bigg(\frac{e^{\gamma_E}\mu}{H}\bigg)-\hat c_{1,1}\bigg]\log\bigg(\frac{a_*}{a(t)}\bigg)\nonumber\\
&\phantom{=}+\bigg[\frac{1}{2}\log\bigg(\frac{e^{\gamma_E}\mu}{H}\bigg)+1\bigg]\log\bigg(\frac{e^{\gamma_E}\mu}{H}\bigg)+\frac{5}{4}-\frac{5\pi}{3\sqrt{3}}+\frac{5\pi^2}{48}\Bigg\}+\frac{\Xi_{1,1}(\vec k)}{6k^3}\,.
\end{flalign}
It contains two unknown quantities, $\hat c_{1,1}$ and $\Xi_{1,1}(\vec k)$, which both need to be determined by matching the above expression to the full-theory power spectrum. 

\subsection{Matching computation}

The renormalised full-theory one-loop power spectrum computed in the present UV- and IR-regularisation schemes can be extracted from \cite{Beneke:2023wmt} by replacing $a(t)\Lambda\rightarrow \Lambda$ and $e^{\gamma_E}\mu_f\rightarrow\mu_f$ in that result,\footnote{The first substitution compensates for the fact that in \cite{Beneke:2023wmt} the IR regulator $a(t)\Lambda$ rather than $\Lambda$ was implemented, while the second is due to an additional factor $e^{\gamma_E}$ multiplying the renormalisation scale $\mu_f$ that was erroneously included.} where $\mu_f$ denotes the full-theory renormalisation scale, resulting in
\begin{flalign}
\cor{}{\phi(t,\vec k)\phi(t,-\vec k)}'_{|\,\Lo(\kappa)}&=\frac{\kappa H^2}{24\pi^2k^3}\Bigg\{\Big[2-\log\bigg(\frac{2e^{\gamma_E}k}{a(t)H}\bigg)\bigg]\bigg[\log\bigg(\frac{\Lambda}{a(t)\mu_f}\bigg)-\delta\hat m^2_{\textrm{fin}}\bigg]\nonumber\\
&\phantom{=}+\frac{1}{2}\log^2\bigg(\frac{2e^{\gamma_E}k}{a(t)H}\bigg)-\frac{7}{3}\log\bigg(\frac{2e^{\gamma_E}k}{a(t)H}\bigg)+\frac{8}{3}-\frac{\pi^2}{24}\Bigg\}\,,
\label{eq::fullthpwr}
\end{flalign}
with $\delta\hat m^2_{\textrm{fin}}$ as defined in \eqref{eq::deltamfin}. The non-linear relation \eqref{eq::nonlinearphi} between full-theory and EFT fields implies the following $d=4-2\ve$-dimensional matching equation between full-theory and EFT correlators, at $\Lo(\kappa)$:
\begin{flalign}
&\cor{}{\phi(t,\vec k)\phi(t,-\vec k)}'_{|\,\Lo(\kappa)}\nonumber\\
&=H^2a(t)^{2\ve}\bigg[\cor{}{\vp_+(t,\vec k)\vp_+(t,-\vec k)}'_{|\,\Lo(\kappa)}+\frac{2c^0_{1,1}}{9}\cor{}{\vp_+(t,\vec k)\vp_+(t,-\vec k)}'_{|\,\Lo(\kappa^0)}\nonumber\\
&\phantom{=}+\frac{c^0_{3,1}a(t)^{2\ve}}{18(3+2\ve)}\Big(\cor{}{\vp^3_+(t,\vec k)\vp_+(t,-\vec k)}'_{|\,\Lo(\kappa^0)}+\cor{}{\vp_+(t,\vec k)\vp^3_+(t,-\vec k)}'_{|\,\Lo(\kappa^0)}\Big)\bigg]\,.
\label{eq::pwrmatch}
\end{flalign}
The notation \eqref{eq::vpnFT} for the Fourier transforms of the composite operators $\vp^n_+$ is used here. We will find that the combination of the last three EFT correlation functions given in the above equation in \eqref{eq::pwrmatch} is finite, so we can set the overall coefficient $a(t)^{2\ve}$ to one. The subscripts of the EFT correlation functions denote at which order in the perturbative expansion in the couplings and initial conditions they should be evaluated.

\subsubsection{Correlators following from the $\vp_-$-field redefinition}

We need to compute
\begin{flalign}
&\cor{}{\vp_+(t,\vec k)\vp_+(t,-\vec k)}'_{|\,\textrm{redef.}}\nonumber\\
&\equiv\frac{2c^0_{1,1}}{9}\cor{}{\vp_+(t,\vec k)\vp_+(t,-\vec k)}'_{|\,\Lo(\kappa^0)}+\frac{c^0_{3,1}a(t)^{2\ve}}{18(3+2\ve)}\Big[\cor{}{\vp^3_+(t,\vec k)\vp_+(t,-\vec k)}'_{|\,\Lo(\kappa^0)}\nonumber\\
&\phantom{\equiv}+\cor{}{\vp_+(t,\vec k)\vp^3_+(t,-\vec k)}'_{|\,\Lo(\kappa^0)}\Big]\,.
\end{flalign}
Since both $c^0_{1,1}$ and $c^0_{3,1}$ are $\Lo(\kappa)$, we evaluate the correlators in the Gaussian approximation. Therefore, the correlation function that multiplies $c^0_{1,1}$ is simply the $\vp_+$-power spectrum,
\begin{equation}
\cor{}{\vp_+(t,\vec k)\vp_+(t,-\vec k)}'_{|\,\Lo(\kappa^0)}=\frac{1}{2k_{\Lambda}^3}\,.
\end{equation}
For the composite operator two-point correlator it is most convenient to start by inspecting their position-space version, which for $\cor{}{\vp^3_+(t,\vec k)\vp_+(t,-\vec k)}'_{|\,\Lo(\kappa^0)}$ reads
\begin{equation}
\cor{}{\vp^3_+(t,\vec x)\vp_+(t,\vec y)}_{|\,\Lo(\kappa^0)}=3\,\cor{}{\vp_+(t,\vec x)\vp_+(t,\vec y)}_{|\,\Lo(\kappa^0)}\times\cor{}{\vp^2_+(t,\vec x)}_{|\,\Lo(\kappa^0)}\,.
\label{eq::pwrredef}
\end{equation}
An analogous expression with $\vec x\leftrightarrow\vec y$ holds for $\cor{}{\vp_+(t,\vec k)\vp^3_+(t,-\vec k)}'_{|\,\Lo(\kappa^0)}$. Since the one-point function
\begin{equation}
\cor{}{\vp^2_+(t,\vec x)}_{|\,\Lo(\kappa^0)}=\frac{1}{2}\int\frac{\der^{d-1}l}{(2\pi)^{d-1}}\frac{1}{l^3_{\Lambda}}
\end{equation}
is a (divergent) constant, the Fourier transform of the product on the right-hand side of  \eqref{eq::pwrredef} is also proportional to the Gaussian $\vp_+$-power spectrum. Putting the pieces together we find
\begin{flalign}
&\cor{}{\vp_+(t,\vec k)\vp_+(t,-\vec k)}'_{|\,\textrm{redef.}}=\frac{c^0_{1,1}}{9k_{\Lambda}^3}\frac{1}{2k^3}+\frac{\tmu^{2\ve}c_{3,1}}{(3+2\ve)}\frac{a(t)^{2\ve}}{2k^3}\int\frac{\der^{d-1}l}{(2\pi)^{d-1}}\frac{1}{l^3_{\Lambda}}\nonumber\\
&=\frac{c^0_{1,1}}{9k^3}+\frac{c_{3,1}}{2k^3(3+2\ve)}\bigg(\frac{\Lambda}{a(t)\mu}\bigg)^{-2\ve}\frac{e^{\ve\gamma_E}}{4\pi^2}\Gamma(\ve)
\nonumber\\[0.1cm]
&=\frac{c_{3,1}}{24\pi^2k^3}\bigg[-2\log\bigg(\frac{\Lambda}{a(t)\mu}\bigg)+\frac{\hat c_{1,1}}{3}-\frac{2}{3}\bigg]\,.
\label{eq:1lpowerfieldredef}
\end{flalign}
To obtain the last line, we inserted the expression \eqref{eq::c11} for the mass counterterm $c^0_{1,1}$.
Even though the individual terms generated by the field redefinition are divergent, their sum is UV finite. This is required by consistency, since one matches renormalised EFT correlators onto renormalised full-theory ones, and no UV divergences on either side of the matching equation can be left over.

\subsubsection{Matching of $\hat c_{1,1}$ and $\Xi_{1,1}$}

We now have all the necessary pieces to carry out the matching procedure. Summing \eqref{eq::pwrmatch} and \eqref{eq:1lpowerfieldredef} and identifying $c_{3,1}=\kappa$ 
by their tree-level matching, we obtain 
\begin{flalign}
&\cor{}{\vp_+(t,\vec k)\vp_+(t,-\vec k)}'_{|\,\Lo(\kappa)}+\cor{}{\vp_+(t,\vec k)\vp_+(t,-\vec k)}'_{|\,\textrm{redef.}}\nonumber\\
&=\frac{\kappa}{24\pi^2k^3}\Bigg\{\bigg[2-\log\bigg(\frac{2e^{\gamma_E}k}{a(t)H}\bigg)\bigg]\bigg[\log\bigg(\frac{\Lambda}{a(t)\mu}\bigg)-\hat c_{1,1}\bigg]+\frac{1}{2}\log^2\bigg(\frac{2e^{\gamma_E}k}{a(t)H}\bigg)-\frac{7}{3}\log\bigg(\frac{2e^{\gamma_E}k}{a(t)H}\bigg)\nonumber\\
&\phantom{=}+\bigg[\frac{4}{3}-\log\bigg(\frac{e^{\gamma_E}\mu}{H}\bigg)-\hat c_{1,1}\bigg]\log\bigg(\frac{2e^{\gamma_E}k}{a_*H}\bigg)+\bigg[1+\frac{1}{2}\log\bigg(\frac{e^{\gamma_E}\mu}{H}\bigg)\bigg]\log\bigg(\frac{e^{\gamma_E}\mu}{H}\bigg)+\frac{7\hat c_{1,1}}{3}\nonumber\\
&\phantom{=}+\frac{41}{36}-\frac{5\pi}{3\sqrt{3}}+\frac{5\pi^2}{48}\Bigg\}+\frac{\Xi_{1,1}(\vec k)}{6k^3}\,.
\end{flalign}
This expression, multiplied by an overall factor of $H^2$, needs to reproduce the renormalised full-theory result \eqref{eq::fullthpwr}. The matching equation \eqref{eq::pwrmatch} determines the finite part of the EFT mass counterterm $\hat c_{1,1}$ and $\Xi_{1,1}(\vec k)$ at the one-loop order to
\begin{equation}
    \hat c_{1,1}=\delta\hat m^2_{\textrm{fin}}+\log\bigg(\frac{\mu_f}{\mu}\bigg)\,,    
\end{equation}
and
\begin{flalign}
\Xi_{1,1}(\vec k)=\,&-\frac{\kappa}{4\pi^2}\,\Bigg\{\bigg[\frac{4}{3}-\log\bigg(\frac{e^{\gamma_E}\mu}{H}\bigg)-\hat c_{1,1}\bigg]\log\bigg(\frac{2e^{\gamma_E}k}{a_*H}\bigg)\nonumber\\
&+\bigg[1+\frac{1}{2}\log\bigg(\frac{e^{\gamma_E}\mu}{H}\bigg)\bigg]\log\bigg(\frac{e^{\gamma_E}\mu}{H}\bigg)+\frac{7\hat c_{1,1}}{3}-\frac{55}{36}-\frac{5\pi}{3\sqrt{3}}+\frac{7\pi^2}{48}\Bigg\}\,.
\label{eq:Xi11-1loop}
\end{flalign}

At this point, we recall that the finite part of the full-theory mass counterterm $\delta\hat m^2_{\textrm{fin}}$ is still undetermined, because we cannot formulate a physically sensible renormalisation scheme based on IR-divergent perturbative results in the full theory. Here we see that 
$\delta\hat m^2_{\textrm{fin}}$ enters $\hat c_{1,1}$ and $\Xi_{1,1}(\vec k)$   via matching, and through them ultimately the non-perturbatively computed correlation functions and in particular the power spectrum, on which a sensible mass renormalisation condition can be imposed. 

Eq.~\eqref{eq:Xi11-1loop} is the first example of a one-loop matching computation to SdSET, which features both time- and momentum-integral divergences. In particular, even though both the full-theory and SdSET two-point functions are IR divergent, which is evident from their explicit dependence on $\Lambda$, the $\Lambda$-dependent terms of the former are faithfully reproduced by the latter. This allowed us to determine the renormalised quantities $\hat c_{1,1}$ and $\Xi_{1,1}(\vec k)$ from these perturbative results. Consequently, the found expressions for these objects are IR-insensitive, and should therefore be independent of the choice of IR-regularisation that was used for the above computation. It would be interesting to check this explicitly with another consistent IR regulator. 

The existence of time- and momentum {\em ultraviolet} divergences of the SdSET two-point function is reflected in \eqref{eq:Xi11-1loop} by logarithms of two types. The logarithm $\ln(2e^{\gamma_E} k/(a_* H))$ arises from the time integral, and it is familiar from the previous two tree-level matching examples. As appropriate for an initial-condition matching coefficient, this logarithm is not parametrically large when the matching time $t_*$ is taken of order of the horizon-crossing time of the mode $k$. The presence of $a_*$ renders the logarithm consistent with dS dilatations. The second logarithm $\ln(e^{\gamma_E}\mu/H)$ is related to momentum-integral UV divergences, 
and it is small when the renormalisation scale is taken to be of order $H$. This is exactly as it should be  \cite{Senatore:2009cf} for an infrared-insensitive quantity such as a matching coefficient. 

\section{Conclusion}

Soft de Sitter Effective Theory (SdSET) provides a compelling framework by which  the stochastic approach~\cite{Starobinsky:1982ee,Starobinsky:1986fx,Starobinsky:1994bd} may be consistently extended beyond the leading (logarithmic) order in the summation of the IR and secular logs~\cite{Cohen:2020php,Cohen:2021fzf}. 
It is a systematic EFT approach based on power-counting, mode separation and symmetries, and a natural candidate to study the infrared physics of quantum fields in de Sitter space.
This article is the first in a series of two which aims at expanding on the pioneering works~\cite{Cohen:2020php,Cohen:2021fzf} with the ultimate goal of formulating 
regularisation, renormalisation, matching and the derivation of the Kramers-Moyal equation in a way that higher-order computations pose technical but no further conceptual complications.  Experience with new frameworks in flat-space QFT shows that this requires computing to high enough orders that subdivergences and the recursive structure of renormalisation become relevant and that issues related to renormalisation-scheme independence of observables can be addressed. 

The set-up of SdSET presented in the first sections of this paper follows closely the physical picture and ideas of the original formulation~\cite{Cohen:2020php,Cohen:2021fzf}, yet started from the derivation of basic ingredients  in order to supplement their reasoning with explicit and technical arguments (for example, concerning the justification of a naively power-counting violating field redefinition) and to set up the regularisation scheme that differs from~\cite{Cohen:2020php,Cohen:2021fzf}.  Specifically, we employ dimensional regularisation for UV divergences together with an evanescent mass term~\cite{Melville:2021lst} that keeps the index of the dS Hankel mode function equal to 
$\frac{3}{2}$, which greatly simplifies the calculation of the finite parts of loop integrals and matching coefficients. IR divergences of momentum integrals are regulated by a ``comoving mass term'' that preserves this feature and simply shifts $k^2\to k^2+\Lambda^2$. It remains an interesting question whether matching can be simplified further by employing dimensional regularisation simultaneously for the UV and IR as is routinely done for matching calculations of quark-gluon amplitudes in flat-space EFTs of the strong interactions. We summarise the main results of this work as follows:
\begin{itemize}
\item The requirement of non-Gaussian initial conditions for SdSET was introduced in \cite{Cohen:2020php} and is physically intuitive as SdSET is valid only at late times 
$-k\eta\ll 1$, when weak correlations have already built up by the early-time evolution in the full theory. We presented a general implementation of these ICs in the form of an additional time-independent but spatially highly non-local contribution to the action, following \cite{Garny:2009ni}. This introduces a subtraction scale factor $a_*$ on top of the scale $\mu$ associated with the dimensional regulator.
\item  We match the tree-level four-point and six-point functions and from them determine the corresponding SdSET Lagrangian couplings and IC functions. These calculations demonstrate that the effective theory correctly reproduces the IR and late-time physics of the massless minimally coupled scalar field, as the matching coefficients are free from IR divergences and secular terms. The matching of the six-point function is non-trivial, since it involves two (nested) time integrals, and checks the recursive subtraction of time-integral divergences, including the previously determined four-point IC counterterms. Eq.~\eqref{eq:full6pt} and \eqref{eq:fdef} in App.~\ref{app::fullpenta} present a fully analytic expression for the late-time asymptotic behaviour of the tree-level six-point function in the massless $\phi^4$ theory obtained via the cosmological expansion by regions \cite{Beneke:2023wmt}.
\item The requirement that a scalar remains massless requires fine-tuning. In this connection, we pay attention to the systematics of mass renormalisation, since unlike flat-space field theories, the mass counterterm in the full theory is not IR-finite and should be determined only after matching to SdSET and resummation. By computing and matching the two-point function (power spectrum) at the one-loop order, we determine  the corresponding two-point SdSET Lagrangian and IC functions, which inherit the mass counterterm from the full theory. This example features one divergent 
time- and one divergent momentum integral and the interplay of both types of divergences. 
Again we find that the matching coefficients have the required properties to qualify as short-distance, resp. early-time quantities.
\item The existence of time- and momentum {\em ultraviolet} divergences of the SdSET two-point function is reflected in IR-insensitive matching coefficients in logarithms of two types, $\ln(k/(a_* H))$ and $\ln(\mu/H)$. The former 
arises for initial-condition matching functions from the time integrals 
and is not parametrically large when the matching time $t_*$ is taken of order of the horizon-crossing time of the mode $k$. The presence of $a_*$ renders the logarithm consistent with dS dilatations. The second logarithm 
is related to momentum integrals, and it is small when $\mu$ is of order $H$, as it should be. These logarithms play an important role for the computation of anomalous dimensions, as will be shown in \cite{paperII}.
\end{itemize}
These results demonstrate that SdSET is indeed the appropriate effective field theory 
that describes the quantum dynamics of superhorizon modes of the massless scalar field. They do not yet establish a framework for resummation. The key insight of the work of Cohen et al.~\cite{Cohen:2020php,Cohen:2021fzf} is that the Kramers-Moyal equation, which generalises the Fokker-Planck equation, should follow the fact that SdSET of the massless scalar contains infinitely many relevant interactions and composite operators, whose 
anomalous dimensions are short-distance quantities that build the Kramers-Moyal coefficients. In the sequel to this paper \cite{paperII}, we therefore consider the renormalisation and matching of correlation functions of composite operators. In particular, we shall determine for the first time the one-loop correction to the 
diffusion coefficient in the Fokker-Planck equation and short-distance, composite-operator matching coefficients, both of which contribute to the long-distance dynamics at next-to-next-to-leading order. 

\subsubsection*{Acknowledgement}

We thank Tim Cohen for many discussions and continuous support, Dan Green for valuable
comments on SdSET, and Victor Gorbenko and Riccardo Rattazzi for instructive questions and comments. This work has been supported in part by the Excellence Cluster ORIGINS funded by the Deutsche Forschungsgemeinschaft under Grant No.~EXC - 2094 - 390783311 and by the Cluster of Excellence Precision Physics, Fundamental Interactions, and Structure of Matter (PRISMA$^+$ EXC 2118/1) funded by the German Research Foundation (DFG) within the German Excellence Strategy (Project ID 390831469). The work of AFS is supported by the grants
EUR2024.153549, CNS2024-154834 and PID2022-139466NB-C21 (FEDER/UE) funded by the Spanish Research Agency (MICIU/AEI/10.13039/5011000110\\33).

\appendix

\section{Canonical transformation in \texorpdfstring{$d$}{d} dimensions}
\label{app::freered}

In this Appendix the explicit form of the field redefinition to obtain the free effective action is derived. The procedure is a generalisation of the one used in \cite{Namjoo:2017nia} to rigorously derive the free effective action describing the non-relativistic limit of a real scalar field theory in Minkowski space. 

The starting point is the free, $d$-dimensional full-theory action, which expressed in terms of the dimensionless time variable $\Ht=Ht$ reads
\begin{equation}
S=\frac{1}{2}\int\der^{d-1}x\der\Ht\;Ha(\Ht)^{d-1}\bigg[\dot\phi^2-\frac{(\p_i\phi)^2}{(a(\Ht)H)^2}-\frac{m^2}{H^2}\phi^2\bigg]\,,
\end{equation}
where 
\begin{equation}
    \dot\phi=\frac{\p\phi}{\p\hat t}\,.
\end{equation}
In the following we will abbreviate $a(\Ht)=a$ and always work with the time variable $\Ht$, dropping the hat for brevity.

The goal is now to replace the field $\phi$, which describes a single scalar degree of freedom and satisfies a second-order differential equation in time, by the two fields $\vp_{\pm}$ of the effective theory, which must therefore be a canonically conjugated 
pair of fields, satisfying first-order field equations, amenable to a systematic expansion for superhorizon modes. The appropriate formalism to achieve this is the canonical one, since there the field and its conjugated momentum $\pi$ are a priori two independent variables, which can be used to define a one-to-one mapping from the pair $\{\phi,\pi\}$ to the pair $\{\vp_+,-2\nu\vp_-\}$. 

The canonically conjugated momentum to $\phi$ is 
$\pi=Ha^{d-1}\dot\phi$, resulting in the Hamiltonian
\begin{equation}
    \tilde\Hd =\pi\dot\phi-\Ld\\
    =\frac{Ha^{d-1}}{2}\bigg[\frac{1}{H^2a^{2d-2}}\pi^2+\frac{(\p_i\phi)^2}{(aH)^2}+\frac{m^2}{H^2}\phi^2\bigg]\,.
\end{equation}
The starting ansatz for the decomposition of the field and its canonically conjugated momentum in terms of $\vp_{\pm}$ is
\begin{align}
\phi&=H^{\frac{d}{2}-1}\Big[(aH)^{-\alpha}D_{\nu}(X)\vp_++(aH)^{-\beta}D_{-\nu}(X)\vp_-\Big]\,,\\
\pi&=-a^{d-1}H^{\frac{d}{2}}\Big[\alpha(aH)^{-\alpha}P_{\nu}(X)\vp_++\beta(aH)^{-\beta}P_{-\nu}(X)\vp_-\Big]\,,
\label{eq:pidef}
\end{align}
where
\begin{equation}
\alpha=\frac{d-1}{2}-\nu\,,\quad\beta=\frac{d-1}{2}+\nu\,,\quad\nu=\sqrt{\bigg(\frac{d-1}{2}\bigg)^2-\frac{m^2}{H^2}}\,,
\end{equation}
and we introduced the abbreviation
\begin{equation}
X\equiv\frac{\p^2_i}{(aH)^2}\,.
\label{eq:Xdef}
\end{equation}
This ansatz factors the leading time dependences $(aH)^{-\alpha}$, $(aH)^{-\beta}$ from the effective fields, while $D_{\pm\nu}$ and $P_{\pm\nu}$ are functions of the differential operator $X$. We assume that they can be defined via series expansions in $X$ with 
$D_{\pm\nu}(0)=P_{\pm\nu}(0)=1$ to recover \eqref{eq::fullphitoEFTphi} at leading order in the gradient expansion.
In the following we will suppress their argument and abbreviate them as $D_{\pm}$, $P_{\pm}$.

The equal-time canonical commutation relations for the pairs $\{\phi,\pi\}$ and $\{\vp_+,-2\nu\vp_-\}$ read
\begin{align}
[\phi(t,\vec x),\pi(t,\vec y)]&=i\delta^{(d-1)}(\vec x-\vec y)\,,\\
-2\nu[\vp_+(t,\vec x),\vp_-(t,\vec y)]&=i\delta^{(d-1)}(\vec x-\vec y)\,,
\end{align}
from which we derive the following constraint equation on $D_{\pm}$ and $P_{\pm}$:
\begin{equation}
\beta P_-D_+-\alpha P_+D_-=2\nu\,.
\label{eq::constraint1}
\end{equation}
Taking the time-derivative gives
\begin{equation}
\beta\dot P_-D_+-\alpha P_+\dot D_-=\alpha\dot P_+D_--\beta P_-\dot D_+\,.\label{eq::constraint2}
\end{equation}

The change of variables from $\{\phi,\pi\}$ to $\{\vp_+,-2\nu\vp_-\}$ is achieved by means of a canonical transformation (following \cite{goldstein}). We first express $\pi$ and $\vp_-$ in terms of $\phi$ and $\vp_+$
\begin{align}
\vp_-&=\frac{(aH)^{\beta}}{H^{\frac{d}{2}-1}D_-}\phi-\frac{(aH)^{2\nu}D_+}{D_-}\vp_+\,,\\
\pi&=a^{d-1}H^{\frac{d}{2}}\bigg[\frac{2\nu(aH)^{-\alpha}}{D_-}\vp_+-\frac{\beta P_-}{H^{\frac{d}{2}-1}D_-}\phi\bigg]\,.
\end{align}
The generating functional $F[\phi,\vp_+,t]$ of the canonical transformation has to satisfy\footnote{We will be slightly sloppy with the notation. For the following  manipulations the distinction between the functional $F$ and the function $F$, written as a product of field operators, is immaterial, as long as we are careful about preserving the ordering of the operators.}
\begin{equation}
\pi[\phi,\vp_+]=\frac{\p F}{\p\phi}\,,\quad -2\nu\vp_-[\phi,\vp_+]=-\frac{\p F}{\p\vp_+}\,,
\end{equation}
which is solved by
\begin{equation}
F[\phi,\vp_+,t]=-a^{d-1}\phi\frac{H\beta P_-}{2D_-}\phi+2\nu\vp_+\frac{(aH)^{\beta}}{H^{\frac{d}{2}-1}D_-}\phi-\nu\vp_+\frac{(aH)^{2\nu}D_+}{D_-}\vp_+\,.
\end{equation}
Its partial time derivative, which only acts on the explicit time-dependence in $a(t)$ and $D_{\pm}$, $P_{\pm}$, but not on the implicit one of the fields $\phi$ and $\vp_+$, reads
\begin{align}
\frac{\p F}{\p t}&=-\frac{\alpha(aH)^{2\nu}}{2}\vp_+\Big[(d-1)P_+D_++\dot P_+D_+-P_+\dot D_+\Big]\vp_+-\frac{1}{2}\vp_+\Big[(d-1)\Big(\alpha P_+D_-\nonumber\\
&\phantom{=}+\beta P_-D_+\Big)+\alpha\Big(\dot P_+D_--P_+\dot D_-\Big)+\beta\Big(\dot P_-D_+-P_-\dot D_+\Big)-4\nu^2\Big]\vp_-\nonumber\\
&\phantom{=}-\frac{\beta(aH)^{-2\nu}}{2}\vp_-\Big[(d-1)P_-D_-+\dot P_-D_--P_-\dot D_-\Big]\vp_-\,,
\end{align}
where we used \eqref{eq::constraint1}, \eqref{eq::constraint2} to bring this expression in a manifestly symmetric form  under $\nu\rightarrow-\nu$. Using again \eqref{eq::constraint1}, \eqref{eq::constraint2}, we find the canonically transformed Hamiltonian expressed only in terms of $\vp_{\pm}$:
\begin{align}
\Hd&=\tilde\Hd+\frac{\p F}{\p t}
\nonumber\\
&=\frac{1}{2}\,\Bigg\{(aH)^{2\nu}\vp_+\bigg[\alpha^2 P_+^2-D_+^2\bigg(X+\nu^2-\bigg(\frac{d-1}{2}\bigg)^{\!2}\,\bigg)-\alpha\Big[(d-1)P_+D_++\dot P_+D_+\nonumber\\
&\phantom{=}-P_+\dot D_+\Big]\bigg]\vp_++(aH)^{-2\nu}\vp_-\bigg[\beta^2 P_-^2-D_-^2\bigg(X+\nu^2-\bigg(\frac{d-1}{2}\bigg)^{\!2}\,\bigg)-\beta\Big[(d-1)P_-D_-\nonumber\\
&\phantom{=}+\dot P_-D_--P_-\dot D_-\Big]\bigg]\vp_-+2\vp_+\bigg[\alpha\beta P_+P_--D_+D_-\bigg(X+\nu^2-\bigg(\frac{d-1}{2}\bigg)^{\!2}\,\bigg)+2\nu^2\nonumber\\
&\phantom{=}-\frac{\alpha}{2}\Big((d-1)P_+D_-+\dot P_+D_--P_+\dot D_-\Big)-\frac{\beta}{2}\Big((d-1)P_-D_++\dot P_-D_+-P_-\dot D_+\Big)\bigg]\vp_-\Bigg\}\,.
\label{eq::cantrafoH}
\end{align}
By construction $\Hd$ satisfies the canonical equations
\begin{equation}
\dot\vp_+=-\frac{1}{2\nu}\frac{\p\Hd}{\p\vp_-}\,,\quad -2\nu\dot\vp_-=-\frac{\p\Hd}{\p\vp_+}\,,
\end{equation}
where the second equation is a copy of the first with the replacements $\vp_+\leftrightarrow\vp_-$ and $\nu\rightarrow-\nu$. We can therefore require the equations of motion for $\vp_{\pm}$ to have the form
\begin{equation}
\dot\vp_{\pm}= \mp\nu E\vp_{\pm}
\end{equation}
with $E$ an operator that depends only on $\nu^2$ and is yet to be determined. The corresponding Hamiltonian is 
\begin{equation}
\Hd= 2\nu^2\vp_+E\vp_-\,.
\end{equation}
Hence the $\varphi_+^2$ and $\varphi_-^2$ terms in \eqref{eq::cantrafoH} must vanish, which implies the differential equations
\begin{equation}
f^2_{\pm}+(d-1)f_{\pm}+\dot f_{\pm}-\bigg[X+\nu^2-\bigg(\frac{d-1}{2}\bigg)^2\bigg]=0,
\label{eq::fde}
\end{equation}
for
\begin{equation}
f_{\pm}\equiv-\bigg(\frac{d-1}{2}\mp \nu\bigg)\frac{P_{\pm}}{D_{\pm}}\,.
\end{equation}
Assuming \eqref{eq::constraint1} and \eqref{eq::fde} the Hamiltonian  simplifies to
\begin{equation}
\Hd=\vp_+\bigg[2\nu^2\bigg(1-\frac{1}{D_+D_-}\bigg)+\nu\frac{\dot D_+D_--D_+\dot D_-}{D_+D_-}\bigg]\vp_-\,.
\label{eq::simplifiedh}
\end{equation}

Due to $D_{\pm\nu}(0)=P_{\pm\nu}(0)=1$, the differential equations \eqref{eq::fde} 
should be solved with the boundary conditions
\begin{equation}
\lim\limits_{X\rightarrow 0}f_{\pm}(X)=-\bigg(\frac{d-1}{2}\mp \nu\bigg)+\Lo(X)\,.
\end{equation}
The solutions are found to be
\begin{align}
f_+(X)&=\frac{\sqrt{X}}{2\nu}\frac{\alpha I_{-\nu-1}\big(\sqrt{X}\big)-\beta I_{-\nu+1}\big(\sqrt{X}\big)}{I_{-\nu}\big(\sqrt{X}\big)}\,,\label{eq:fplus}\\
f_-(X)&=-\frac{\sqrt{X}}{2\nu}\frac{\beta I_{\nu-1}\big(\sqrt{X}\big)-\alpha I_{\nu+1}\big(\sqrt{X}\big)}{I_{\nu}\big(\sqrt{X}\big)}\,,\label{eq:fminus}
\end{align}
where $I_{\nu}(z)$ is the modified Bessel function of the first kind. Combining this result with \eqref{eq::constraint1} we further find
\begin{align}
D_+D_-&=\Gamma(1-\nu)\Gamma(1+\nu)I_{\nu}\big(\sqrt{X}\big)I_{-\nu}\big(\sqrt{X}\big)\,,\label{eq:prod1}\\
\alpha\beta P_+P_-&=\frac{X}{4}\,\Gamma(\nu)\Gamma(-\nu)\,\Big[\alpha I_{-\nu-1}\big(\sqrt{X}\big)-\beta I_{-\nu+1}\big(\sqrt{X}\big)\Big]\Big[\beta I_{\nu-1}\big(\sqrt{X}\big)-\alpha I_{\nu+1}\big(\sqrt{X}\big)\Big]\,.
\nonumber\\[-0.2cm]
\label{eq:prod2}
\end{align}
Notice that while the above results feature $\sqrt{X}$, the series expansion depends only on $X$. These expressions should therefore be understood as compact representations of the underlying infinite series in $X$.

At this point the constraints on $D_{\pm}$ and $P_{\pm}$ have been exhausted, but only the products $D_+D_-$ and $P_+P_-$ are uniquely determined, not the operators $D_{\pm}$ and $P_{\pm}$ themselves. We can redefine
\begin{equation}
D_{\pm}(X)=g_{\pm}(X)\check D_{\pm}(X)\,,\quad P_{\pm}(X)=g_{\pm}(X)\check P_{\pm}(X)\,,
\end{equation}
where $g_{\pm}$ absorbs any residual freedom left in the choice of $D_{\pm}$ and $P_{\pm}$. We must require that both $g_{\pm}$ and $\check D_{\pm}$, $\check P_{\pm}$ be 1 at leading power in the expansion in $X$, and further that
\begin{equation}
g_+g_-=1\,,
\end{equation}
in order to still satisfy the above constraints. We can therefore write the $g_{\pm}$ as exponentials
\begin{equation}
g_{\pm}(X)=\exp\Big[\pm\nu\gamma(X)\Big]\,,\quad \gamma(X)=0+\Lo(X)\,,
\end{equation}
with $\gamma$ a function that only depends on $\nu^2$. Once we choose $\check D_{\pm}$, the $\check P_{\pm}$ are fixed by the above equations, or vice-versa. We find that the operators
\begin{align}
\check D_{\pm}(X)&={_0F_1}\bigg(1\mp\nu;\frac{X}{4}\bigg)\,,\label{vanishd}\\
\check P_{\pm}(X)&={_0F_1}\bigg(\!\mp\nu;\frac{X}{4}\bigg)+\frac{\beta X}{4\alpha\nu(\nu\mp1)}\,{_0F_1}\bigg(2\mp\nu;\frac{X}{4}\bigg)\,,
\end{align}
where ${_0F_1}(a;z)$ denotes the confluent hypergeometric function 
\begin{equation}
{_0F_1}(a;z)\equiv\sum_{k=0}^{\infty}\frac{z^k}{k!(a)_k}\,,\quad (a)_n\equiv\prod_{k=0}^{n-1}(a-k)\,,
\end{equation}
meet the necessary requirements: they both satisfy \eqref{eq:fplus} and \eqref{eq:fminus}, or, equivalently, \eqref{eq:prod1} and \eqref{eq:prod2}, and their series expansions for small $X$ start with 1. Substituting the expressions for $D_{\pm}$ into \eqref{eq::simplifiedh} we find that all that is left is
\begin{equation}
\Hd=2\nu^2\vp_+\dot \gamma(X)\vp_-\,.
\label{eq::Hdgamma}
\end{equation}
To write the action as the integral of a Lagrangian we eliminate the canonically conjugated momentum $-2\nu\vp_-$ in favour of $\dot\vp_+$ by means of a Legendre transformation, and get
\begin{align}
S&=\int\der^{d-1}x\der t\Big[-2\nu\dot\vp_+\vp_--\Hd\Big]\nonumber\\
&=-\nu\int\der^{d-1}x\der t\;\bigg[\dot\vp_+\vp_--\vp_+\dot\vp_-+2\nu\vp_+\dot\gamma(X)\vp_-\bigg]\,.
\end{align}
The equations of motion of $\vp_{\pm}$ are fully determined by the form of $\gamma$
\begin{equation}
\dot\vp_{\pm}=\mp\nu\dot\gamma(X)\vp_{\pm}\,.
\end{equation}

Differently than in flat spacetime, energy is not conserved in de Sitter space, which means that $\p\Hd/\p t=0$ is not required. Therefore, we cannot constrain the functional form of $\gamma$ further, which leaves us a considerable amount of freedom. However, we can choose the form of $\gamma$ most suited for our purposes by means of the following argument: taking the limit $m/H\rightarrow \infty$ should recover the free action for the heavy scalar field in Minkowski space derived in \cite{Namjoo:2017nia}, with the replacement $\p_i\rightarrow \p_i/a(t)$. By letting the mass of the field become infinitely large compared to the Hubble parameter the effects from spacetime curvature become negligible. We therefore expect that the superhorizon modes approach the non-relativistic modes in flat space. In this limit, the parameter $\nu$ goes to 
$i m/H$. We can take this limit by analytically continuing the results obtained above to 
complex $\nu$. With this in mind, we choose
\begin{equation}
\gamma(X)=\sqrt{1+\frac{X}{\nu^2}}-\log\left(\frac{1}{2}+\frac{1}{2}\sqrt{1+\frac{X}{\nu^2}}\right)-1\,,
\label{eq:finalgamma}
\end{equation}
which results in the free SdSET action
\begin{equation}\label{eq:fullfreeS}
S=-\nu\int\der^{d-1}x\der t\;\bigg\{\dot\vp_+\vp_--\vp_+\dot\vp_-+2\nu\vp_+\bigg[1-\sqrt{1+\frac{\p^2_i}{(\nu aH)^2}}\bigg]\vp_-\bigg\}\,.
\end{equation}
This leads to the equations of motion
\begin{equation}
    \dot\vp_{\pm}=\mp\nu\bigg[1-\sqrt{1+\frac{\p_i^2}{(\nu a(t)H)^2}}\bigg]\vp_{\pm}\,.
    \label{eq::appEOM}
\end{equation}
The form of the action agrees with the one of \cite{Cohen:2020php} once the square root is expanded the to first order in the gradient terms. The exact field redefinitions with this choice of $\gamma$ read
\begin{align}
\phi(t,\vec x)&=H^{\frac{d}{2}-\alpha-1}{_0F_1}\bigg(1-\nu;\frac{X}{4}\bigg)e^{-\alpha t+\nu\gamma(X)}\vp_+(t,\vec x)\nonumber\\
&\phantom{=}+H^{\frac{d}{2}-\beta-1}{_0F_1}\bigg(1+\nu;\frac{X}{4}\bigg)e^{-\beta t-\nu\gamma(X)}\vp_-(t,\vec x)\,,\label{eq:phiredef}\\
\pi(t,\vec x)&=-H^{\frac{d}{2}-\alpha}\bigg[{_0F_1\bigg(-\nu;\frac{X}{4}\bigg)}+\frac{\beta X}{4\alpha\nu(\nu-1)}{_0F_1\bigg(2-\nu;\frac{X}{4}\bigg)}\bigg]e^{\beta  t+\nu\gamma(X)}\vp_+(t,\vec x)\nonumber\\
&\phantom{=}-H^{\frac{d}{2}-\beta}\bigg[{_0F_1\bigg(\nu;\frac{X}{4}\bigg)}+\frac{\alpha X}{4\beta\nu(\nu+1)}{_0F_1}\bigg(2+\nu;\frac{X}{4}\bigg)\bigg]e^{\alpha t-\nu\gamma(X)}\vp_-(t,\vec x)\,.\label{eq:piredef}
\end{align}

Comparing the leading time dependence in the exponentials of these expressions to  equations (7a) and (7b) in \cite{Namjoo:2017nia} in the limit $\nu\to i m/H$  we see that the modes $\vp_{\pm}$ take the roles of the non-relativistic flat space modes $\psi$, $\psi^*$ with correspondence 
$\vp_-\rightarrow\psi$, $\vp_+\rightarrow\psi^*$, 
and from \eqref{eq::appEOM} we indeed recover the form of the equations of motion presented in \cite{Namjoo:2017nia} with $\p_i\rightarrow\p_i/a(t)$ and $\vp_{\pm}$:
\begin{equation}
i\frac{\p\vp_{\pm}}{\p t}
=\mp m\bigg[\sqrt{1-\frac{\p^2_i}{(a(t)m)^2}}-1\bigg]\vp_{\pm}\,.
\end{equation}

We conclude the discussion by remarking that different choices of $\gamma$ determine which part of the time dependence of the full-theory field is contained in the effective fields, and which part is removed from the effective description. Indeed, a similar, but far more constrained redundancy is present in the case of the non-relativistic scalar field in flat spacetime. The field redefinition for the NREFT presented in \cite{Namjoo:2017nia} corresponds, in our language, to making the minimal choice $\gamma=0$. In the present case, the same choice would lead to a vanishing Hamiltonian, see \eqref{eq::Hdgamma}, which means that the complete time evolution of the full-theory fields has been pulled out of the $\vp_{\pm}$, yielding a static version of the SdSET. 


\section{Isometries of dS space and active transformation of the SdSET fields}
\label{app::dSiso}

In this Appendix we briefly review the isometry group of dS space, and derive the active transformation of the effective fields under these isometries.

The metric of $d$-dimensional dS space is invariant under the action of the symmetry group SO$(1,d)$, the Lorentz group in $d+1$-dimensions. It contains spatial translations and rotations, dilatations and special conformal transformations. In conformal coordinates, its generators read
\begin{flalign}
P_i&=\p_i\,,\hspace{2.3cm} R^i_{\;j}=x^i\p_j-x_j\p^i\,,
\nonumber\\
D&=-\eta\p_{\eta}-x^i\p_i\,,\quad K^i=2x^i\eta\p_{\eta}+2x^ix^j\p_j+(\eta^2-\vec x^2)\p^i\,.
\end{flalign}

We focus on the dilatations and special conformal transformations. 
Infinitesimal dilatations act on $x^{\mu}$ as
\begin{equation}
x^{\mu}\rightarrow {x'}^{\mu}=(1+\epsilon D)x^{\mu} = (1-\epsilon) x^\mu\,,
\end{equation}
where $\epsilon\ll1$ parametrises the transformation. 
The transformation of $\eta$ immediately implies 
\begin{equation}
a(\eta)\rightarrow a(\eta')=(1+\epsilon )a(\eta) 
\end{equation}
for the scale factor. It can also be convenient to work with large dilatation transformations, which amounts to replacing $(1\pm\epsilon)\rightarrow\xi^{\pm 1}$ in the above equations, where $\xi$ denotes the large dilatation factor. 
The special conformal transformations 
\begin{equation}
x^{\mu}\rightarrow {x'}^{\mu}=(1+\vec b\cdot \vec K)x^{\mu}\,,
\end{equation}
where $\vec b$ is a three-vector which parametrises the transformation,
act differently on the components of $x^{\mu}$. Writing this equation in components one finds
\begin{flalign}
\eta\rightarrow\eta'&=(1+2\vec b\cdot\vec x)\eta\label{eq::SCTeta}\,,\\
\vec x\rightarrow\vec x'&=\vec x+2(\vec b\cdot\vec x)\vec x+(\eta^2-\vec x^2)\vec b\,.
\end{flalign}
Under special conformal transformations the temporal and spatial components of the vector $x^{\mu}$ get mixed, similarly to Lorentz boosts in Minkowski space.

Eqs.~\eqref{eq::vptrafo1}-\eqref{eq::vptrafo3} in the main text specify how the effective fields and their arguments transform simultaneously. However, it is also useful to know their active transformation, that is,  how the fields $\vp'_{\pm}(t,\vec x)$ and $\vp_{\pm}(t,\vec x)$ at the same spacetime point are related to each other.  To this end, we expand
\begin{equation}
\vp'_{\pm}(t',\vec x')=\vp'_{\pm}(t,\vec x)+({x'}^{\mu}-x^{\mu})\p_{\mu}\vp'_{\pm}(t,\vec x)+\Lo(\p^2\vp'_{\pm})\,.
\end{equation}
to linear order, insert the expressions \eqref{eq::vptrafo1}-\eqref{eq::vptrafo3} for $\vp'_{\pm}(t',\vec x')$ and invert the resulting equation for $\vp'_{\pm}(t,\vec x)$ perturbatively in powers of the parameters of the transformations. To linear order this amounts to replacing $({x'}^{\mu}-x^{\mu})\p_{\mu}\vp'_{\pm}\rightarrow({x'}^{\mu}-x^{\mu})\p_{\mu}\vp_{\pm}$, 
and results in the following infinitesimal active dilatation and special conformal transformations, respectively:
\begin{flalign}
\vp'_{\pm}(t,\vec x)&=\bigg[1+\bigg(\frac{d-1}{2}\mp\nu\bigg)\delta+\delta\Big[-\p_t+\vec x\cdot\vec \p\Big]\bigg]\vp_{\pm}(t,\vec x)\,,\\
\vp'_{\pm}(t,\vec x)&=\Big[1-(d-1\mp2\nu)\vec b\cdot\vec x+2(\vec b\cdot\vec x)\p_t\nonumber\\
&\phantom{=}-\Big(2(\vec b\cdot\vec x)\vec x+[(a(t)H)^{-2}-\vec x^2]\vec b\Big)\cdot\vec \p\Big]\vp_{\pm}(t,\vec x)\,.
\end{flalign}

\section{Field redefinition to remove super-leading interactions}
\label{app::intredef}

We wish to remove interaction terms of the form
\begin{equation}
S_{\textrm{int},\,\vp^{2n+2}_+}=-\int\der^{d-1}x\der t\;a(t)^{2n\ve+3}\,C_n\vp^{2n+2}_+\,,
\label{eq::offending}
\end{equation}
where $C_n$ is a constant. We can do so by redefining the field $\vp_-$ as
\begin{equation}
\vp_-\rightarrow\vp_-+a(t)^{2n\ve+3}\,r_n\vp^{2n+1}_+\,,
\end{equation}
with $r_n$ to be determined. 
This induces new $\vp^n_+$-terms through the kinetic term
\begin{flalign}
-3\int\der^{d-1}x\der t\;\dot\vp_+\vp_-&\rightarrow -3\int\der^{d-1}x\der t\;\dot\vp_+\bigg[\vp_-+\sum_{n=0}^{\infty}a(t)^{2n\ve+3}r_n\vp^{2n+1}_+\bigg]\nonumber\\
&=\,-3\int\der^{d-1}x\der t\;\bigg[\dot\vp_+\vp_-+\sum_{n=0}^{\infty}a(t)^{2n\ve+3}\frac{r_n}{2n+2}\frac{\der}{\der t}\vp^{2n+2}_+\bigg]\,.
\end{flalign}
The second term together with \eqref{eq::offending} can be written as 
a total derivative, which can be dropped from the action, 
\begin{flalign}
&-\int\der^{d-1}x\der t\;a(t)^{2n\ve+3}\bigg[\frac{3r_n}{2n+2}\frac{\der}{\der t}\vp^{2n+2}_++C_n\vp^{2n+2}_+\bigg]\nonumber\\
&=\,-\int\der^{d-1}x\der t\;\frac{\der}{\der t}\bigg[\frac{C_n}{2n\ve+3}a(t)^{2n\ve+3}\vp^{2n+2}_+\bigg]
\end{flalign}
provided we choose
\begin{equation}
r_n=\frac{(2n+2)C_n}{3(2n\ve+3)}\,.
\label{eq::masterredef}
\end{equation}
This formula can be used to determine the appropriate field redefinition iteratively in 
powers of $\vp_\pm$, starting from the lowest interaction terms, which are bilinear in $\vp_{\pm}$, and keeping track of the terms containing higher powers of $\vp_{\pm}$ which are generated by this redefinition. It is important to account for these newly generated terms to obtain the correct field redefinition in higher powers. We consider terms up to $\vp^6_{\pm}$ in the following.

\subsection{Elimination of \texorpdfstring{$\varphi^2_+$}{phi-plus squared}-terms}
\label{sec:elvp2}

We start from 
\begin{equation}
S_{\textrm{int,}\vp^2_+}=-\int\der^{d-1}x\der t\;a(t)^3\,\frac{c^0_{2,0}}{2}\vp^2_+\,.
\end{equation}
Using $c^0_{2,0}=H^3c^0_{1,1}$ from \eqref{eq::RPIcouplings} and comparing to the form of \eqref{eq::offending} we identify
\begin{equation}
C_0=\frac{c^0_{1,1}H^3}{2}\,,
\end{equation}
which yields  
\begin{equation}
r_0=\frac{c^0_{1,1}H^3}{9}
\end{equation}
from \eqref{eq::masterredef}. This corresponds to the redefiniton
\begin{equation}
\vp_-\rightarrow\vp_-+\frac{c^0_{1,1}(a(t)H)^3}{9}\,\vp_+\,.
\end{equation}
This shift then induces new terms. In particular, since the field redefinition is naively power-counting violating, one 
needs to keep track of subleading-power terms in the action prior to the field redefinition. For instance, from the sub-leading term $\vp^2_-$ one obtains 
a shift in the coefficient of the bilinear term $\vp_+\vp_-$,
\begin{equation}
a(t)^{-3}\,\frac{c^0_{0,2}}{2}\vp^2_-\rightarrow \frac{H^3c^0_{0,2}c^0_{1,1}}{9}
\,\vp_+\vp_-\,,
\label{eq::bilinearsuppressed}
\end{equation}
proportional to the product of $c^0_{0,2}$, $c^0_{1,1}$. However, since the two couplings are linearly related by \eqref{eq::RPIcouplings}, we know that they must have the same parametric size, which we assume to be small. Their product is therefore suppressed with respect to $c^0_{1,1}$, which also multiplies the operator $\vp_+\vp_-$ in the SdSET action. Since in this paper the bilinear interaction term is matched only at leading non-vanishing order in the couplings, we may safely neglect the correction \eqref{eq::bilinearsuppressed}. However, when computing higher-order corrections to the $\vp_+ \vp_-$ coupling, these newly induced terms must be included in the field redefinition. 

Below we will proceed similarly, and only retain the terms generated by the non-linear field redefinition which are needed for the computations in the main text, ignoring terms that are further suppressed. We use the following criterion to decide which terms to keep: since we are matching onto $\kappa\phi^4$ theory, we can expect the quartic EFT coupling $c^0_{3,1}$ to be $\mathcal{O}(\kappa)$. Its leading term is determined by tree-level matching of the four-point function. Analogously, the leading contribution to all higher couplings $c^0_{2n-m,m}$ for fixed $n$ are determined by the tree-level matching computation of correlation functions involving $2n$ fields. The cost of increasing $n$ by one unit is the insertion of at least one more quartic EFT interaction $c^0_{3,1}\vp_+^3\vp_-$.   
The coupling $c^0_{1,1}$ does not fit into this scheme, since it is generated 
only from the one-loop order.\footnote{However, see comment on $c^0_{2n-1,1}$, $n>2$, in footnote~\ref{ft:coupling}.} This translates to the parametric estimates
\begin{equation}
    c^0_{1,1}\sim c^0_{3,1}\,,\quad c^0_{2n-m,m}\sim (c^0_{3,1})^{n-1}\,.
\end{equation}

\subsection{Elimination of \texorpdfstring{$\vp^4_+$}{phi-plus 4}-terms} 

The next interaction term to be removed is
\begin{equation}
S_{\textrm{int,}\vp^4_+}=-\int\der^{d-1}x\der t\;a(t)^{3+2\ve}\frac{c^0_{4,0}}{4!}\vp^4_+\,.
\end{equation}
Using \eqref{eq::offending}, as well as \eqref{eq::RPIcouplings} to relate $c^0_{4,0}$ to $c^0_{3,1}$, we identify
\begin{equation}
 C_1=\frac{H^3c^0_{3,1}}{4!}\,,
 \end{equation}
 which leads to
 \begin{equation}
 r_1=\frac{H^3c^0_{3,1}}{18(3+2\ve)}\,,
 \end{equation}
 and the redefinition
\begin{equation}
\vp_-\rightarrow\vp_-+(a(t)H)^3\bigg[\frac{c^0_{1,1}}{9}\vp_++\frac{c^0_{3,1}a(t)^{2\ve}}{18(3+2\ve)}\vp^3_+\bigg]\,.
\end{equation}
This redefinition induces a contribution to the $\vp^6_+$ interaction that is as important as $c^0_{6,0}$,
\begin{equation}
a(t)^{2\ve}\frac{c^0_{3,1}}{6}\vp^3_+\vp_-\rightarrow a(t)^{3+4\ve}\frac{H^3(c^0_{3,1})^2}{108(3+2\ve)}\vp^6_+\,,
\end{equation}
and we therefore need to keep track of it.

\subsection{Elimination of \texorpdfstring{$\vp^6_+$}{phi-plus 6}-terms}

We now need to eliminate the interaction terms
\begin{equation}
S_{\textrm{int,}\vp^6_+}=-\int\der^{d-1}x\der t\;a(t)^{3+4\ve}\bigg[\frac{c^0_{6,0}}{6!}+\frac{H^3(c^0_{3,1})^2}{108(3+2\ve)}\bigg]\vp^6_+\,,
\end{equation}
so we identify
\begin{equation}
C_2=H^3\bigg[\frac{c^0_{5,1}}{6!}+\frac{(c^0_{3,1})^2}{108(3+2\ve)}\bigg]\,,
\end{equation}
where we used \eqref{eq::RPIcouplings} again to express $c^0_{6,0}$ in terms of $c^0_{5,1}$. We therefore need to pick
\begin{equation}
r_2=\frac{2H^3}{3+4\ve}\bigg[\frac{c^0_{5,1}}{6!}+\frac{(c^0_{3,1})^2}{108(3+2\ve)}\bigg]\,.
\end{equation}
Putting all previous terms together, we obtain the field redefinition
\begin{equation}
\vp_-\rightarrow\vp_-+(a(t)H)^3\bigg[\frac{c^0_{1,1}}{9}\vp_++\frac{c^0_{3,1}a(t)^{2\ve}}{18(3+2\ve)}\vp^3_++\frac{2a(t)^{4\ve}}{3+4\ve}\bigg[\frac{c^0_{5,1}}{6!}+\frac{(c^0_{3,1})^2}{108(3+2\ve)}\bigg]\vp^5_+\bigg]\,,
\end{equation}
as given in \eqref{eq::vpmredef} of the main text. 


\section{Canonical SdSET commutation relations in the presence of the IR regulator}
\label{app:comm}

In this Appendix we collect the details pertaining to the statement that the 
canonical commutation relations of the SdSET fields are unaffected by the 
$\Lambda$ IR-regulator discussed in \secref{sec:SdSETreg}. 

We define the mode-function decomposition of the field operators $\vp_{\pm}$ starting from the standard decomposition of the full-theory field operator 
\begin{equation}
\phi(t,\vec x)=\int\frac{\der^{d-1}k}{(2\pi)^{d-1}}\Big[e^{i\vec k\cdot\vec x}\phi_{\vec k}(t)a_{\vec k}+e^{-i\vec k\cdot\vec x}\phi^*_{\vec k}(t)a^{\dagger}_{\vec k}\Big]\,,
\label{eq::phidecomp}
\end{equation}
where the creation and annihilation operators $a^{\dagger}_{\vec k}$, $a_{\vec k}$ satisfy the canonical commutation relations
\begin{flalign}
[a_{\vec k},a_{\vec p}]&=[a^{\dagger}_{\vec k},a^{\dagger}_{\vec p}]=0\,,\label{eq::acomm1}\\
[a_{\vec k},a^{\dagger}_{\vec p}]&=(2\pi)^{d-1}\delta^{(d-1)}(\vec k-\vec p)\,.\label{eq::acomm2}
\end{flalign}
Plugging \eqref{eq::phidecomp} into the leading-power relation between $\phi$ and $\vp_{\pm}$,
\begin{equation}
\lim\limits_{k/(a(t)H)\rightarrow0}\phi(t,\vec x)=H^{\frac{d}{2}-1}\Big[(a(t)H)^{-\alpha}\vp_+(t,\vec x)+(a(t)H)^{-\beta}\vp_-(t,\vec x)\Big]\,,
\label{eq::appfieldred}
\end{equation}
leads to
\begin{flalign}
\vp_+(t,\vec x)&=\int\frac{\der^{d-1}k}{(2\pi)^{d-1}}\;\Big[e^{i\vec k\cdot\vec x}\vp_{+\vec k}(t)a_{\vec k}+e^{-i\vec k\cdot\vec x}\vp^*_{+\vec k}(t)a^{\dagger}_{\vec k}\Big]\,,\\
\vp_-(t,\vec x)&=\int\frac{\der^{d-1}k}{(2\pi)^{d-1}}\;\Big[e^{i\vec k\cdot\vec x}\vp_{-\vec k}(t)a_{\vec k}+e^{-i\vec k\cdot\vec x}\vp^*_{-\vec k}(t)a^{\dagger}_{\vec k}\Big]\,.
\end{flalign}
In the last two equations we introduced the mode functions $\vp_{\pm\vec k}(t)$ of the SdSET fields. 

Eq.~\eqref{eq::appfieldred} implies the analogous relation
\begin{equation}
\lim\limits_{k/(a(t)H)\rightarrow0}\phi_{\vec k}(t)=H^{\frac{d}{2}-1}\Big[(a(t)H)^{-\alpha}\vp_{+\vec k}(t)+(a(t)H)^{-\beta}\vp_{-\vec k}(t)\Big]\,,
\label{eq::moderelation}
\end{equation}
between the mode functions of the full-theory and effective fields. We expand the full-theory mode function in the presence of the IR regulator, which reads 
\begin{equation}
    \phi_{\vec{k}}(t) = \sqrt{\frac{\pi}{4H}}\,a(t)^{-\frac{d-1}{2}} H_\nu^{(1)}\bigg(\frac{k_{\Lambda}}{a(t)H}\bigg)\,,
\end{equation}
and keep only the two leading terms proportional to $(a(t)H)^{-\frac{d-1}{2}\pm\nu}$ for $k_{\Lambda}/(a(t)H)\ll1$. Comparing powers of $a(t)H$ in the resulting expression with \eqref{eq::moderelation} we find the leading-power effective mode functions
\begin{equation}
\vp_{+\vec k}=-\frac{i}{\sqrt{2}k^{\frac{3}{2}}_{\Lambda}}\,,\qquad \vp_{-\vec k}=\frac{k^{\frac{3}{2}}_{\Lambda}}{3\sqrt{2}}\,,
\end{equation}
where we set $\nu=3/2$. With these results and \eqref{eq::acomm1}, \eqref{eq::acomm2} we find
\begin{equation}
[\vp_+(t,\vec x),\vp_+(t,\vec y)]=\int\frac{\der^{d-1}k}{(2\pi)^{d-1}}e^{i\vec k\cdot(\vec x-\vec y)}\Big[\vp_{+\vec k}\vp^*_{+\vec k}-\vp^*_{+\vec k}\vp_{+\vec k}\Big] = 0\,,
\end{equation}
and the same holds for the commutator $[\vp_-(t,\vec x),\vp_-(t,\vec y)]$, while 
\begin{equation}
[\vp_+(t,\vec x),\vp_-(t,\vec y)]=\int\frac{\der^{d-1}k}{(2\pi)^{d-1}}e^{i\vec k\cdot(\vec x-\vec y)}\Big[\vp_{+\vec k}\vp^*_{-\vec k}-\vp^*_{+\vec k}\vp_{-\vec k}\Big] =-\frac{i}{3}\delta^{(d-1)}(\vec x-\vec y)\,,\quad
\end{equation}
which proves the statement in the main text.


\section{Tree-level six-point function in the full theory}
\label{app::fullpenta}

\begin{figure}[t]
\centering
\includegraphics[width=0.45\textwidth]{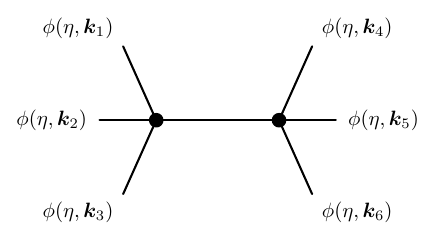}
\caption{The diagram contributing to $P(\eta;\vec k_1,...,\vec k_6)$. Only the diagram with (++)-type Schwinger-Keldysh vertices is shown.}
\label{fig:penta}
\end{figure}

In this Appendix we compute the late-time limit of the full-theory tree-level six-point function
\begin{flalign}
&\cor{}{\phi(\eta,\vec k_1)\phi(\eta,\vec k_2)\phi(\eta,\vec k_3)\phi(\eta,\vec k_4)\phi(\eta,\vec k_5)\phi(\eta,\vec k_6)}'\nonumber\\
&=-\kappa^2\int_{-\infty}^{\eta}\frac{\der \eta_1}{(-H\eta_1)^d}\int_{-\infty}^{\eta}\frac{\der \eta_2}{(-H\eta_2)^d}\;\Bigg[\,G_{++}(\eta,\vec k_1;\eta_1,\vec k_1)G_{++}(\eta,\vec k_2;\eta_1,\vec k_2)G_{++}(\eta,\vec k_3;\eta_1,\vec k_3)\nonumber\\
&\phantom{=}\times \bigg(G_{++}(\eta_1,\vec K_{123};\eta_2,\vec K_{123})G_{++}(\eta,\vec k_4;\eta_2,\vec k_4)G_{++}(\eta,\vec k_5;\eta_2,\vec k_5)G_{++}(\eta,\vec k_6;\eta_2,\vec k_6)\nonumber\\
&\phantom{=}-G_{+-}(\eta_1,\vec K_{123};\eta_2,\vec K_{123})G_{--}(\eta,\vec k_4;\eta_2,\vec k_4)G_{--}(\eta,\vec k_5;\eta_2,\vec k_5)G_{--}(\eta,\vec k_6;\eta_2,\vec k_6)\bigg)\nonumber\\
&\phantom{=}\hskip1cm+G_{--}(\eta,\vec k_1;\eta_1,\vec k_1)G_{--}(\eta,\vec k_2;\eta_1,\vec k_2)G_{--}(\eta,\vec k_3;\eta_1,\vec k_3)\,\nonumber\\
&\phantom{=}\times\bigg(G_{--}(\eta_1,\vec K_{123};\eta_2,\vec K_{123}) G_{--}(\eta,\vec k_4;\eta_2,\vec k_4)G_{--}(\eta,\vec k_5;\eta_2,\vec k_5)G_{--}(\eta,\vec k_6;\eta_2,\vec k_5)\nonumber\\
&\phantom{=}-G_{-+}(\eta_1,\vec K_{123};\eta_2,\vec K_{123})G_{++}(\eta,\vec k_4;\eta_2,\vec k_4)G_{++}(\eta,\vec k_5;\eta_2,\vec k_5)G_{++}(\eta,\vec k_6;\eta_2,\vec k_6)\bigg)\,\Bigg]\nonumber\\
&\phantom{=}+\textrm{all combinations of the permutations $(\vec k_1\leftrightarrow\vec k_3)$, $(\vec k_2\leftrightarrow\vec k_4)$, $(\vec k_3\leftrightarrow\vec k_6)$}\,,
\label{eq::psstart}
\end{flalign}
where we introduced the abbreviation
\begin{equation}
\vec K_{123}\equiv\sum_{i=1}^3\vec k_i=-\sum_{i=4}^6\vec k_i\,.
\end{equation}
Since the other permutations are trivially obtained by exchanging the external momenta we now focus on evaluating the part in square brackets in \eqref{eq::psstart}, denoted by $P(\eta;\vec k_1,...,\vec k_6)$ in the following. The $(++)$-vertex SK diagram is shown in \figref{fig:penta} as a representative of the four possible combinations of filled (+)
and empty ($-$) SK vertices.

Plugging the four SK propagators for the massless, minimally coupled scalar field into \eqref{eq::psstart} and resolving the (anti-)time orderings we find
\begin{flalign}
&P(\eta;\vec k_1,...,\vec k_6)\nonumber\\
&=-\frac{\kappa^2H^{6-4\ve}(-H\eta)^{-6\ve}}{32(k_1k_2k_3k_4k_5k_6K)^3}\Bigg\{\int_{-\infty}^{\eta}\frac{\der\eta_1}{(-\eta_1)^{4+2\ve}}\int_{-\infty}^{\eta}\frac{\der\eta_2}{(-\eta_2)^{4+2\ve}}\;\Im\bigg[\,e^{iK(\eta_1-\eta_2)}(-i+K\eta_2)\nonumber\\
&\phantom{=}\times(i+K\eta_1)\prod_{j=1}^3e^{ik_j(\eta_1-\eta)}(-i+k_j\eta)(i+k_j\eta_1)\bigg]\Im\bigg[\prod_{n=4}^6e^{ik_n(\eta-\eta_2)}(-i+k_n\eta_2)(i+k_n\eta)\bigg]\nonumber\\
&\phantom{=}-\int_{-\infty}^{\eta}\frac{\der\eta_1}{(-\eta_1)^{4+2\ve}}\int_{-\infty}^{\eta_1}\frac{\der\eta_2}{(-\eta_2)^{4+2\ve}}\;\Im\Big[\,e^{iK(\eta_2-\eta_1)}(-i+K\eta_1)(i+K\eta_2)\Big]\nonumber\\
&\phantom{=}\times\Im\bigg[\prod_{j=1}^3e^{ik_j(\eta_1-\eta)}(-i+k_j\eta)(i+k_j\eta_1)\prod_{n=4}^6e^{ik_n(\eta_2-\eta)}(-i+k_n\eta)(i+k_n\eta_2)\bigg]\Bigg\}\,.
\label{eq::tlps}
\end{flalign}
Since we are interested in the late-time limit, the computation can be simplified using the method of regions. The region decomposition of each of the two time integrals appearing in this expression follows the rules presented in \cite{Beneke:2023wmt}. Therefore, following the nomenclature introduced in this reference, 
both time integrals possess an early-time and a late-time region. This leads to a total of four regions, which can be reduced to three independent ones using the symmetry of \eqref{eq::tlps} under $\{\vec k_1,\vec k_2,\vec k_3\}\leftrightarrow\{\vec k_4,\vec k_5,\vec k_6\}$ leaving: the region in which both integration variables satisfy $\eta_{1,2}\sim\eta$ (``late-late region"), the two regions in which one integration variable satisfies $\eta_{1,2}\ll\eta$ and the other $\eta_{2,1}\sim\eta$ (``early-late regions"), which are related to each other by the mentioned symmetry, and the region in which both satisfy $\eta_{1,2}\ll\eta$ (``early-early region").

The new feature with respect to the computations presented in  \cite{Beneke:2023wmt} is that we encounter for the first time diagrams involving two vertices, which lead to both nested and factorised time integrals. However, this does not lead to conceptual complications, since the region decomposition is mirrored by the SdSET computation of the six-point function in \secref{sec:penta}.

\subsection{Late-late region}
\label{pentaspecss}

In this region both integration variables in \eqref{eq::tlps} satisfy $\eta_{1,2}\sim\eta$, and the exponential functions can be expanded in their small exponents before 
integrating. We do so to high enough order to ensure that no $\Lo(\eta^0)$ terms 
are missed after the product of both integrals has been evaluated.
The results of the single integrals contain $\Lo(\eta^{-3})$-terms, which are singular in the late-time limit. Therefore, the integrands of the factorised integrals 
have to be expanded up to $\Lo(\eta_i^2)$, which yields $\Lo(\eta^3)$-terms after integration, and thus possibly $\Lo(\eta^0)$-contributions after being multiplied with an $\Lo(\eta^{-3})$-term. For the non-factorised integrals appearing in \eqref{eq::tlps} we need to expand the innermost integrand to $\Lo(\eta_2^2)$, and it then suffices to expand the result of the inner integration to $\Lo(\eta_1^{-1})$. Evaluating the resulting integrals using the formula
\begin{equation}
\int_{-\infty}^{\eta}\der\eta'\;(-\eta')^a=-\frac{(-\eta)^{a+1}}{a+1}
\end{equation}
we find, using the abbreviations \eqref{eq:kns},  
\begin{flalign}
&P_{\textrm{ll}}(\eta;\vec k_1,...,\vec k_6)\nonumber\\
&=\frac{H^{6-4\ve}(-H\eta)^{-6\ve}\kappa^2}{576(k_1k_2k_3k_4k_5k_6K)^3}(-\tmu\eta)^{-4\ve}\bigg\{k^3_{123}\bigg[\frac{1}{4\ve^2}\,k^3_{K456}-\frac{1}{2\ve}\bigg(K^3+\frac{2}{3}k^3_{456}\bigg)+\frac{7}{9}K^3+\frac{1}{3}k^3_{456}\bigg]\nonumber\\
&\phantom{=}+\{\vec k_1,\vec k_2,\vec k_3\}\leftrightarrow\{\vec k_4,\vec k_5,\vec k_6\}\bigg\}\,.
\label{pssoft}
\end{flalign}
We note that the region expansion introduces a double pole from $\eta_i=-\infty$, which will cancel with the other regions.

\subsection{Early-late regions}

To compute the mixed-region integrals it is more convenient to start from \eqref{eq::psstart} rather than \eqref{eq::tlps}, let $\eta\rightarrow0$ in the upper integration boundary for the early-time integral and expand the integrand in small $-k_j\eta$ before 
performing the late-time integral. Since we have one early- and one late-time integral, this unambiguously fixes the (anti-)time ordering appearing in the propagators connecting the two vertices, and we can resolve them over the full integration ranges of 
$\eta_{1,2}$ yielding
\begin{flalign}
&P_{\textrm{el}}(\eta;\vec k_1,...,\vec k_6)\nonumber\\
&=\frac{\kappa^2H^{6-4\ve}(-H\eta)^{-6\ve}}{32(k_1k_2k_3k_4k_5k_6K)^3}\,\Bigg\{\int_{-\infty}^0\frac{\der \eta_1}{(-\eta_1)^{4+2\ve}}\int_{-\infty}^{\eta}\frac{\der \eta_2}{(-\eta_2)^{4+2\ve}}\;\Im\bigg[e^{iK(\eta_1-\eta_2)}(-i+K\eta_2)(i+K\eta_1)\nonumber\\
&\phantom{=}\times\prod_{j=1}^3e^{ik_j(\eta_1-\eta)}(-i+k_j\eta)(i+k_j\eta_1)\bigg]\Im\bigg[\prod_{j=4}^6e^{ik_j(\eta_2-\eta)}(-i+k_j\eta)(i+k_j\eta_2)\bigg]\bigg|_{\substack{\eta_1\ll\eta\\ \eta_2\sim\eta}}\nonumber\\
&\phantom{=}+\{\vec k_1,\vec k_2,\vec k_3\}\leftrightarrow\{\vec k_4,\vec k_5,\vec k_6\}\Bigg\}\,.
\end{flalign}
{}From the intermediate result
\begin{flalign}
&\int_{-\infty}^0\frac{\der \eta_1}{(-\eta_1)^{4+2\ve}}\int_{-\infty}^{\eta}\frac{\der \eta_2}{(-\eta_2)^{4+2\ve}}\;\Im\bigg[e^{iK(\eta_1-\eta_2)}(-i+K\eta_2)(i+K\eta_1)\nonumber\\
&\phantom{=}\times\prod_{j=1}^3e^{ik_j(\eta_1-\eta)}(-i+k_j\eta)(i+k_j\eta_1)\bigg]\Im\bigg[\prod_{j=4}^6e^{ik_j(\eta_2-\eta)}(-i+k_j\eta)(i+k_j\eta_2)\bigg]\bigg|_{\substack{\eta_1\ll\eta\\ \eta_2\sim\eta}}\nonumber\\
&=\bigg[-\frac{(-\eta)^{-2\ve}}{2\ve(3+2\ve)}\sum_{j=4}^6k^3_j\bigg]\Im\bigg[\int_{-\infty}^0\frac{\der \eta_1}{(-\eta_1)^{4+2\ve}}\;e^{iK\eta_1}(i+K\eta_1)\prod_{j=1}^3e^{ik_j\eta_1}(i+k_j\eta_1)\bigg]
\label{eq::pentaELintermediate}
\end{flalign}
we see that \eqref{eq::pentaELintermediate} factorises into the product of an $\eta$-dependent term featuring a pole, and a $d$-dimensional time integral. The structure of the result in this region matches the corresponding piece \eqref{eq::EFTpentamixed} of the EFT computation. The former term has the correct $\eta$- and momentum dependence to be generated by a tree-level four-point SdSET vertex. The $\ve$-expansion of the latter gives rise to the terms $[K^3+\sum_{i=1}^3k^3_i][\xi_{3,1}+\Xi_{3,1}+\Lo(\ve)]$, which, up to $\Lo(\ve)$-terms, are generated by the insertion of the bare IC function $\Xi^0_{3,1}$ in the SdSET six-point function.
Evaluating the remaining early-time integrals using the formula
\begin{equation}
\int_{-\infty}^0\der\eta'\,(-\eta')^ae^{\pm ip\eta'}=(\pm ip)^{-a-1}\Gamma(a+1)\,,\quad p>0\,,
\label{eq::factearlyint}
\end{equation}
and expanding the result in $\ve$ we find
\begin{flalign}
&P_{\textrm{el}}(t;\vec k_1,...,\vec k_6)\nonumber\\
&=-\frac{\kappa^2H^{6-4\ve}(-H\eta)^{-6\ve}}{576(k_1k_2k_3k_4k_5k_6K)^3}(-\tmu\eta)^{-2\ve}\,\Bigg\{\bigg(\frac{e^{\gamma_E}k_{K456}}{\tmu}\bigg)^{\!2\ve}k^3_{123}\,\Bigg[\frac{1}{2\ve^2}\,k^3_{K456}\nonumber\\
&\phantom{=}-\frac{3}{\ve}\bigg[\,\frac{1}{9}k^3_{K456}+\frac{Kk_4k_5k_6}{k_{K456}}-k_{K456}\bigg(Kk_{456}+k_4k_{56}+k_5k_6-\frac{4}{9}(k_{K456})^2\bigg)\bigg]\nonumber\\
&\phantom{=}+2\bigg[\,\frac{1}{3}k^3_{K456}+\frac{Kk_4k_5k_6}{k_{K456}}+\frac{1}{3}\Big[5k^2_{K456}k_{K456}-2\Big(K[k_4k_{56}+k_5k_6]+k_4k_5k_6\Big)\Big]\bigg]\nonumber\\
&\phantom{=}-\frac{\pi^2}{12}\,k^3_{K456}\Bigg]+\{\vec k_1,\vec k_2,\vec k_3\}\leftrightarrow\{\vec k_4,\vec k_5,\vec k_6\}\Bigg\}\,.
\end{flalign}

\subsection{Early-early region}

In the region where both integration variables satisfy $\eta_{1,2}\ll\eta$ we let $\eta\rightarrow0$ in the upper integration boundaries 
and in the integrands appearing in \eqref{eq::tlps}. This leads to the time-independent expression
\begin{flalign}
&P_{\textrm{ee}}(\vec k_1,...,\vec k_6)\nonumber\\
&=-\frac{\kappa^2H^{2(d-1)}(-H\eta)^{3(d-4)}}{32(k_1k_2k_3k_4k_5k_6K)^3}\,\Bigg\{\int_{-\infty}^0\frac{\der\eta_1}{(-\eta_1)^{8-d}}\int_{-\infty}^0\frac{\der\eta_2}{(-\eta_2)^{8-d}}\,\Im\bigg[e^{iK(\eta_1-\eta_2)}(-i+K\eta_2)\nonumber\\
&\phantom{=}\times (i+K\eta_1)(-i)^3\prod_{j=1}^3e^{ik_j\eta_1}(i+k_j\eta_1)\bigg]\,\Im\bigg[i^3\prod_{j=4}^6e^{-ik_j\eta_2}(-i+k_j\eta_2)\bigg]\nonumber\\
&\phantom{=}-\int_{-\infty}^0\frac{\der\eta_1}{(-\eta_1)^{8-d}}\int_{-\infty}^{\eta_1}\frac{\der\eta_2}{(-\eta_2)^{8-d}}\;\Im\Big[e^{iK(\eta_2-\eta_1)}(-i+K\eta_1)(i+K\eta_2)\Big]\nonumber\\
&\phantom{=}\times\Im\bigg[(-i)^6\prod_{j=1}^3e^{ik_j\eta_1}(i+k_j\eta_1)\prod_{n=4}^6e^{ik_n\eta_2}(i+k_n\eta_2)\bigg]\Bigg\}\,.
\label{eq::pentaee}
\end{flalign}
As was the case in the late-late region, we have to evaluate factorised, as well as nested time integrals. The factorised integrals appearing in the first summand of \eqref{eq::pentaee} can be evaluated using \eqref{eq::factearlyint}. The nested integrals appearing in the second summand of \eqref{eq::pentaee} can be reduced to the master integral
\begin{flalign}
&\int_{-\infty}^0\frac{\der\eta_1}{(-\eta_1)^{4+2\ve}}\;e^{\pm ip_1\eta_1}\int_{-\infty}^{\eta_1}\frac{\der\eta_2}{(-\eta_2)^{4+2\ve}}\;e^{\pm ip_2\eta_2}\nonumber\\
&=\frac{(\pm i)^{4\ve}p^{6+4\ve}_2}{3+2\ve}\Gamma(-6-4\ve)\,{_2F_1}\bigg(-6-4\ve,-3-2\ve;-2-2\ve;-\frac{p_1}{p_2}\bigg)\,,\quad p_{1,2}>0\,,
\label{eq::nestedint}
\end{flalign}
where ${_2F_1}(a,b;c;z)$ is Gauss' hypergeometric function. By using the identity
\begin{equation}
(\pm i+k_j\eta_{1,2})e^{\pm i(p_1\eta_1+p_2\eta_2)}=\pm i\bigg(1- k_j\frac{\p}{\p p_{1,2}}\bigg)e^{\pm i(p_1\eta_1+p_2\eta_2)}
\end{equation}
and pulling the differential operators acting on the exponentials out of the integrals, the various terms making up the second summand of \eqref{eq::pentaee} can be generated.
The result is then expanded in $\ve$ with the help of the \texttt{Mathematica} package \texttt{HypExp 2}~\cite{Huber:2007dx}.
We find 
\begin{flalign}
&P_{\textrm{ee}}(\vec k_1,...,\vec k_6)\nonumber\\
&=\frac{H^{6-4\ve}\kappa^2(-H\eta)^{-6\ve}}{576(k_1k_2k_3k_4k_5k_6K)^3}\bigg(\frac{e^{2\gamma_E}k_{K123}k_{K456}}{\tmu^2}\bigg)^{\!2\ve}\,\Bigg\{\frac{1}{4\ve^2}\,k^3_{123}k^3_{K456}\nonumber\\
&\phantom{=}+\frac{1}{2\ve}\,\Bigg[K^3\bigg(\frac{1}{3}k^3_{123}+\frac{1}{2}\Big[k^3_{123}-k^3_{456}\Big]\log\bigg(\frac{k_{K456}}{k_{K123}}\bigg)\bigg)-6k^3_{123}\bigg(\frac{Kk_4k_5k_6}{k_{K456}}\nonumber\\
&\qquad\quad-k_{K456}\bigg(Kk_{456}+k_4k_{56}+k_5k_6-\frac{4}{9}(k_{K456})^2\bigg)\bigg)\Bigg]
+f(\vec k_1,...,\vec k_6)\nonumber\\
&\phantom{=}+\{\vec k_1,\vec k_2,\vec k_3\}\leftrightarrow\{\vec k_4,\vec k_5,\vec k_6\}\Bigg\}
\label{eq:6ptfull}
\end{flalign}
where we defined the finite part 
\begin{flalign}
&f(\vec k_1,...,\vec k_6)\nonumber\\
&\equiv\frac{1}{2}\bigg[k^3_{K123}\Big[K^3-k^3_{456}\Big]\Li_2\bigg(-\frac{k_{K123}}{k_{456}-K}\bigg)-\Big[K^3-k^3_{123}\Big]k^3_{K456}\Li_2\bigg(-\frac{k_{123}-K}{k_{K456}}\bigg)\bigg]\nonumber\\
&\phantom{=}+\frac{1}{4}\bigg[\Big[k^3_{123}k^3_{456}-K^6\Big]\log\bigg(\frac{(k_{K123})^2}{k_{K456}(k_{456}-K)}\bigg)\log\bigg(\frac{k_{456}-K}{k_{K456}}\bigg)\nonumber\\
&\phantom{=}\;\quad\quad+K^3\Big[k^3_{123}-k^3_{456}\Big]\log^2\bigg(\frac{k_{K123}}{k_{456}-K}\bigg)\bigg]\nonumber\\
&\phantom{=}+\frac{\pi^2}{12}\bigg[3K^6+2K^3k^3_{123}-k^3_{123}k^3_{456}\bigg]\nonumber\\
&\phantom{=}+\bigg[-3Kk_4k_5k_6\bigg(\frac{k^3_{123}-K^3}{k_{K456}}+\frac{k^3_{K123}}{k_{456}-K}\bigg)+2K^5k_{456}-2Kk^3_{123}\bigg(k^2_{456}-k_4k_{56}\nonumber\\
&\phantom{=}\qquad-k_5k_6+\frac{7}{6}K^2\bigg)+K^3\Big[k^3_{456}+2(k^2_{456}k_{456}-k_4k_5k_6)\Big]\bigg]\log\bigg(\frac{k_t}{k_{K123}}\bigg)\nonumber\\
&\phantom{=}+9\bigg[\frac{Kk_1k_2k_3}{k_{K123}}-k_{K123}\bigg(Kk_{123}+k_1k_{23}+k_2k_3-\frac{4}{9}(k_{K123})^2\bigg)\bigg]\nonumber\\
&\phantom{==}\times\bigg[\frac{Kk_4k_5k_6}{k_{K456}}-k_{K456}\bigg(Kk_{456}+k_4k_{56}+k_5k_6-\frac{4}{9}(k_{K456})^2\bigg)\bigg]\nonumber\\
&\phantom{=}+\frac{2}{9}\,k^3_{K123}k_{K456}\Big[13k^2_{K456}-Kk_{456}-k_4k_{56}-k_5k_6\Big]\nonumber\\
&\phantom{=}+\frac{1}{18k_{K123}k_{K456}}\Bigg[-84K^8-300K^7k_{123}-12K^6\Big[19k^2_{123}+18k_{123}k_{456}+17\Big(k_1k_{23}+k_2k_3\Big)\Big]\nonumber\\
&\phantom{==}-12K^5\,\bigg(\frac{109}{6}k^3_{123}+\Big[23k^2_{123}+16\Big(k_1k_{23}+k_2k_3\Big)\Big]k_{456}+12\Big[k^2_1k_{23}+k^2_2k_{13}+k^2_3k_{23}\Big]\nonumber\\
&\phantom{==}-6k_1k_2k_3\bigg)\nonumber\\
&\phantom{==}-2K^4\bigg(61k^4_{123}+133k^3_{123}k_{456}+115\Big[k^3_1k_{23}+k^3_2k_{13}+k^3_3k_{12}\Big]+33k^2_{123}k^2_{456}\nonumber\\
&\phantom{==}+108\Big[k^2_1k^2_{23}+k^2_2k^2_3\Big]+24k^2_{123}\Big[k_4k_{56}+k_5k_6\Big]+90\Big[k^2_1k_{23}+k^2_2k_{13}+k^2_3k_{12}+k_4k_5k_6\Big]k_{456}\nonumber\\
&\phantom{==}-54k_{123}k_4k_5k_6-3\Big[k_1k_{23}+k_2k_3\Big]\Big[k_4k_{56}+k_5k_6\Big]\bigg)\nonumber\\
&\phantom{==}-\frac{2K^3}{k_t}\bigg(61k^5_{123}k_{456}+k^4_{123}\Big[103k^2_{456}+152\Big(k_4k_{56}+k_5k_6\Big)\Big]\nonumber\\
&\phantom{===}+176\Big[k^4_1k_{23}+k^4_2k_{13}+k^4_3k_{12}\Big]k_{456}+42k^3_{123}k^3_{456}\nonumber\\
&\phantom{===}+\Big[k^3_1\Big(223k^2_{23}+320k_2k_3\Big)+k^3_2\Big(223k^2_{13}+320k_1k_3\Big)+k^3_3\Big(223k^2_{12}+320k_1k_2\Big)\Big]k_{456}\nonumber\\
&\phantom{===}+\Big[k^3_1k_{23}+k^3_2k_{13}+k^3_3k_{12}\Big]\Big[175k^2_{456}+296\Big(k_4k_{56}+k_5k_6\Big)\Big]\nonumber\\
&\phantom{===}+k^3_{123}\Big[90\Big(k^2_4k_{56}+k^2_5k_{46}+k^2_6k_{45}\Big)+72k_4k_5k_6\Big]+\Big[144\Big(k^2_1k^2_{23}+k^2_2k^2_3\Big)k_{456}\nonumber\\
&\phantom{===}+288\Big(k^2_1(k^2_2k_3+k^2_3k_2)+k^2_2k^2_3k_1\Big)+108\Big(k^2_{123}+2(k_1k_{23}+k_2k_3)\Big)k_4k_5k_6\Big]k_{456}\nonumber\\
&\phantom{===}+\frac{27}{2}\Big[2\Big(k^2_1k_{23}+k^2_2k_{13}+k^2_3k_{12}\Big)+k_1k_2k_3\Big]\Big[2\Big(k^2_4k_{56}+k^2_5k_{46}+k^2_6k_{45}\Big)+k_4k_5k_6\Big]\nonumber\\
&\phantom{===}-\frac{81}{2}k_1k_2k_3k_4k_5k_6\bigg)\nonumber\\
&\phantom{=}-18K^2k^3_{123}k^3_{456}\Bigg]\,.
\label{eq:fdef}
\end{flalign}
In the above expression $\Li_2(z)$ denotes the dilogarithm function, and $k_t\equiv\sum_{j=1}^6k_j$.

\vskip0.3cm\noindent
The sum of all regions is free of $\ve$-poles and gives \eqref{eq:full6pt} in the main text.


\section{Loop integrals for the one-loop SdSET power spectrum}
\label{app::pwrloops}

In this Appendix we provide some details on the evaluation of the loop integrals relevant to the insertion of the non-Gaussian initial condition $\Xi_{3,1}$ into the SdSET power spectrum discussed in Sec.~\ref{sec:pwr}.

To evaluate the integral 
\begin{equation}
\int\frac{\der^{d-1}l}{(2\pi)^{d-1}}\frac{1}{l^3_{\Lambda}}\log\bigg(\frac{2e^{\gamma_E}(k_{\Lambda}+l_{\Lambda})}{a_*H}\bigg)
\end{equation}
we first write
\begin{equation}
\log\bigg(\frac{2e^{\gamma_E}(k_{\Lambda}+l_{\Lambda})}{a_*H}\bigg)=\frac{\der}{\der u}\bigg|_{u=0}\bigg(\frac{2e^{\gamma_E}(k_{\Lambda}+l_{\Lambda})}{a_*H}\bigg)^{\!u}\,,
\end{equation}
then take the derivative out of the integral and employ the Mellin-Barnes representation
\begin{equation}
\frac{1}{(k_{\Lambda}+l_{\Lambda})^{-u}}=\frac{1}{\Gamma(-u)}\frac{1}{2\pi i}\int_{-i\infty}^{i\infty}\der z\;\Gamma(-u+z)\Gamma(-z)\frac{l_{\Lambda}^z}{k_{\Lambda}^{-u+z}}\,,
\label{eq::MBpwrlog}
\end{equation}
which yields the simple integral
\begin{equation}
\int\frac{\der^{d-1}l}{(2\pi)^{d-1}}\frac{1}{[l^2+\Lambda^2]^{\frac{3-z}{2}}}=\frac{\Lambda^{z-2\ve}\Gamma(-\frac{z}{2}+\ve)}{(4\pi)^{\frac{3}{2}-\ve}\Gamma(\frac{3-z}{2})}\,.
\end{equation}
After substituting this result into the remaining Mellin-Barnes integral, we close the contour in the right complex half-plane and pick up the residues at poles that survive the $\Lambda\rightarrow0$ limit. Finally, we take the derivative of the result with respect to $u$, then set $u=0$ and expand the resulting expression in $\ve$. At this point, we can also let $k_{\Lambda}\rightarrow k$, since the result features no inverse powers of $\Lambda$. In this way, we find
\begin{flalign}
&\bigg(\frac{e^{\gamma_E}\Lambda^2}{4\pi}\bigg)^{\ve}\!\int\!\frac{\der^{d-1}l}{(2\pi)^{d-1}}\frac{1}{l^3_{\Lambda}}\log\bigg(\frac{2e^{\gamma_E}(k_{\Lambda}+l_{\Lambda})}{a_*H}\bigg)\nonumber\\
&=\frac{1}{4\pi^2}\,\bigg\{\frac{1}{2\ve^2}+\frac{1}{\ve}\bigg[1+\log\bigg(\frac{e^{\gamma_E}\Lambda}{a_*H}\bigg)\bigg]+\log^2\bigg(\frac{2k}{\Lambda}\bigg)-2\log\bigg(\frac{2k}{\Lambda}\bigg)+2+\frac{5\pi^2}{24}\bigg\}\,.
\end{flalign}
For the rational part \eqref{eq:1lXi31rational} of the integrand we use the  Mellin-Barnes representations
\begin{flalign}
\frac{1}{(k_{\Lambda}+l_{\Lambda})^2}&=\frac{1}{\Gamma(2)}\frac{1}{2\pi i}\int_{-i\infty}^{i\infty}\der z\;\Gamma(2+z)\Gamma(-z)\frac{l_{\Lambda}^z}{k_{\Lambda}^{2+z}}\,,\\
\frac{1}{k_{\Lambda}^3+l_{\Lambda}^3}&=\frac{1}{\Gamma(1)}\frac{1}{2\pi i}\int_{-i\infty}^{i\infty}\der z\;\Gamma(1+z)\Gamma(-z)\frac{l_{\Lambda}^{3z}}{k_{\Lambda}^{3+3z}}\,,
\end{flalign}
and proceeding as before obtain
\begin{flalign}
\bigg(\frac{e^{\gamma_E}\Lambda^2}{4\pi}\bigg)^{\!\ve}\int\frac{\der^{d-1}l}{(2\pi)^{d-1}}\frac{1}{l^3_{\Lambda}(l_{\Lambda}+k_{\Lambda})^2}&=\frac{1}{2\pi^2k^2}\bigg[\log\bigg(\frac{2k}{\Lambda}\bigg)-2\bigg]\,,\\
\bigg(\frac{e^{\gamma_E}\Lambda^2}{4\pi}\bigg)^{\!\ve}\int\frac{\der^{d-1}l}{(2\pi)^{d-1}}\frac{1}{l^3_{\Lambda}(l^3_{\Lambda}+k^3_{\Lambda})}&=\frac{1}{2\pi^2k^3}\bigg[\log\bigg(\frac{2k}{\Lambda}\bigg)-1\bigg]\,,\\
\bigg(\frac{e^{\gamma_E}\Lambda^2}{4\pi}\bigg)^{\!\ve}\int\frac{\der^{d-1}l}{(2\pi)^{d-1}}\frac{1}{l^2_{\Lambda}(l^3_{\Lambda}+k^3_{\Lambda})}&=\frac{1}{3\sqrt{3}\pi k^2}\,,\\
\bigg(\frac{e^{\gamma_E}\Lambda^2}{4\pi}\bigg)^{\!\ve}\int\frac{\der^{d-1}l}{(2\pi)^{d-1}}\frac{1}{l_{\Lambda}(l^3_{\Lambda}+k^3_{\Lambda})}&=\frac{1}{3\sqrt{3}\pi k}\,.
\end{flalign}
To carry out the analytic continuation of the Mellin-Barnes integrands in the limit $\ve\rightarrow0$ we made use of the \texttt{Mathematica} package \texttt{MB} \cite{Czakon:2005rk}. The relevant residues were identified by hand and computed using the built-in \texttt{Mathematica} function.


\bibliography{Bibliography}{}
\end{document}